\documentclass[12pt,a4paper]{article}
\usepackage[a4paper,top=3cm,bottom=3cm,left=2cm,right=3cm,bindingoffset=5mm]{geometry}
\usepackage[utf8]{inputenc}
\usepackage[bottom]{footmisc}
\pdfoutput=1 

\usepackage{tikz}
\usepackage{jheppub}
\usepackage{xcolor}
\usepackage[makeroom]{cancel}
\usepackage{amsmath}
\usepackage{amsthm}
\usepackage{amssymb}
\usepackage{physics}
\usepackage{comment}
\usepackage{graphicx}
\usepackage{float}
\usepackage{hyperref}
\usepackage{amsfonts}
\usepackage{mathrsfs}
\usepackage{xcolor}
\usepackage{subcaption}
\usepackage{simpler-wick}
\usepackage{mathtools}
\usepackage{makecell}

\hypersetup{
colorlinks=true,         
linkcolor=blue,          
citecolor=red,        
urlcolor=blue            
}

\newcommand{\smatrixmeasure}{\int_{\eval{\phi}_{\mathcal{I}^-} = \varphi'}^{\eval{\phi}_{\mathcal{I}^+} = \varphi} \mD \phi \, e^{i S_{dS_d}[\phi]}}
\newcommand{\mIp}{{\mathcal{I}^+}}
\newcommand{\mIm}{{\mathcal{I}^-}}
\newcommand{\mD}{\mathcal{D}}
\newcommand{\rmint}{{\rm int}}
\newcommand{\tx}{\tilde{x}}

\newcommand{\pd}{\overset{\leftrightarrow}{\partial}}

\title{Path integral games with de Sitter $\alpha$-vacua}

\author{Noah Miller}

\affiliation{School of Natural Sciences, Institute for Advanced Study,
1 Einstein Drive, Princeton, NJ 08540 USA}
\affiliation{Princeton Gravity Initiative, Princeton University,
Jadwin Hall, Washington Road, Princeton NJ 08544, USA}
\emailAdd{noahmiller@ias.edu}

\preprint{}

\abstract{

The $\alpha$-vacua are a 1-parameter family of quantum field vacua in de Sitter space which are invariant under the isometry group $SO(1,d)$. In this work give a path integral construction of the de Sitter $\alpha$-vacua. We explain that these states can be prepared by acting on the Bunch-Davies vacuum with a certain non-local charge operator. While most conserved charges live on a single codimension-1 manifold, we show that this particular charge lives on a pair of two codimension-1 manifolds which are antipodal mirrors of each other. The rules for the manipulation of this charge as an insertion in the path integral are explained. We further explain how this charge can be used to solve for the wavefunctionals of the $\alpha$-vacua at $\mIp$ (in the regime that $\alpha$ is small) by deforming the equator of de Sitter space to $\mathcal{I}^+$ / $\mathcal{I}^-$.

We also discuss the special $\alpha$-vacua known as the ``in'' and ``out'' vacua, for both heavy and light scalars. It is well known that the ``in'' and ``out'' vacua are equal in odd dimensions, but we also show that they are equal in even dimensions when $\sqrt{(d-1)^2/4 - m^2}$ is a half-integer.

Finally, we investigate the question of whether the $\alpha$-vacua can be constructed in interacting quantum field theories. Using the fact that the antipodal charge operator is only conserved in a free theory, we give a symmetry based argument that in general they cannot be, explaining why interactions and $SO(1,d)$ invariance of the vacua are in tension.

}

\begin{document}

\maketitle 

\section{Introduction}

A free quantum field theory of a scalar field in Minkowski space or Anti-de Sitter space has only a single vacuum state that is invariant under the full isometry group of the spacetime. The situation is different in de Sitter space, where there exists a 1-parameter family of states, called the $\alpha$-vacua, which are invariant under the full de Sitter isometry group $SO(1,d)$ \cite{Chernikov:1968zm,schomblond1976conditions,Mottola:1984ar,Allen:1985ux,Burges:1984qm}. The special case $\ket{\alpha = 0}$ is the Bunch-Davies vacuum \cite{Bunch:1978yq}, also called the euclidean vacuum or the Hartle-Hawking vacuum \cite{Hartle:1983ai}.

These states have become a somewhat niche topic of recurring interest since their discovery in 1968 \cite{Chernikov:1968zm}. Because the $\alpha$-vacua are homogeneous and isotropic, many works have considered the possibility that the $\alpha$-vacua (or states which look like the $\alpha$-vacua above some short distance cut off) could have been initial conditions for the inflaton field in the early universe, and investigated what high energy physics would be imprinted in these states
\cite{Kaloper:2002cs,deBoer:2004nd,Danielsson:2002kx,Danielsson:2002qh,Easther:2001fi,Easther:2001fz,Easther:2002xe,Collins:2005cm}.  In quantum cosmology, \cite{Chen:2024ckx} found $\alpha$-like states emerging in a system of two dS spacetimes connected by a euclidean wormhole. The $\alpha$-vacua were also conjectured to be dual to marginal deformations of a dual CFT in the context of the dS/CFT correspondence \cite{Bousso:2001mw, Ng:2012xp, Spradlin:2001nb}. In the field of celestial holography, analog $\alpha$-vacua have recently been defined in flat space which are not invariant under the full Poincaré group but are invariant under the Lorentz subgroup \cite{Melton:2023dee}. See \cite{Joung:2006gj,Joung:2007je} for a group theory construction of the $\alpha$-vacua. See also \cite{Kanno:2014lma,Polyakov:2007mm,Nguyen:2017ggc,Chopping:2024oiu,Ansari:2024pgq,Cotler:2023xku}.

The purpose of this note is, simply, to understand how the $\alpha$-vacua are constructed using the Feynman path integral.

In 1984, it was noted by Burges\footnote{ who referenced an unpublished note by Gross and Witten} \cite{Burges:1984qm} that there exists a certain non-local charge operator $\hat{Q}^A$ which generates the $\alpha$-vacua from the Bunch-Davies vacuum via $\ket{\alpha} = e^{i \alpha \hat{Q}^A} \ket{0}$. However, something was conspicuously missing from Burges' paper which motivated this present investigation: it was not explained how the codimension-1 surface on which this charge was defined could be continuously deformed, in the way that surfaces can naturally be deformed when charges are conserved.

Let us be more precise. If we cover de Sitter space $dS_d = \mathbb{R} \times S_{d-1}$ with global coordinates $(t,\Omega)$, with $\Omega \in S_{d-1}$, as
\begin{equation}
    ds^2 = -dt^2 + \cosh^2(t) d \Omega_{d-1}^2
\end{equation}
then Burges gave an expression for the charge that only holds on the $t = 0$ slice:
\begin{equation}
    \hat{Q}^A = \int d^{d-1} \Omega \, \hat{\phi}(0,\Omega) \hat{\pi}(0,\Omega^A).
\end{equation}
Here $\Omega^A$ is the point antipodal to $\Omega$ on the $S_{d-1}$ sphere. In this work we will explain how this charge can be defined beyond the $t = 0$ slice. In fact, we will see that this charge actually lives on \textit{two} antipodal slices in spacetime, and will study the consequences of this fact. This fact did not appear in Burges' original analysis because the $t = 0$ slice happens to be its own antipodal surface.

In section \ref{sec2} we review global coordinates in de Sitter spacetime and in its euclidean analytic continuation.

In section \ref{sec3} we review the path integral construction of the Bunch-Davies vacuum for a free scalar on a de Sitter background using the euclidean path integral on the half-sphere. We explain why this state is invariant under the dS isometry group. 

In section \ref{sec4} we discuss the well known fact that annihilation operators constructed out of ``euclidean modes'' (solutions to the dS Klein-Gordon equation which are non-singular when analytically continued to the lower half euclidean sphere) annihilate the Bunch-Davies state. We prove this fact with Noether's theorem, which we are able to do because annihilation operators are the Noether charges of a linear symmetry that only exists in free theories. 

In section \ref{sec5} we discuss the antipodal map in de Sitter space, and use it to construct a non-local symmetry variation $\delta \phi = \epsilon \, \phi^A$ of the free scalar field. We calculate the associated classical conserved current using Noether's theorem. This conserved current is unusual in that it is non-local, in the sense that the current evaluated a point in de Sitter space depends on the value of the field and its derivatives at that point and at the antipodal point. After quantization, we show that this classical conserved current is indeed promoted to a conserved insertion in the path integral à la a Ward–Takahashi identity, although the proof of this is slightly subtler than one might expect. 

In section \ref{sec6} we explain ``why'' extra dS isometry invariant states can be constructed aside from the Bunch-Davies vacuum. To briefly outline the argument, one would expect that a dS isometry invariant state could be constructed by acting on the Bunch-Davies state by some conserved charge, producing a state like $\hat{Q} \ket{0}$ for some $\hat{Q}$ which is defined as the integral over a codimension-1 surface of a locally-defined conserved current. However, the Bunch-Davies state is prepared by a path integral over the half-sphere, and the half-sphere is a simply-connected space, meaning the surface on which the conserved charge lives (the equator) is able to be contracted to zero, implying $\hat{Q} \ket{0} = 0$. However, due to the strange nature of the non-local ``antipodal'' charge $\hat{Q}^A$, the equator together with its antipodal mirror are not able to be contracted on the lower half-sphere, so $\hat{Q}^A \ket{0} \neq 0$ and we can construct a family of dS-invariant states $\ket{\alpha} = e^{i \alpha \hat{Q}^A} \ket{0}$.

In section \ref{sec7} we explain which $\alpha \in \mathbb{C}$ result in $\ket{\alpha}$ being a normalizable state.

In section \ref{sec8} we turn our attention to the so-called ``in'' and ``out'' vacuum states, which are special $\alpha$-vacua that contain no particles in the infinite past/infinite future. Using a simple analytic argument, we are able to calculate which values of $\alpha$ produce these states. The argument uses the fact that the antipodal map in de Sitter space can be continuously connected to the identity via an analytic continuation through the euclidean section $t \mapsto t + i \pi$. This argument works for all spacetime dimensions $d$ and mass  $m$. While it is well known that for heavy masses $m > \tfrac{d-1}{2}$ the in vacuum equals the out vacuum in odd dimensions \cite{Bousso:2001mw,Lagogiannis:2011st}, we show that for light masses $m < \tfrac{d-1}{2}$ the in vacuum also equals the out vacuum when $\sqrt{(d-1)^2/4 - m^2}$ is a half-integer. This includes the massless and conformally coupled scalars. 

In section \ref{sec9} we discuss the $\alpha$-vacua wavefunctionals at the infinite future of de Sitter space $\mathcal{I}^+$. We use a path integral argument to find the operation which maps the Bunch-Davies wavefunctional at $\mathcal{I}^+$ to the $\alpha$-vacuum wavefunctional at $\mathcal{I}^+$ when $\alpha$ is small. If $\alpha$ is finite, operator ordering problems prevent us from using our path integral argument to solve for the wavefunctional at $\mathcal{I}^+$, which we explain.

In section \ref{sec10} we address the question of whether or not the $\alpha$-vacua can be defined in interacting field theories. We argue that they cannot be. This is congruous with the findings of the previous works \cite{Collins:2003zv,Danielsson:2002mb,Einhorn:2002nu,Einhorn:2003xb,Goldstein:2003ut,Goldstein:2003qf,Goldstein:2005re,Akhmedov:2013vka,Kaloper:2002cs,Banks:2002nv,Akhmedov:2022uug}.

\section{Global coordinates in lorentzian and euclidean de Sitter}\label{sec2}

Let us define a $d+1$ dimensional Minkowski embedding space with coordinates $X^M$ for $M = 0, \ldots, d$ and metric $ds^2 = -(d X^0)^2 + (d X^1)^2 + \ldots + (d X^d)^2 = \eta_{MN} dX^M dX^N$. $d$-dimensional de Sitter space $dS_d$ can be defined as a hyperboloid cut out of the embedding space
\begin{equation}
    -(X^0)^2+ (X^1)^2 + \ldots + (X^d)^2 = 1
\end{equation}
where we have set the de Sitter length scale to 1. Euclidean de Sitter space can then be defined via the analytic continuation
\begin{equation}\label{Xtilde}
    X^0 = -i \tilde{X}^{0}
\end{equation}
where $\tilde{X}^{0}$ is taken to be real. Euclidean de Sitter space is a $d$ dimensional sphere $S_d$
\begin{equation}
    (\tilde{X}^{0})^2+ (X^1)^2 + \ldots + (X^d)^2 = 1
\end{equation}
in an embedding space with metric $ds^2 = (d \tilde{X}^{0})^2 + (dX^1)^2 + \ldots + (dX^d)^2$.
Let us also define the lower half of the sphere $S_{d,-} \subset S_d$ to be
\begin{equation}
    S_{d,-} \equiv \{ \tilde{X}^0, \ldots, X^d \; | \;  \tilde{X}^0 \leq 0, (\tilde{X}^{0})^2+ \ldots + (X^d)^2 = 1 \}
\end{equation}
and the $X^0 = \tilde{X}^0 = 0$ ``equator'' $E$ to be
\begin{equation}
    E \equiv \{ X^0, \ldots, X^d \; | \;  X^0 = 0, (X^{1})^2+ \ldots + (X^d)^2 = 1 \}.
\end{equation}
The equator $E$ can be understood as a subset of either the lorentzian $dS_d$ or the euclidean $S_d$.

\begin{figure}
    \centering
    \tikzset{every picture/.style={line width=0.75pt}} 

\begin{tikzpicture}[x=0.75pt,y=0.75pt,yscale=-0.9,xscale=0.9]

\draw   (483.93,118.84) .. controls (483.93,99.61) and (499.52,84.02) .. (518.75,84.02) .. controls (537.98,84.02) and (553.57,99.61) .. (553.57,118.84) .. controls (553.57,138.07) and (537.98,153.66) .. (518.75,153.66) .. controls (499.52,153.66) and (483.93,138.07) .. (483.93,118.84) -- cycle ;
\draw  [draw opacity=0][line width=1.5]  (553.57,118.84) .. controls (553.57,124.35) and (537.98,128.81) .. (518.75,128.81) .. controls (499.52,128.81) and (483.93,124.35) .. (483.93,118.84) -- (518.75,118.84) -- cycle ; \draw  [line width=1.5]  (553.57,118.84) .. controls (553.57,124.35) and (537.98,128.81) .. (518.75,128.81) .. controls (499.52,128.81) and (483.93,124.35) .. (483.93,118.84) ;  
\draw  [draw opacity=0][dash pattern={on 5.63pt off 4.5pt}][line width=1.5]  (483.93,118.84) .. controls (483.93,113.34) and (499.52,108.88) .. (518.75,108.88) .. controls (537.98,108.88) and (553.57,113.34) .. (553.57,118.84) -- (518.75,118.84) -- cycle ; \draw  [dash pattern={on 5.63pt off 4.5pt}][line width=1.5]  (483.93,118.84) .. controls (483.93,113.34) and (499.52,108.88) .. (518.75,108.88) .. controls (537.98,108.88) and (553.57,113.34) .. (553.57,118.84) ;  
\draw    (51,32) -- (51,72) ;
\draw [shift={(51,29)}, rotate = 90] [fill={rgb, 255:red, 0; green, 0; blue, 0 }  ][line width=0.08]  [draw opacity=0] (8.93,-4.29) -- (0,0) -- (8.93,4.29) -- cycle    ;
\draw    (94,72) -- (51,72) ;
\draw [shift={(97,72)}, rotate = 180] [fill={rgb, 255:red, 0; green, 0; blue, 0 }  ][line width=0.08]  [draw opacity=0] (8.93,-4.29) -- (0,0) -- (8.93,4.29) -- cycle    ;
\draw    (39.3,96.3) -- (51,72) ;
\draw [shift={(38,99)}, rotate = 295.71] [fill={rgb, 255:red, 0; green, 0; blue, 0 }  ][line width=0.08]  [draw opacity=0] (8.93,-4.29) -- (0,0) -- (8.93,4.29) -- cycle    ;
\draw    (421,32) -- (421,72) ;
\draw [shift={(421,29)}, rotate = 90] [fill={rgb, 255:red, 0; green, 0; blue, 0 }  ][line width=0.08]  [draw opacity=0] (8.93,-4.29) -- (0,0) -- (8.93,4.29) -- cycle    ;
\draw    (464,72) -- (421,72) ;
\draw [shift={(467,72)}, rotate = 180] [fill={rgb, 255:red, 0; green, 0; blue, 0 }  ][line width=0.08]  [draw opacity=0] (8.93,-4.29) -- (0,0) -- (8.93,4.29) -- cycle    ;
\draw    (409.3,96.3) -- (421,72) ;
\draw [shift={(408,99)}, rotate = 295.71] [fill={rgb, 255:red, 0; green, 0; blue, 0 }  ][line width=0.08]  [draw opacity=0] (8.93,-4.29) -- (0,0) -- (8.93,4.29) -- cycle    ;
\draw  [draw opacity=0][line width=1.5]  (220.7,118.95) .. controls (220.7,124.46) and (205.11,128.92) .. (185.88,128.92) .. controls (166.65,128.92) and (151.06,124.46) .. (151.06,118.95) -- (185.88,118.95) -- cycle ; \draw  [line width=1.5]  (220.7,118.95) .. controls (220.7,124.46) and (205.11,128.92) .. (185.88,128.92) .. controls (166.65,128.92) and (151.06,124.46) .. (151.06,118.95) ;  
\draw  [draw opacity=0][dash pattern={on 5.63pt off 4.5pt}][line width=1.5]  (151.06,118.95) .. controls (151.06,113.45) and (166.65,108.99) .. (185.88,108.99) .. controls (205.11,108.99) and (220.7,113.45) .. (220.7,118.95) -- (185.88,118.95) -- cycle ; \draw  [dash pattern={on 5.63pt off 4.5pt}][line width=1.5]  (151.06,118.95) .. controls (151.06,113.45) and (166.65,108.99) .. (185.88,108.99) .. controls (205.11,108.99) and (220.7,113.45) .. (220.7,118.95) ;  
\draw    (118,48.88) .. controls (160.91,86.24) and (160.91,153.07) .. (118,189.03) ;
\draw    (253.76,48.87) .. controls (210.85,86.24) and (210.85,153.07) .. (253.76,189.03) ;
\draw  [draw opacity=0][line width=0.75]  (253.76,45.96) .. controls (253.76,45.96) and (253.76,45.96) .. (253.76,45.96) .. controls (253.76,55.09) and (223.33,62.5) .. (185.79,62.5) .. controls (148.25,62.5) and (117.82,55.09) .. (117.82,45.96) -- (185.79,45.96) -- cycle ; \draw  [line width=0.75]  (253.76,45.96) .. controls (253.76,45.96) and (253.76,45.96) .. (253.76,45.96) .. controls (253.76,55.09) and (223.33,62.5) .. (185.79,62.5) .. controls (148.25,62.5) and (117.82,55.09) .. (117.82,45.96) ;  
\draw  [draw opacity=0][line width=0.75]  (117.82,48.88) .. controls (117.82,48.88) and (117.82,48.88) .. (117.82,48.87) .. controls (117.82,41.17) and (148.25,34.92) .. (185.79,34.92) .. controls (223.33,34.92) and (253.76,41.17) .. (253.76,48.87) -- (185.79,48.87) -- cycle ; \draw  [line width=0.75]  (117.82,48.88) .. controls (117.82,48.88) and (117.82,48.88) .. (117.82,48.87) .. controls (117.82,41.17) and (148.25,34.92) .. (185.79,34.92) .. controls (223.33,34.92) and (253.76,41.17) .. (253.76,48.87) ;  
\draw  [draw opacity=0][line width=0.75]  (253.94,189.03) .. controls (253.94,189.03) and (253.94,189.03) .. (253.94,189.03) .. controls (253.94,198.17) and (223.51,205.57) .. (185.97,205.57) .. controls (148.43,205.57) and (118,198.17) .. (118,189.03) -- (185.97,189.03) -- cycle ; \draw  [line width=0.75]  (253.94,189.03) .. controls (253.94,189.03) and (253.94,189.03) .. (253.94,189.03) .. controls (253.94,198.17) and (223.51,205.57) .. (185.97,205.57) .. controls (148.43,205.57) and (118,198.17) .. (118,189.03) ;  
\draw  [draw opacity=0][dash pattern={on 4.5pt off 4.5pt}][line width=0.75]  (117.82,189.03) .. controls (117.82,189.03) and (117.82,189.03) .. (117.82,189.03) .. controls (117.82,181.32) and (148.25,175.08) .. (185.79,175.08) .. controls (223.33,175.08) and (253.76,181.32) .. (253.76,189.03) -- (185.79,189.03) -- cycle ; \draw  [dash pattern={on 4.5pt off 4.5pt}][line width=0.75]  (117.82,189.03) .. controls (117.82,189.03) and (117.82,189.03) .. (117.82,189.03) .. controls (117.82,181.32) and (148.25,175.08) .. (185.79,175.08) .. controls (223.33,175.08) and (253.76,181.32) .. (253.76,189.03) ;  
\draw  [color={rgb, 255:red, 255; green, 255; blue, 255 }  ,draw opacity=1 ][fill={rgb, 255:red, 255; green, 255; blue, 255 }  ,fill opacity=1 ] (589.5,115.75) .. controls (589.5,110.37) and (593.87,106) .. (599.25,106) .. controls (604.63,106) and (609,110.37) .. (609,115.75) .. controls (609,121.13) and (604.63,125.5) .. (599.25,125.5) .. controls (593.87,125.5) and (589.5,121.13) .. (589.5,115.75) -- cycle ;

\draw (556.08,117.62) node [anchor=west] [inner sep=0.75pt]    {$E$};
\draw (525.02,160.91) node [anchor=north west][inner sep=0.75pt]  [font=\footnotesize]  {$S_{d,-}$};
\draw (525.46,80.56) node [anchor=south west] [inner sep=0.75pt]  [font=\footnotesize]  {$S_{d,+}$};
\draw (421,25.6) node [anchor=south] [inner sep=0.75pt]  [font=\footnotesize]  {$\tilde{X}^{0}$};
\draw (36,102.4) node [anchor=north east] [inner sep=0.75pt]  [font=\footnotesize]  {$X^{a}$};
\draw (51,25.6) node [anchor=south] [inner sep=0.75pt]  [font=\footnotesize]  {$X^{0}$};
\draw (406,102.4) node [anchor=north east] [inner sep=0.75pt]  [font=\footnotesize]  {$X^{a}$};
\draw (227.43,117.03) node [anchor=west] [inner sep=0.75pt]    {$E$};

\end{tikzpicture}
    \caption{\label{lorentzianandeuclidean}Lorentzian de Sitter space $dS_d$ and euclidean de Sitter space $S_d$. The equator $E$, defined at $X^0 = \tilde{X}^0 = 0$, exists on both. The halves of the euclidean sphere above and below $E$ are denoted $S_{d,+}$ and $S_{d,-}$, respectively.}
\end{figure}
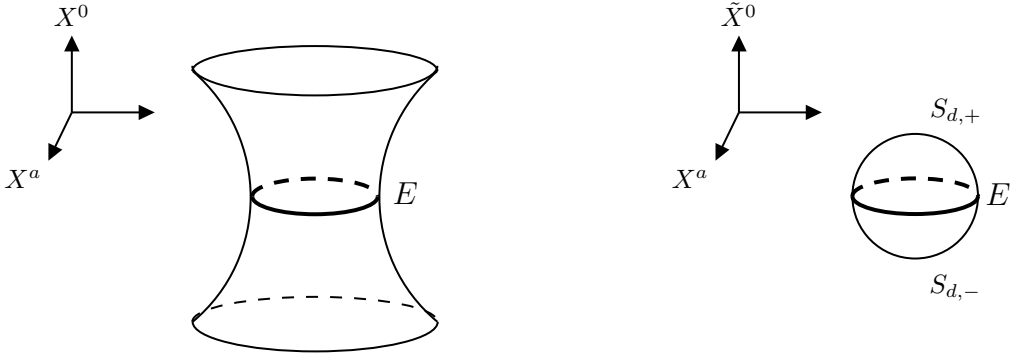

Let $\Omega$ denote a set of sphere coordinates which cover the unit sphere $S_{d-1}$ with metric $d \Omega_{d-1}^2$, and let $n^a(\Omega)$ with $a = 1, \ldots, d$ be the unit vector on the sphere parameterized by $\Omega$ with $n(\Omega)^2 = 1$. Then $dS_d$ may be parameterized with global coordinates $x^\mu = (t,\Omega)$, for $\mu = 0, \ldots, d-1$, by
\begin{equation}\label{XM}
    \begin{aligned}
        X^0(t,\Omega) &= \sinh(t) \\
        X^a(t,\Omega) &= \cosh(t) n^a(\Omega)
    \end{aligned}
\end{equation}
with metric
\begin{equation}
    ds^2 = - dt^2 + \cosh^2(t) d \Omega^2_{d-1}.
\end{equation}

If we analytically continue $t = -i \tilde{t}$ and take $\tilde{t}$ to be real, we make $X^0$ imaginary just as in \eqref{Xtilde}. Euclidean de Sitter space $S_d$ is then parameterized as
\begin{equation}
    \begin{aligned}
        \tilde{X}^0(- i\tilde{t}, \Omega) &= \sin(\tilde{t}) \\
        X^a(- i \tilde{t},\Omega) &= \cos(\tilde{t}) n^a(\Omega)
    \end{aligned}
\end{equation}
where $\tilde{t} \in [-\pi/2, \pi/2]$. $\tilde{t} = -\pi/2$ is the south pole and $\tilde{t} = \pi/2$ is the north pole. The euclidean metric is
\begin{equation}
        ds^2 = d\tilde{t}^2 + \cos^2(\tilde{t}) d \Omega^2_{d-1}.
\end{equation}
We denote $\tx^\mu = (\tilde{t},\Omega)$.

\section{The Bunch-Davies state}\label{sec3}

In this section we review the path integral construction of the Bunch-Davies state in $dS_d$. The euclidean action for a free scalar field $\phi$ of mass $m$ on a spacetime patch of the euclidean sphere $U \subset S_d$ is
\begin{equation}\label{SfreeU}
    S_U[\phi] = \frac{1}{2}  \int_U d^d \tx \sqrt{g} \left( ( \nabla \phi)^2 +  m^2 \phi^2 \right).
\end{equation}

The Bunch-Davies state, denoted as $\ket{0}$, can be defined on the $\tilde{t} = 0$ time slice $E$ as the euclidean path integral over the half-sphere with open boundary conditions. If $\varphi(\Omega)$ is some function on the equatorial $d-1$ sphere, then let us define the field ``position eigenvector'' state $\ket{\varphi}$ to be the state that satisfies
\begin{equation}
    \hat{\phi}(0,\Omega) \ket{\varphi} = \varphi(\Omega) \ket{\varphi}.
\end{equation}
The Bunch-Davies wavefunctional evaluated at $\varphi$ is then given by the path integral over all fields $\phi$ on the half-sphere which are equal to $\varphi$ on $E$. See figure \ref{pic3fig}.
\begin{equation}
    \Psi^E_0[\varphi] = \braket{\varphi}{0} = \int_{\eval{\phi}_E = \, \varphi } \mD \phi \; e^{- S_{S_{d,-}}[\phi]}.
\end{equation}

\begin{figure}
    \centering
    \tikzset{every picture/.style={line width=0.75pt}} 

\begin{tikzpicture}[x=0.75pt,y=0.75pt,yscale=-1,xscale=1]

\draw  [draw opacity=0][line width=0.75]  (393,59.52) .. controls (393,59.52) and (393,59.52) .. (393,59.52) .. controls (393,63.93) and (377.27,67.5) .. (357.87,67.5) .. controls (338.47,67.5) and (322.74,63.93) .. (322.74,59.52) -- (357.87,59.52) -- cycle ; \draw  [line width=0.75]  (393,59.52) .. controls (393,59.52) and (393,59.52) .. (393,59.52) .. controls (393,63.93) and (377.27,67.5) .. (357.87,67.5) .. controls (338.47,67.5) and (322.74,63.93) .. (322.74,59.52) ;  
\draw  [draw opacity=0] (392.57,59.84) .. controls (392.57,79.13) and (376.94,94.76) .. (357.66,94.76) .. controls (338.37,94.76) and (322.74,79.13) .. (322.74,59.84) -- (357.66,59.84) -- cycle ; \draw   (392.57,59.84) .. controls (392.57,79.13) and (376.94,94.76) .. (357.66,94.76) .. controls (338.37,94.76) and (322.74,79.13) .. (322.74,59.84) ;  
\draw  [draw opacity=0][line width=0.75]  (322.74,59.52) .. controls (322.74,59.52) and (322.74,59.52) .. (322.74,59.52) .. controls (322.74,55.11) and (338.47,51.53) .. (357.87,51.53) .. controls (377.27,51.53) and (393,55.11) .. (393,59.52) -- (357.87,59.52) -- cycle ; \draw  [line width=0.75]  (322.74,59.52) .. controls (322.74,59.52) and (322.74,59.52) .. (322.74,59.52) .. controls (322.74,55.11) and (338.47,51.53) .. (357.87,51.53) .. controls (377.27,51.53) and (393,55.11) .. (393,59.52) ;  

\draw (357.9,49.99) node [anchor=south] [inner sep=0.75pt]    {$\varphi $};
\draw (317.74,66.52) node [anchor=east] [inner sep=0.75pt]    {$\Psi^E _{0}[ \varphi ] =$};

\end{tikzpicture}
    \caption{\label{pic3fig} Definition of the Bunch-Davies wavefunctional $\Psi^E_0[\varphi]$ as a euclidean path integral on the half-sphere with boundary conditions $\eval{\phi}_{E} = \varphi$.}
\end{figure}

The group of isometries of $dS_d$ is $SO(1,d)$. Let us define the Lie algebra $\mathfrak{so}(1,d)$ vector fields as
\begin{equation}
    L^{MN} \equiv \eta^{NP} X^M \pdv{X^P} -  \eta^{MP}X^N \pdv{X^P}
\end{equation}
and denote $\hat{L}^{MN}$ to be the corresponding quantum operators. Generators of the form $L^{0M}$ are ``boost'' vector fields, while generators of the form $L^{MN}$ with $M, N = 1, \ldots, d$ are ``spatial rotation'' vector fields which generate a $SO(d) \subset SO(1,d)$ subgroup. Only the spatial rotations map $E$ into itself.

One might wonder if $\ket{0}$ is invariant under all of these symmetries. Clearly, $\ket{0}$ is invariant under the $SO(d)$ subgroup because the euclidean half-sphere itself is symmetric under these rotations. But what about the boost generators $\hat{L}^{0M}$? Do they annihilate $\ket{0}$? Without loss of generality we focus on $\hat{L}^{01}$.
\begin{figure}[b]
    \centering
    \tikzset{every picture/.style={line width=0.75pt}} 

\begin{tikzpicture}[x=0.75pt,y=0.75pt,yscale=-0.75,xscale=0.75]

\draw   (483.93,118.84) .. controls (483.93,99.61) and (499.52,84.02) .. (518.75,84.02) .. controls (537.98,84.02) and (553.57,99.61) .. (553.57,118.84) .. controls (553.57,138.07) and (537.98,153.66) .. (518.75,153.66) .. controls (499.52,153.66) and (483.93,138.07) .. (483.93,118.84) -- cycle ;
\draw    (81,119) -- (38,119) ;
\draw [shift={(84,119)}, rotate = 180] [fill={rgb, 255:red, 0; green, 0; blue, 0 }  ][line width=0.08]  [draw opacity=0] (6.25,-3) -- (0,0) -- (6.25,3) -- cycle    ;
\draw    (118,47.88) .. controls (160.91,85.24) and (160.91,152.07) .. (118,188.03) ;
\draw    (253.76,47.87) .. controls (210.85,85.24) and (210.85,152.07) .. (253.76,188.03) ;
\draw  [draw opacity=0][line width=0.75]  (253.76,47.88) .. controls (253.76,52.44) and (223.33,56.15) .. (185.79,56.15) .. controls (148.25,56.15) and (117.82,52.44) .. (117.82,47.88) -- (185.79,47.88) -- cycle ; \draw  [line width=0.75]  (253.76,47.88) .. controls (253.76,52.44) and (223.33,56.15) .. (185.79,56.15) .. controls (148.25,56.15) and (117.82,52.44) .. (117.82,47.88) ;  
\draw    (38,76) -- (38,119) ;
\draw [shift={(38,73)}, rotate = 90] [fill={rgb, 255:red, 0; green, 0; blue, 0 }  ][line width=0.08]  [draw opacity=0] (6.25,-3) -- (0,0) -- (6.25,3) -- cycle    ;
\draw    (439,119) -- (396,119) ;
\draw [shift={(442,119)}, rotate = 180] [fill={rgb, 255:red, 0; green, 0; blue, 0 }  ][line width=0.08]  [draw opacity=0] (6.25,-3) -- (0,0) -- (6.25,3) -- cycle    ;
\draw    (396,76) -- (396,119) ;
\draw [shift={(396,73)}, rotate = 90] [fill={rgb, 255:red, 0; green, 0; blue, 0 }  ][line width=0.08]  [draw opacity=0] (6.25,-3) -- (0,0) -- (6.25,3) -- cycle    ;
\draw  [draw opacity=0][line width=0.75]  (118,47.88) .. controls (118,47.88) and (118,47.87) .. (118,47.87) .. controls (118,43.31) and (148.43,39.6) .. (185.97,39.6) .. controls (223.51,39.6) and (253.94,43.31) .. (253.94,47.87) -- (185.97,47.87) -- cycle ; \draw  [line width=0.75]  (118,47.88) .. controls (118,47.88) and (118,47.87) .. (118,47.87) .. controls (118,43.31) and (148.43,39.6) .. (185.97,39.6) .. controls (223.51,39.6) and (253.94,43.31) .. (253.94,47.87) ;  
\draw  [draw opacity=0][line width=0.75]  (253.76,188.88) .. controls (253.76,188.88) and (253.76,188.88) .. (253.76,188.88) .. controls (253.76,193.44) and (223.33,197.15) .. (185.79,197.15) .. controls (148.25,197.15) and (117.82,193.44) .. (117.82,188.88) -- (185.79,188.88) -- cycle ; \draw  [line width=0.75]  (253.76,188.88) .. controls (253.76,188.88) and (253.76,188.88) .. (253.76,188.88) .. controls (253.76,193.44) and (223.33,197.15) .. (185.79,197.15) .. controls (148.25,197.15) and (117.82,193.44) .. (117.82,188.88) ;  
\draw  [draw opacity=0][dash pattern={on 4.5pt off 4.5pt}][line width=0.75]  (118,188.88) .. controls (118,184.31) and (148.43,180.6) .. (185.97,180.6) .. controls (223.51,180.6) and (253.94,184.31) .. (253.94,188.88) -- (185.97,188.88) -- cycle ; \draw  [dash pattern={on 4.5pt off 4.5pt}][line width=0.75]  (118,188.88) .. controls (118,184.31) and (148.43,180.6) .. (185.97,180.6) .. controls (223.51,180.6) and (253.94,184.31) .. (253.94,188.88) ;  
\draw  [fill={rgb, 255:red, 0; green, 0; blue, 0 }  ,fill opacity=1 ] (515.25,118.84) .. controls (515.25,116.91) and (516.82,115.34) .. (518.75,115.34) .. controls (520.68,115.34) and (522.25,116.91) .. (522.25,118.84) .. controls (522.25,120.78) and (520.68,122.34) .. (518.75,122.34) .. controls (516.82,122.34) and (515.25,120.78) .. (515.25,118.84) -- cycle ;
\draw  [fill={rgb, 255:red, 0; green, 0; blue, 0 }  ,fill opacity=1 ] (182.56,118.37) .. controls (182.56,116.44) and (184.12,114.87) .. (186.06,114.87) .. controls (187.99,114.87) and (189.56,116.44) .. (189.56,118.37) .. controls (189.56,120.31) and (187.99,121.87) .. (186.06,121.87) .. controls (184.12,121.87) and (182.56,120.31) .. (182.56,118.37) -- cycle ;
\draw  [draw opacity=0] (562.25,94.07) .. controls (566.42,101.38) and (568.8,109.83) .. (568.8,118.84) .. controls (568.8,130.11) and (565.08,140.5) .. (558.8,148.86) -- (518.75,118.84) -- cycle ; \draw   (562.25,94.07) .. controls (566.42,101.38) and (568.8,109.83) .. (568.8,118.84) .. controls (568.8,130.11) and (565.08,140.5) .. (558.8,148.86) ;  
\draw    (562,94) .. controls (564.23,90.89) and (564.47,96.48) .. (560.62,91.41) ;
\draw [shift={(559,89)}, rotate = 58.57] [fill={rgb, 255:red, 0; green, 0; blue, 0 }  ][line width=0.08]  [draw opacity=0] (10.72,-5.15) -- (0,0) -- (10.72,5.15) -- (7.12,0) -- cycle    ;
\draw  [draw opacity=0] (478.69,148.85) .. controls (472.42,140.49) and (468.7,130.1) .. (468.7,118.84) .. controls (468.7,109.46) and (471.29,100.68) .. (475.78,93.17) -- (518.75,118.84) -- cycle ; \draw   (478.69,148.85) .. controls (472.42,140.49) and (468.7,130.1) .. (468.7,118.84) .. controls (468.7,109.46) and (471.29,100.68) .. (475.78,93.17) ;  
\draw    (478.69,148.85) .. controls (480.83,145.85) and (478.58,147.25) .. (480.81,151.68) ;
\draw [shift={(482.38,154.19)}, rotate = 233.54] [fill={rgb, 255:red, 0; green, 0; blue, 0 }  ][line width=0.08]  [draw opacity=0] (10.72,-5.15) -- (0,0) -- (10.72,5.15) -- (7.12,0) -- cycle    ;
\draw  [draw opacity=0] (259.44,159.18) .. controls (250.81,149.34) and (245.5,135.97) .. (245.5,121.25) .. controls (245.5,108.19) and (249.67,96.21) .. (256.64,86.8) -- (295.5,121.25) -- cycle ; \draw   (259.44,159.18) .. controls (250.81,149.34) and (245.5,135.97) .. (245.5,121.25) .. controls (245.5,108.19) and (249.67,96.21) .. (256.64,86.8) ;  
\draw  [draw opacity=0] (117.63,88.58) .. controls (123.83,97.7) and (127.5,109.01) .. (127.5,121.25) .. controls (127.5,135.74) and (122.36,148.91) .. (113.97,158.7) -- (77.5,121.25) -- cycle ; \draw   (117.63,88.58) .. controls (123.83,97.7) and (127.5,109.01) .. (127.5,121.25) .. controls (127.5,135.74) and (122.36,148.91) .. (113.97,158.7) ;  
\draw    (257.83,85.34) .. controls (260.05,82.22) and (255.54,88.22) .. (259.46,83.58) ;
\draw [shift={(261.33,81.34)}, rotate = 129.47] [fill={rgb, 255:red, 0; green, 0; blue, 0 }  ][line width=0.08]  [draw opacity=0] (10.72,-5.15) -- (0,0) -- (10.72,5.15) -- (7.12,0) -- cycle    ;
\draw    (113.97,158.7) .. controls (114.45,157.61) and (117.77,153.83) .. (112.5,159.89) ;
\draw [shift={(110.55,162.14)}, rotate = 311.02] [fill={rgb, 255:red, 0; green, 0; blue, 0 }  ][line width=0.08]  [draw opacity=0] (10.72,-5.15) -- (0,0) -- (10.72,5.15) -- (7.12,0) -- cycle    ;

\draw (36,73) node [anchor=east] [inner sep=0.75pt]  [font=\footnotesize]  {$X^{0}$};
\draw (84,113.6) node [anchor=south east] [inner sep=0.75pt]  [font=\footnotesize]  {$X^{1}$};
\draw (394,73) node [anchor=east] [inner sep=0.75pt]  [font=\footnotesize]  {$\tilde{X}^{0}$};
\draw (442,113.6) node [anchor=south east] [inner sep=0.75pt]  [font=\footnotesize]  {$X^{1}$};
\draw (189,22) node    {$L^{01}$};
\draw (518,23) node    {$-iL^{01}$};

\end{tikzpicture}
    \caption{\label{figrotation} The Killing vector field $L^{01}$ generates a boost in lorentzian de Sitter and a rotation in euclidean de Sitter.}
\end{figure}
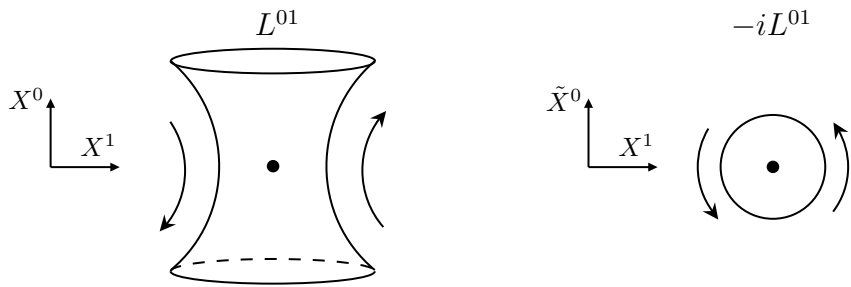
Notice that in the euclidean section, the vector field is
\begin{equation}
    - i L^{01} = - \tilde{X}^0 \pdv{X^1} + X^1 \pdv{\tilde{X}^0}
\end{equation}
and generates a $U(1)$ rotation in the $\tilde{X}^0$-$X^1$ plane. See figure \ref{figrotation}. This means that, if $\epsilon$ is a tiny angle, then in euclidean de Sitter space the operator $1 - \epsilon \, \hat{L}^{01}$ has the effect of rotating the path integral by an angle of $\epsilon$ in said plane. (For a review on how the quantum operators corresponding to spacetime vector fields shift the spacetime boundary of path integration, see appendix \ref{appA}.)

However, due to the $SO(d+1)$ symmetry of $S_d$, we can trivially see that the rotated path integral equals the original path integral, and thus $\hat{L}^{01} \ket{0} = 0$. See figure \ref{figrotpi}. If our path integral had, say, contained insertions on the lower half-sphere, then the state would not have been annihilated by $\hat{L}^{01}$. See figure \ref{figrotpiins}.

\begin{figure}
    \centering
    \tikzset{every picture/.style={line width=0.75pt}} 

\begin{tikzpicture}[x=0.75pt,y=0.75pt,yscale=-1,xscale=1]

\draw  [draw opacity=0][line width=0.75]  (450,138.76) .. controls (450,141.05) and (434.27,142.91) .. (414.87,142.91) .. controls (395.47,142.91) and (379.74,141.05) .. (379.74,138.76) -- (414.87,138.76) -- cycle ; \draw  [line width=0.75]  (450,138.76) .. controls (450,141.05) and (434.27,142.91) .. (414.87,142.91) .. controls (395.47,142.91) and (379.74,141.05) .. (379.74,138.76) ;  
\draw  [draw opacity=0] (449.57,139.09) .. controls (449.57,139.09) and (449.57,139.09) .. (449.57,139.09) .. controls (449.57,139.09) and (449.57,139.09) .. (449.57,139.09) .. controls (449.57,158.37) and (433.94,174) .. (414.66,174) .. controls (395.37,174) and (379.74,158.37) .. (379.74,139.09) -- (414.66,139.09) -- cycle ; \draw   (449.57,139.09) .. controls (449.57,139.09) and (449.57,139.09) .. (449.57,139.09) .. controls (449.57,139.09) and (449.57,139.09) .. (449.57,139.09) .. controls (449.57,158.37) and (433.94,174) .. (414.66,174) .. controls (395.37,174) and (379.74,158.37) .. (379.74,139.09) ;  
\draw  [draw opacity=0][line width=0.75]  (379.74,138.76) .. controls (379.74,138.76) and (379.74,138.76) .. (379.74,138.76) .. controls (379.74,136.46) and (395.47,134.6) .. (414.87,134.6) .. controls (434.27,134.6) and (450,136.46) .. (450,138.76) -- (414.87,138.76) -- cycle ; \draw  [line width=0.75]  (379.74,138.76) .. controls (379.74,138.76) and (379.74,138.76) .. (379.74,138.76) .. controls (379.74,136.46) and (395.47,134.6) .. (414.87,134.6) .. controls (434.27,134.6) and (450,136.46) .. (450,138.76) ;  
\draw  [draw opacity=0][dash pattern={on 4.5pt off 4.5pt}] (379.74,138.76) .. controls (379.74,138.76) and (379.74,138.76) .. (379.74,138.76) .. controls (379.74,119.48) and (395.37,103.84) .. (414.66,103.84) .. controls (433.94,103.84) and (449.57,119.48) .. (449.57,138.76) -- (414.66,138.76) -- cycle ; \draw  [dash pattern={on 4.5pt off 4.5pt}] (379.74,138.76) .. controls (379.74,138.76) and (379.74,138.76) .. (379.74,138.76) .. controls (379.74,119.48) and (395.37,103.84) .. (414.66,103.84) .. controls (433.94,103.84) and (449.57,119.48) .. (449.57,138.76) ;  
\draw  [draw opacity=0][line width=0.75]  (308.97,124.65) .. controls (309.89,126.75) and (296.24,134.78) .. (278.48,142.58) .. controls (260.71,150.38) and (245.56,154.99) .. (244.64,152.89) -- (276.81,138.77) -- cycle ; \draw  [line width=0.75]  (308.97,124.65) .. controls (309.89,126.75) and (296.24,134.78) .. (278.48,142.58) .. controls (260.71,150.38) and (245.56,154.99) .. (244.64,152.89) ;  
\draw  [draw opacity=0] (308.71,125.12) .. controls (308.71,125.12) and (308.71,125.12) .. (308.71,125.12) .. controls (308.71,125.12) and (308.71,125.12) .. (308.71,125.12) .. controls (316.46,142.78) and (308.43,163.38) .. (290.77,171.13) .. controls (273.12,178.88) and (252.52,170.85) .. (244.77,153.19) -- (276.74,139.16) -- cycle ; \draw   (308.71,125.12) .. controls (308.71,125.12) and (308.71,125.12) .. (308.71,125.12) .. controls (308.71,125.12) and (308.71,125.12) .. (308.71,125.12) .. controls (316.46,142.78) and (308.43,163.38) .. (290.77,171.13) .. controls (273.12,178.88) and (252.52,170.85) .. (244.77,153.19) ;  
\draw  [draw opacity=0][line width=0.75]  (244.64,152.89) .. controls (244.64,152.89) and (244.64,152.89) .. (244.64,152.89) .. controls (243.72,150.79) and (257.37,142.76) .. (275.13,134.97) .. controls (292.9,127.17) and (308.05,122.55) .. (308.97,124.65) -- (276.81,138.77) -- cycle ; \draw  [line width=0.75]  (244.64,152.89) .. controls (244.64,152.89) and (244.64,152.89) .. (244.64,152.89) .. controls (243.72,150.79) and (257.37,142.76) .. (275.13,134.97) .. controls (292.9,127.17) and (308.05,122.55) .. (308.97,124.65) ;  
\draw  [draw opacity=0][dash pattern={on 4.5pt off 4.5pt}] (244.64,152.89) .. controls (244.64,152.89) and (244.64,152.89) .. (244.64,152.89) .. controls (236.89,135.24) and (244.92,114.64) .. (262.57,106.89) .. controls (280.23,99.14) and (300.83,107.17) .. (308.58,124.82) -- (276.61,138.86) -- cycle ; \draw  [dash pattern={on 4.5pt off 4.5pt}] (244.64,152.89) .. controls (244.64,152.89) and (244.64,152.89) .. (244.64,152.89) .. controls (236.89,135.24) and (244.92,114.64) .. (262.57,106.89) .. controls (280.23,99.14) and (300.83,107.17) .. (308.58,124.82) ;  
\draw  [draw opacity=0][dash pattern={on 0.84pt off 2.51pt}][line width=0.75]  (311.87,139.16) .. controls (311.87,141.45) and (296.14,143.31) .. (276.74,143.31) .. controls (257.34,143.31) and (241.61,141.45) .. (241.61,139.16) -- (276.74,139.16) -- cycle ; \draw  [dash pattern={on 0.84pt off 2.51pt}][line width=0.75]  (311.87,139.16) .. controls (311.87,141.45) and (296.14,143.31) .. (276.74,143.31) .. controls (257.34,143.31) and (241.61,141.45) .. (241.61,139.16) ;  
\draw  [draw opacity=0][dash pattern={on 0.84pt off 2.51pt}][line width=0.75]  (241.68,138.77) .. controls (241.68,138.77) and (241.68,138.77) .. (241.68,138.77) .. controls (241.68,136.48) and (257.4,134.62) .. (276.81,134.62) .. controls (296.21,134.62) and (311.93,136.48) .. (311.93,138.77) -- (276.81,138.77) -- cycle ; \draw  [dash pattern={on 0.84pt off 2.51pt}][line width=0.75]  (241.68,138.77) .. controls (241.68,138.77) and (241.68,138.77) .. (241.68,138.77) .. controls (241.68,136.48) and (257.4,134.62) .. (276.81,134.62) .. controls (296.21,134.62) and (311.93,136.48) .. (311.93,138.77) ;  

\draw (218.36,134.4) node [anchor=east] [inner sep=0.75pt]    {$-\left( \epsilon \hat{L}^{01} \Psi^E_{0}\right)[ \varphi ] \ \ =$};
\draw (422.69,125.43) node [anchor=east] [inner sep=0.75pt]    {$\varphi $};
\draw (276.08,121.58) node [anchor=east] [inner sep=0.75pt]  [rotate=-336]  {$\varphi $};
\draw (316,124.4) node [anchor=north west][inner sep=0.75pt]    {$\epsilon $};
\draw (350.22,137.2) node    {$-$};
\draw (484.02,136.2) node    {$=\ \ 0$};

\end{tikzpicture}
    \caption{\label{figrotpi} Proof of the de Sitter invariance of the Bunch-Davies state. $\epsilon \hat{L}^{01}$ rotates the euclidean half-sphere by angle $\epsilon$, which does not actually change the path integral.}
\end{figure}

\begin{figure}
    \centering
    \tikzset{every picture/.style={line width=0.75pt}} 

\begin{tikzpicture}[x=0.75pt,y=0.75pt,yscale=-1,xscale=1]

\draw  [draw opacity=0][line width=0.75]  (411,72.76) .. controls (411,72.76) and (411,72.76) .. (411,72.76) .. controls (411,75.05) and (395.27,76.91) .. (375.87,76.91) .. controls (356.47,76.91) and (340.74,75.05) .. (340.74,72.76) -- (375.87,72.76) -- cycle ; \draw  [line width=0.75]  (411,72.76) .. controls (411,72.76) and (411,72.76) .. (411,72.76) .. controls (411,75.05) and (395.27,76.91) .. (375.87,76.91) .. controls (356.47,76.91) and (340.74,75.05) .. (340.74,72.76) ;  
\draw  [draw opacity=0] (410.57,73.09) .. controls (410.57,92.37) and (394.94,108) .. (375.66,108) .. controls (356.37,108) and (340.74,92.37) .. (340.74,73.09) -- (375.66,73.09) -- cycle ; \draw   (410.57,73.09) .. controls (410.57,92.37) and (394.94,108) .. (375.66,108) .. controls (356.37,108) and (340.74,92.37) .. (340.74,73.09) ;  
\draw  [draw opacity=0][line width=0.75]  (340.74,72.76) .. controls (340.74,72.76) and (340.74,72.76) .. (340.74,72.76) .. controls (340.74,70.46) and (356.47,68.6) .. (375.87,68.6) .. controls (395.27,68.6) and (411,70.46) .. (411,72.76) -- (375.87,72.76) -- cycle ; \draw  [line width=0.75]  (340.74,72.76) .. controls (340.74,72.76) and (340.74,72.76) .. (340.74,72.76) .. controls (340.74,70.46) and (356.47,68.6) .. (375.87,68.6) .. controls (395.27,68.6) and (411,70.46) .. (411,72.76) ;  
\draw  [draw opacity=0][dash pattern={on 4.5pt off 4.5pt}] (340.74,72.76) .. controls (340.74,72.76) and (340.74,72.76) .. (340.74,72.76) .. controls (340.74,53.48) and (356.37,37.84) .. (375.66,37.84) .. controls (394.94,37.84) and (410.57,53.48) .. (410.57,72.76) -- (375.66,72.76) -- cycle ; \draw  [dash pattern={on 4.5pt off 4.5pt}] (340.74,72.76) .. controls (340.74,72.76) and (340.74,72.76) .. (340.74,72.76) .. controls (340.74,53.48) and (356.37,37.84) .. (375.66,37.84) .. controls (394.94,37.84) and (410.57,53.48) .. (410.57,72.76) ;  
\draw  [draw opacity=0][line width=0.75]  (269.97,58.65) .. controls (269.97,58.65) and (269.97,58.65) .. (269.97,58.65) .. controls (270.89,60.75) and (257.24,68.78) .. (239.48,76.58) .. controls (221.71,84.38) and (206.56,88.99) .. (205.64,86.89) -- (237.81,72.77) -- cycle ; \draw  [line width=0.75]  (269.97,58.65) .. controls (269.97,58.65) and (269.97,58.65) .. (269.97,58.65) .. controls (270.89,60.75) and (257.24,68.78) .. (239.48,76.58) .. controls (221.71,84.38) and (206.56,88.99) .. (205.64,86.89) ;  
\draw  [draw opacity=0] (269.71,59.12) .. controls (269.71,59.12) and (269.71,59.12) .. (269.71,59.12) .. controls (277.46,76.78) and (269.43,97.38) .. (251.77,105.13) .. controls (234.12,112.88) and (213.52,104.85) .. (205.77,87.19) -- (237.74,73.16) -- cycle ; \draw   (269.71,59.12) .. controls (269.71,59.12) and (269.71,59.12) .. (269.71,59.12) .. controls (277.46,76.78) and (269.43,97.38) .. (251.77,105.13) .. controls (234.12,112.88) and (213.52,104.85) .. (205.77,87.19) ;  
\draw  [draw opacity=0][line width=0.75]  (205.64,86.89) .. controls (205.64,86.89) and (205.64,86.89) .. (205.64,86.89) .. controls (204.72,84.79) and (218.37,76.76) .. (236.13,68.97) .. controls (253.9,61.17) and (269.05,56.55) .. (269.97,58.65) -- (237.81,72.77) -- cycle ; \draw  [line width=0.75]  (205.64,86.89) .. controls (205.64,86.89) and (205.64,86.89) .. (205.64,86.89) .. controls (204.72,84.79) and (218.37,76.76) .. (236.13,68.97) .. controls (253.9,61.17) and (269.05,56.55) .. (269.97,58.65) ;  
\draw  [draw opacity=0][dash pattern={on 4.5pt off 4.5pt}] (205.64,86.89) .. controls (197.89,69.24) and (205.92,48.64) .. (223.57,40.89) .. controls (241.23,33.14) and (261.83,41.17) .. (269.58,58.82) -- (237.61,72.86) -- cycle ; \draw  [dash pattern={on 4.5pt off 4.5pt}] (205.64,86.89) .. controls (197.89,69.24) and (205.92,48.64) .. (223.57,40.89) .. controls (241.23,33.14) and (261.83,41.17) .. (269.58,58.82) ;  
\draw  [draw opacity=0][dash pattern={on 0.84pt off 2.51pt}][line width=0.75]  (272.87,73.16) .. controls (272.87,73.16) and (272.87,73.16) .. (272.87,73.16) .. controls (272.87,75.45) and (257.14,77.31) .. (237.74,77.31) .. controls (218.34,77.31) and (202.61,75.45) .. (202.61,73.16) -- (237.74,73.16) -- cycle ; \draw  [dash pattern={on 0.84pt off 2.51pt}][line width=0.75]  (272.87,73.16) .. controls (272.87,73.16) and (272.87,73.16) .. (272.87,73.16) .. controls (272.87,75.45) and (257.14,77.31) .. (237.74,77.31) .. controls (218.34,77.31) and (202.61,75.45) .. (202.61,73.16) ;  
\draw  [draw opacity=0][dash pattern={on 0.84pt off 2.51pt}][line width=0.75]  (202.68,72.77) .. controls (202.68,72.77) and (202.68,72.77) .. (202.68,72.77) .. controls (202.68,70.48) and (218.4,68.62) .. (237.81,68.62) .. controls (257.21,68.62) and (272.93,70.48) .. (272.93,72.77) -- (237.81,72.77) -- cycle ; \draw  [dash pattern={on 0.84pt off 2.51pt}][line width=0.75]  (202.68,72.77) .. controls (202.68,72.77) and (202.68,72.77) .. (202.68,72.77) .. controls (202.68,70.48) and (218.4,68.62) .. (237.81,68.62) .. controls (257.21,68.62) and (272.93,70.48) .. (272.93,72.77) ;  
\draw  [fill={rgb, 255:red, 0; green, 0; blue, 0 }  ,fill opacity=1 ] (220.43,94.69) .. controls (220.43,93.96) and (221.02,93.37) .. (221.75,93.37) .. controls (222.47,93.37) and (223.06,93.96) .. (223.06,94.69) .. controls (223.06,95.41) and (222.47,96) .. (221.75,96) .. controls (221.02,96) and (220.43,95.41) .. (220.43,94.69) -- cycle ;
\draw  [fill={rgb, 255:red, 0; green, 0; blue, 0 }  ,fill opacity=1 ] (238.93,104.69) .. controls (238.93,103.96) and (239.52,103.37) .. (240.25,103.37) .. controls (240.97,103.37) and (241.56,103.96) .. (241.56,104.69) .. controls (241.56,105.41) and (240.97,106) .. (240.25,106) .. controls (239.52,106) and (238.93,105.41) .. (238.93,104.69) -- cycle ;
\draw  [fill={rgb, 255:red, 0; green, 0; blue, 0 }  ,fill opacity=1 ] (255.43,94.69) .. controls (255.43,93.96) and (256.02,93.37) .. (256.75,93.37) .. controls (257.47,93.37) and (258.06,93.96) .. (258.06,94.69) .. controls (258.06,95.41) and (257.47,96) .. (256.75,96) .. controls (256.02,96) and (255.43,95.41) .. (255.43,94.69) -- cycle ;
\draw    (239,124) -- (239,115) ;
\draw [shift={(239,112)}, rotate = 90] [fill={rgb, 255:red, 0; green, 0; blue, 0 }  ][line width=0.08]  [draw opacity=0] (5.36,-2.57) -- (0,0) -- (5.36,2.57) -- cycle    ;
\draw    (375,123.5) -- (375,114.5) ;
\draw [shift={(375,111.5)}, rotate = 90] [fill={rgb, 255:red, 0; green, 0; blue, 0 }  ][line width=0.08]  [draw opacity=0] (5.36,-2.57) -- (0,0) -- (5.36,2.57) -- cycle    ;
\draw  [fill={rgb, 255:red, 0; green, 0; blue, 0 }  ,fill opacity=1 ] (355.93,94.69) .. controls (355.93,93.96) and (356.52,93.37) .. (357.25,93.37) .. controls (357.97,93.37) and (358.56,93.96) .. (358.56,94.69) .. controls (358.56,95.41) and (357.97,96) .. (357.25,96) .. controls (356.52,96) and (355.93,95.41) .. (355.93,94.69) -- cycle ;
\draw  [fill={rgb, 255:red, 0; green, 0; blue, 0 }  ,fill opacity=1 ] (374.43,104.69) .. controls (374.43,103.96) and (375.02,103.37) .. (375.75,103.37) .. controls (376.47,103.37) and (377.06,103.96) .. (377.06,104.69) .. controls (377.06,105.41) and (376.47,106) .. (375.75,106) .. controls (375.02,106) and (374.43,105.41) .. (374.43,104.69) -- cycle ;
\draw  [fill={rgb, 255:red, 0; green, 0; blue, 0 }  ,fill opacity=1 ] (390.93,94.69) .. controls (390.93,93.96) and (391.52,93.37) .. (392.25,93.37) .. controls (392.97,93.37) and (393.56,93.96) .. (393.56,94.69) .. controls (393.56,95.41) and (392.97,96) .. (392.25,96) .. controls (391.52,96) and (390.93,95.41) .. (390.93,94.69) -- cycle ;

\draw (383.69,60.43) node [anchor=east] [inner sep=0.75pt]    {$\varphi $};
\draw (237.08,55.58) node [anchor=east] [inner sep=0.75pt]  [rotate=-336]  {$\varphi $};
\draw (277,58.4) node [anchor=north west][inner sep=0.75pt]    {$\epsilon $};
\draw (311.22,71.2) node    {$-$};
\draw (445.02,70.2) node    {$\neq \ \ 0$};
\draw (238.97,130.35) node  [font=\scriptsize] [align=left] {insertions};
\draw (375,129.5) node  [font=\scriptsize] [align=left] {insertions};

\end{tikzpicture}
    \caption{\label{figrotpiins} If we were to add some insertions into the half-sphere path integral, the corresponding state would not be annihilated by $\hat{L}^{01}$ because the rotation of the equator by a tiny angle $\epsilon$ would not leave the path integral invariant.}
\end{figure}

We have now seen that the Bunch-Davies state is invariant under the full dS isometry group. This is a special property, and one may wonder if there are more states invariant under this group. As we will see, the $\alpha$-vacua are a 1-parmeter family of states which are also dS-invariant.

\section{Noether's theorem and annihilation operators}\label{sec4}

In this section we review some properties of annihilation operators, euclidean modes, and their relation to the Bunch-Davies state.

Consider a function $f$ on $dS_d$ which satisfies the free de Sitter Klein-Gordon equation of motion
\begin{equation}\label{feom}
    (\nabla^2 - m^2) f = 0.
\end{equation}
$f$ can also be understood as a function on the sphere $S_d$ upon analytic continuation.

On the euclidean sphere $S_d$, the eigenvalues of the Laplace operator $\nabla^2$ are given by $-l ( l + d - 1)$ for all $l = 0, 1, 2 \ldots$. Notice, then, that $\nabla^2$ only has negative eigenvalues while any $f$ which solves \eqref{feom} will be an eigenfunction of $\nabla^2$ with the positive eigenvalue $m^2$. This implies that, strictly speaking, there exists \textit{no} such $f$ which is defined on the whole sphere. Any $f$ that solves \eqref{feom} must be ill-defined at some points on $S_d$, where there will be non-analyticities such as poles or  branch cuts.

If the analytic continuation of a function $f_e$ that solves \eqref{feom} is well-defined on the whole lower half-sphere $S_{d,-}$, we call it a ``euclidean mode.'' As a consequence of the previous discussion, all euclidean modes must necessarily have non-analyticities or singularities in the upper half-sphere $S_{d,+}$.

Every solution $f$ to the Klein-Gordon equation can be used to define an associated ``annihilation operator'' $\hat{a}(f)$. It turns out that any annihilation operator built out of a euclidean mode will annihilate the Bunch-Davies state, $\hat{a}(f_e)\ket{0} = 0$. We will now present a proof of this fact using Noether's theorem. 

Classically, the free euclidean action of a scalar field $S_U[\phi]$ possesses a symmetry
\begin{equation}\label{varf}
    \delta \phi = \epsilon f
\end{equation}
where $\epsilon$ is a tiny constant and $f$ is any function which satisfies \eqref{feom}. This symmetry only exists because the free equation of motion is linear, and any two solutions can be added together to produce a new solution.

We will now preform the Noether trick to find the conserved current corresponding to this symmetry. To do this, we promote $\epsilon$ to an infinitesimal function $\varepsilon(\tx)$ which vanishes at the boundary of $U$, and consider the change in the action corresponding to $\delta \phi(\tx)  = \varepsilon(\tx) f(\tx)$. It turns out to be
\begin{equation}
    \delta S_U[\phi] = \int_U d^d \tx \sqrt{g} \left(  ( \partial^\mu \varepsilon ) ( f \pd_\mu \phi) - (\varepsilon \, \phi) ( \nabla^2 f - m^2 f)   \right) 
\end{equation}
where $f_1 \pd_\mu f_2 \equiv f_1 (\partial_\mu f_2) - (\partial_\mu f_1) f_2$. Because $f$ satisfies \eqref{feom} and $\varepsilon$ is arbitrary, this implies that the Noether current
\begin{equation}
    J(f)_\mu \equiv f \pd_\mu \phi
\end{equation}
is conserved on shell, satisfying $\nabla_\mu J(f)^\mu = \frac{1}{\sqrt{g}} \partial_\mu( \sqrt{g} J(f)^\mu)= 0$. Let us define the associated conserved charge
\begin{equation}
    a_{\Sigma}(f) \equiv i \int_\Sigma d \Sigma^\mu J(f)_\mu.
\end{equation}
which is defined on a codimension-$1$ surface $\Sigma$. This conserved charge is the classical version of the annihilation operator corresponding to the mode $f$. (We note that the standard definition of the annihilation operator differs from the one above by an extra complex conjugation, replacing $f$ above with $f^*$.)

We now turn to the quantum path integral on the lower half euclidean sphere. If we define $\phi' = \phi + \varepsilon f_e$, then $\mD \phi' = \mD \phi$ because the addition of $\varepsilon f_e$ simply shifts the measure. Crucially, because $f_e$ is a euclidean mode, the change in measure $\phi' = \phi + \varepsilon f_e$ is well defined everywhere on the lower half-sphere.

Using this symmetry variation, one can derive a corresponding quantum Ward identity in the usual way, and it will read
\begin{equation}
    \int_{\eval{\phi}_E = \, \varphi } \mD \phi \;  e^{- S_{S_{d,-}}[\phi]} \nabla_\mu J(f_e)^\mu(\tx) = 0
\end{equation}
where $\tx$ can be any point in $S_{d,-}$.

The above equation implies that, if one inserts the charge $a_\Sigma(f_e)$ into the path integral, $\Sigma$ can be deformed without changing the path integral.

Let us now define the quantum annihilation operator\footnote{In this paper we use hatted symbols to denote operators acting on states, such as $\hat a(f)$, and non-hatted symbols to denote insertions in the path integral, like $a_\Sigma(f)$.}
\begin{equation}
    \hat{a}(f) \equiv i \int_E d E^\mu f \pd_\mu \hat\phi
\end{equation}
where we set the codimension-1 surface $\Sigma$ to be the equator $E$.

\begin{figure}
    \centering
    \tikzset{every picture/.style={line width=0.75pt}} 

\begin{tikzpicture}[x=0.75pt,y=0.75pt,yscale=-1,xscale=1]

\draw  [draw opacity=0][line width=1.5]  (393,59.52) .. controls (393,59.52) and (393,59.52) .. (393,59.52) .. controls (393,63.93) and (377.27,67.5) .. (357.87,67.5) .. controls (338.47,67.5) and (322.74,63.93) .. (322.74,59.52) -- (357.87,59.52) -- cycle ; \draw  [line width=1.5]  (393,59.52) .. controls (393,59.52) and (393,59.52) .. (393,59.52) .. controls (393,63.93) and (377.27,67.5) .. (357.87,67.5) .. controls (338.47,67.5) and (322.74,63.93) .. (322.74,59.52) ;  
\draw  [draw opacity=0] (392.57,59.84) .. controls (392.57,79.13) and (376.94,94.76) .. (357.66,94.76) .. controls (338.37,94.76) and (322.74,79.13) .. (322.74,59.84) -- (357.66,59.84) -- cycle ; \draw   (392.57,59.84) .. controls (392.57,79.13) and (376.94,94.76) .. (357.66,94.76) .. controls (338.37,94.76) and (322.74,79.13) .. (322.74,59.84) ;  
\draw  [draw opacity=0][line width=1.5]  (322.74,59.52) .. controls (322.74,59.52) and (322.74,59.52) .. (322.74,59.52) .. controls (322.74,55.11) and (338.47,51.53) .. (357.87,51.53) .. controls (377.27,51.53) and (393,55.11) .. (393,59.52) -- (357.87,59.52) -- cycle ; \draw  [line width=1.5]  (322.74,59.52) .. controls (322.74,59.52) and (322.74,59.52) .. (322.74,59.52) .. controls (322.74,55.11) and (338.47,51.53) .. (357.87,51.53) .. controls (377.27,51.53) and (393,55.11) .. (393,59.52) ;  
\draw  [draw opacity=0][line width=0.75]  (502.5,58.52) .. controls (502.5,58.52) and (502.5,58.52) .. (502.5,58.52) .. controls (502.5,58.52) and (502.5,58.52) .. (502.5,58.52) .. controls (502.5,62.93) and (486.77,66.5) .. (467.37,66.5) .. controls (447.97,66.5) and (432.24,62.93) .. (432.24,58.52) -- (467.37,58.52) -- cycle ; \draw  [line width=0.75]  (502.5,58.52) .. controls (502.5,58.52) and (502.5,58.52) .. (502.5,58.52) .. controls (502.5,58.52) and (502.5,58.52) .. (502.5,58.52) .. controls (502.5,62.93) and (486.77,66.5) .. (467.37,66.5) .. controls (447.97,66.5) and (432.24,62.93) .. (432.24,58.52) ;  
\draw  [draw opacity=0] (502.07,58.84) .. controls (502.07,78.13) and (486.44,93.76) .. (467.16,93.76) .. controls (447.87,93.76) and (432.24,78.13) .. (432.24,58.84) -- (467.16,58.84) -- cycle ; \draw   (502.07,58.84) .. controls (502.07,78.13) and (486.44,93.76) .. (467.16,93.76) .. controls (447.87,93.76) and (432.24,78.13) .. (432.24,58.84) ;  
\draw  [draw opacity=0][line width=0.75]  (432.24,58.52) .. controls (432.24,58.52) and (432.24,58.52) .. (432.24,58.52) .. controls (432.24,54.11) and (447.97,50.53) .. (467.37,50.53) .. controls (486.77,50.53) and (502.5,54.11) .. (502.5,58.52) -- (467.37,58.52) -- cycle ; \draw  [line width=0.75]  (432.24,58.52) .. controls (432.24,58.52) and (432.24,58.52) .. (432.24,58.52) .. controls (432.24,54.11) and (447.97,50.53) .. (467.37,50.53) .. controls (486.77,50.53) and (502.5,54.11) .. (502.5,58.52) ;  
\draw  [draw opacity=0][line width=1.5]  (494,79.75) .. controls (494,79.75) and (494,79.75) .. (494,79.75) .. controls (494,82.37) and (481.85,84.5) .. (466.87,84.5) .. controls (451.89,84.5) and (439.74,82.37) .. (439.74,79.75) -- (466.87,79.75) -- cycle ; \draw  [line width=1.5]  (494,79.75) .. controls (494,79.75) and (494,79.75) .. (494,79.75) .. controls (494,82.37) and (481.85,84.5) .. (466.87,84.5) .. controls (451.89,84.5) and (439.74,82.37) .. (439.74,79.75) ;  
\draw  [draw opacity=0][dash pattern={on 5.63pt off 4.5pt}][line width=1.5]  (439.74,79.75) .. controls (439.74,79.75) and (439.74,79.75) .. (439.74,79.75) .. controls (439.74,77.13) and (451.89,75) .. (466.87,75) .. controls (481.85,75) and (494,77.13) .. (494,79.75) -- (466.87,79.75) -- cycle ; \draw  [dash pattern={on 5.63pt off 4.5pt}][line width=1.5]  (439.74,79.75) .. controls (439.74,79.75) and (439.74,79.75) .. (439.74,79.75) .. controls (439.74,77.13) and (451.89,75) .. (466.87,75) .. controls (481.85,75) and (494,77.13) .. (494,79.75) ;  

\draw (357.9,48.99) node [anchor=south] [inner sep=0.75pt]    {$\varphi $};
\draw (317.74,57.52) node [anchor=east] [inner sep=0.75pt]    {$a_{E}( f_{e})$};
\draw (420.24,68.02) node [anchor=east] [inner sep=0.75pt]    {$=$};
\draw (467.4,47.99) node [anchor=south] [inner sep=0.75pt]    {$\varphi $};
\draw (505.21,68.02) node [anchor=west] [inner sep=0.75pt]    {$=\ 0$};

\end{tikzpicture}
    \caption{\label{figfecontract} If $f_e$ has no singularities on $S_{d,-}$, then $\hat{a}(f_e)$ annihilates $\ket{0}$ because the equator is contractible.}
\end{figure}

Because $E$ is contractable on the half-sphere, the annihilation operator $\hat a(f_e)$ annihilates the Bunch-Davies state, as shown in figure \ref{figfecontract}.
\begin{equation}
    (\hat a(f_e) \Psi_0^E) [\varphi] = \int_{\eval{\phi}_E = \, \varphi } \mD \phi \;  e^{- S_{S_{d,-}}[\phi]} a_E(f_e) = 0.
\end{equation}
This completes the proof.

\section{The antipodal symmetry, classically and quantumly}\label{sec5}

One can define an antipodal map on de Sitter space, which takes the form $X \mapsto - X$ in the embedding space. This map commutes with the dS isometry group $SO(1,d)$.

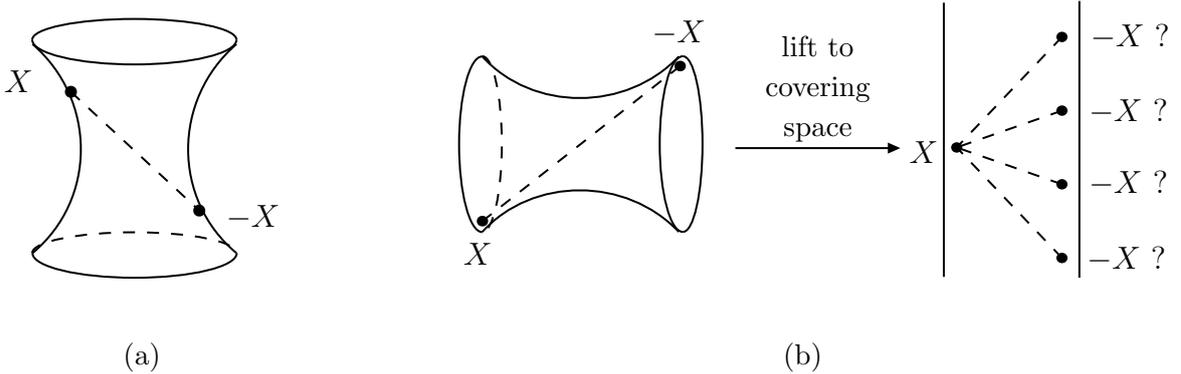
\begin{figure}
     \centering
     \begin{subfigure}{0.25\textwidth}
         \centering
         \tikzset{every picture/.style={line width=0.75pt}} 

\begin{tikzpicture}[x=0.75pt,y=0.75pt,yscale=-0.75,xscale=0.75]

\draw    (118,48.88) .. controls (160.91,86.24) and (160.91,153.07) .. (118,189.03) ;
\draw    (253.76,48.87) .. controls (210.85,86.24) and (210.85,153.07) .. (253.76,189.03) ;
\draw  [draw opacity=0][line width=0.75]  (253.76,45.96) .. controls (253.76,45.96) and (253.76,45.96) .. (253.76,45.96) .. controls (253.76,55.09) and (223.33,62.5) .. (185.79,62.5) .. controls (148.25,62.5) and (117.82,55.09) .. (117.82,45.96) -- (185.79,45.96) -- cycle ; \draw  [line width=0.75]  (253.76,45.96) .. controls (253.76,45.96) and (253.76,45.96) .. (253.76,45.96) .. controls (253.76,55.09) and (223.33,62.5) .. (185.79,62.5) .. controls (148.25,62.5) and (117.82,55.09) .. (117.82,45.96) ;  
\draw  [draw opacity=0][line width=0.75]  (117.82,45.96) .. controls (117.82,45.96) and (117.82,45.96) .. (117.82,45.96) .. controls (117.82,38.25) and (148.25,32) .. (185.79,32) .. controls (223.33,32) and (253.76,38.25) .. (253.76,45.96) -- (185.79,45.96) -- cycle ; \draw  [line width=0.75]  (117.82,45.96) .. controls (117.82,45.96) and (117.82,45.96) .. (117.82,45.96) .. controls (117.82,38.25) and (148.25,32) .. (185.79,32) .. controls (223.33,32) and (253.76,38.25) .. (253.76,45.96) ;  
\draw  [draw opacity=0][line width=0.75]  (253.94,189.03) .. controls (253.94,189.03) and (253.94,189.03) .. (253.94,189.03) .. controls (253.94,198.17) and (223.51,205.57) .. (185.97,205.57) .. controls (148.43,205.57) and (118,198.17) .. (118,189.03) -- (185.97,189.03) -- cycle ; \draw  [line width=0.75]  (253.94,189.03) .. controls (253.94,189.03) and (253.94,189.03) .. (253.94,189.03) .. controls (253.94,198.17) and (223.51,205.57) .. (185.97,205.57) .. controls (148.43,205.57) and (118,198.17) .. (118,189.03) ;  
\draw  [draw opacity=0][dash pattern={on 4.5pt off 4.5pt}][line width=0.75]  (117.82,189.03) .. controls (117.82,189.03) and (117.82,189.03) .. (117.82,189.03) .. controls (117.82,181.32) and (148.25,175.08) .. (185.79,175.08) .. controls (223.33,175.08) and (253.76,181.32) .. (253.76,189.03) -- (185.79,189.03) -- cycle ; \draw  [dash pattern={on 4.5pt off 4.5pt}][line width=0.75]  (117.82,189.03) .. controls (117.82,189.03) and (117.82,189.03) .. (117.82,189.03) .. controls (117.82,181.32) and (148.25,175.08) .. (185.79,175.08) .. controls (223.33,175.08) and (253.76,181.32) .. (253.76,189.03) ;  
\draw  [dash pattern={on 4.5pt off 4.5pt}]  (143.56,80.87) -- (229,160.5) ;
\draw [shift={(229,160.5)}, rotate = 42.98] [color={rgb, 255:red, 0; green, 0; blue, 0 }  ][fill={rgb, 255:red, 0; green, 0; blue, 0 }  ][line width=0.75]      (0, 0) circle [x radius= 3.35, y radius= 3.35]   ;
\draw [shift={(143.56,80.87)}, rotate = 42.98] [color={rgb, 255:red, 0; green, 0; blue, 0 }  ][fill={rgb, 255:red, 0; green, 0; blue, 0 }  ][line width=0.75]      (0, 0) circle [x radius= 3.35, y radius= 3.35]   ;

\draw (120.26,74.03) node [anchor=east] [inner sep=0.75pt]    {$X$};
\draw (245,154.57) node [anchor=north west][inner sep=0.75pt]    {$-X$};

\end{tikzpicture}
         \caption{\label{fig:1a} }
     \end{subfigure} \hfill
     \begin{subfigure}{0.6\textwidth}
         \centering
         \tikzset{every picture/.style={line width=0.75pt}} 

\begin{tikzpicture}[x=0.75pt,y=0.75pt,yscale=-0.65,xscale=0.7]

\draw    (309.03,62.08) .. controls (271.66,104.99) and (204.84,104.99) .. (168.87,62.08) ;
\draw    (309.03,197.85) .. controls (271.66,154.94) and (204.84,154.94) .. (168.87,197.85) ;
\draw  [draw opacity=0][line width=0.75]  (311.95,197.85) .. controls (311.95,197.85) and (311.95,197.85) .. (311.95,197.85) .. controls (311.95,197.85) and (311.95,197.85) .. (311.95,197.85) .. controls (302.81,197.85) and (295.41,167.42) .. (295.41,129.88) .. controls (295.41,92.34) and (302.81,61.91) .. (311.95,61.91) -- (311.95,129.88) -- cycle ; \draw  [line width=0.75]  (311.95,197.85) .. controls (311.95,197.85) and (311.95,197.85) .. (311.95,197.85) .. controls (311.95,197.85) and (311.95,197.85) .. (311.95,197.85) .. controls (302.81,197.85) and (295.41,167.42) .. (295.41,129.88) .. controls (295.41,92.34) and (302.81,61.91) .. (311.95,61.91) ;  
\draw  [draw opacity=0][line width=0.75]  (311.95,61.91) .. controls (319.66,61.91) and (325.91,92.34) .. (325.91,129.88) .. controls (325.91,167.42) and (319.66,197.85) .. (311.95,197.85) -- (311.95,129.88) -- cycle ; \draw  [line width=0.75]  (311.95,61.91) .. controls (319.66,61.91) and (325.91,92.34) .. (325.91,129.88) .. controls (325.91,167.42) and (319.66,197.85) .. (311.95,197.85) ;  
\draw  [draw opacity=0][line width=0.75]  (168.87,198.02) .. controls (168.87,198.02) and (168.87,198.02) .. (168.87,198.02) .. controls (159.74,198.02) and (152.33,167.59) .. (152.33,130.05) .. controls (152.33,92.52) and (159.74,62.08) .. (168.87,62.08) -- (168.87,130.05) -- cycle ; \draw  [line width=0.75]  (168.87,198.02) .. controls (168.87,198.02) and (168.87,198.02) .. (168.87,198.02) .. controls (159.74,198.02) and (152.33,167.59) .. (152.33,130.05) .. controls (152.33,92.52) and (159.74,62.08) .. (168.87,62.08) ;  
\draw  [draw opacity=0][dash pattern={on 4.5pt off 4.5pt}][line width=0.75]  (168.87,61.91) .. controls (176.58,61.91) and (182.83,92.34) .. (182.83,129.88) .. controls (182.83,167.42) and (176.58,197.85) .. (168.87,197.85) -- (168.87,129.88) -- cycle ; \draw  [dash pattern={on 4.5pt off 4.5pt}][line width=0.75]  (168.87,61.91) .. controls (176.58,61.91) and (182.83,92.34) .. (182.83,129.88) .. controls (182.83,167.42) and (176.58,197.85) .. (168.87,197.85) ;  
\draw  [dash pattern={on 4.5pt off 4.5pt}]  (310,69.5) -- (169,189.5) ;
\draw [shift={(169,189.5)}, rotate = 139.6] [color={rgb, 255:red, 0; green, 0; blue, 0 }  ][fill={rgb, 255:red, 0; green, 0; blue, 0 }  ][line width=0.75]      (0, 0) circle [x radius= 3.35, y radius= 3.35]   ;
\draw [shift={(310,69.5)}, rotate = 139.6] [color={rgb, 255:red, 0; green, 0; blue, 0 }  ][fill={rgb, 255:red, 0; green, 0; blue, 0 }  ][line width=0.75]      (0, 0) circle [x radius= 3.35, y radius= 3.35]   ;
\draw    (498,20.5) -- (498,232.67) ;
\draw    (594.67,19.17) -- (594.67,233.33) ;
\draw  [fill={rgb, 255:red, 0; green, 0; blue, 0 }  ,fill opacity=1 ] (503.92,132.5) .. controls (503.92,130.71) and (505.37,129.25) .. (507.17,129.25) .. controls (508.96,129.25) and (510.42,130.71) .. (510.42,132.5) .. controls (510.42,134.29) and (508.96,135.75) .. (507.17,135.75) .. controls (505.37,135.75) and (503.92,134.29) .. (503.92,132.5) -- cycle ;
\draw  [fill={rgb, 255:red, 0; green, 0; blue, 0 }  ,fill opacity=1 ] (579,47) .. controls (579,45.21) and (580.46,43.75) .. (582.25,43.75) .. controls (584.04,43.75) and (585.5,45.21) .. (585.5,47) .. controls (585.5,48.79) and (584.04,50.25) .. (582.25,50.25) .. controls (580.46,50.25) and (579,48.79) .. (579,47) -- cycle ;
\draw  [fill={rgb, 255:red, 0; green, 0; blue, 0 }  ,fill opacity=1 ] (579,104) .. controls (579,102.21) and (580.46,100.75) .. (582.25,100.75) .. controls (584.04,100.75) and (585.5,102.21) .. (585.5,104) .. controls (585.5,105.79) and (584.04,107.25) .. (582.25,107.25) .. controls (580.46,107.25) and (579,105.79) .. (579,104) -- cycle ;
\draw  [fill={rgb, 255:red, 0; green, 0; blue, 0 }  ,fill opacity=1 ] (579,218) .. controls (579,216.21) and (580.46,214.75) .. (582.25,214.75) .. controls (584.04,214.75) and (585.5,216.21) .. (585.5,218) .. controls (585.5,219.79) and (584.04,221.25) .. (582.25,221.25) .. controls (580.46,221.25) and (579,219.79) .. (579,218) -- cycle ;
\draw  [fill={rgb, 255:red, 0; green, 0; blue, 0 }  ,fill opacity=1 ] (579,161) .. controls (579,159.21) and (580.46,157.75) .. (582.25,157.75) .. controls (584.04,157.75) and (585.5,159.21) .. (585.5,161) .. controls (585.5,162.79) and (584.04,164.25) .. (582.25,164.25) .. controls (580.46,164.25) and (579,162.79) .. (579,161) -- cycle ;
\draw  [dash pattern={on 4.5pt off 4.5pt}]  (582.25,47) -- (507.17,132.5) ;
\draw  [dash pattern={on 4.5pt off 4.5pt}]  (582.25,104) -- (507.17,132.5) ;
\draw  [dash pattern={on 4.5pt off 4.5pt}]  (582.25,161) -- (507.17,132.5) ;
\draw  [dash pattern={on 4.5pt off 4.5pt}]  (582.25,218) -- (507.17,132.5) ;
\draw    (349.33,133) -- (464,133) ;
\draw [shift={(467,133)}, rotate = 180] [fill={rgb, 255:red, 0; green, 0; blue, 0 }  ][line width=0.08]  [draw opacity=0] (8.93,-4.29) -- (0,0) -- (8.93,4.29) -- cycle    ;

\draw (165.8,203.9) node [anchor=north] [inner sep=0.75pt]    {$X$};
\draw (309.03,55.68) node [anchor=south] [inner sep=0.75pt]    {$-X$};
\draw (483.8,125.57) node [anchor=north] [inner sep=0.75pt]    {$X$};
\draw (600.72,46.7) node [anchor=west] [inner sep=0.75pt]    {$-X\ ?$};
\draw (599.39,104.7) node [anchor=west] [inner sep=0.75pt]    {$-X\ ?$};
\draw (599.39,160.03) node [anchor=west] [inner sep=0.75pt]    {$-X\ ?$};
\draw (598.06,218.7) node [anchor=west] [inner sep=0.75pt]    {$-X\ ?$};
\draw (408.17,130) node [anchor=south] [inner sep=0.75pt]  [font=\small] [align=left] {\begin{minipage}[lt]{64.47pt}\setlength\topsep{0pt}
\begin{center}
lift to\\covering space
\end{center}

\end{minipage}};

\end{tikzpicture}
         \caption{\label{fig:1b} }
     \end{subfigure}
     \caption{\label{fig:anti} (a) The dS antipodal map. (b) In AdS there is no antipodal map. If one attempts to define $-X$ as the antipode of $X$ in the embedding space hyperboloid, there will not be a well defined meaning to $-X$ in the universal cover of the hyperboloid, as it could be one of an infinite number of possibilities.}
\end{figure}

Interestingly, one cannot define an analogous antipodal map in Anti-de Sitter space. This is essentially because Anti-de Sitter space is defined as the \textit{universal cover} of the hyperboloid $-(X^0)^2 - (X^1)^2 + (X^2)^2 + \ldots + (X^d)^2 = - 1$ in flat space with signature $(2,d-1)$, which has closed timelike curves and must be ``unwrapped.'' Once unwrapped, however, the point $-X$ has an infinite number of representatives in the covering space, and as a result there is no way to continuously lift the antipodal map.\footnote{Furthermore, a $SO(1,d-1)$ invariant antipodal map cannot be defined in euclidean Anti-de Sitter space (EAdS) either. EAdS can be defined as the ``top cap'' of the hyperboloid $-(X^0)^2 + (\tilde{X}^1)^2 + \ldots + (X^d)^2 = - 1$. The natural embedding space antipodal map $X \mapsto - X$ maps the ``top cap'' with $X^0 > 0$ into the ``bottom cap'' with $X^0 < 0$, and thus does not map EAdS into itself.} If one tries to adopt an ad-hoc discontinuous definition of the antipodal map on the universal cover, said map will not be $SO(2,d-1)$ invariant. See figure \ref{fig:anti}.

Going back to de Sitter, where points are parameterized by $x = (t, \Omega) \in dS_{d}$ with $\Omega \in S_{d-1}$, let us define $\Omega^A$ to be the point on the unit $S_{d-1}$ sphere antipodal to $\Omega$. For each $x$ let us define the antipodal point $x^A$ to be
\begin{equation}
    x^A \equiv (-t, \Omega^A).
\end{equation}
For a point in euclidean de Sitter space $\tx = (\tilde{t}, \Omega) \in S_d$, we similarly define
\begin{equation}
    \tx^A \equiv (-\tilde{t}, \Omega^A).
\end{equation}

For a function $\phi$ on de Sitter space (either lorentzian or euclidean) denote $\phi^A$ to be the antipodally-flipped function
\begin{equation}
    \phi^A(t,\Omega) \equiv \phi(-t,\Omega^A), \hspace{1 cm} \phi^A(-i \tilde{t}, \Omega) = \phi(i \tilde{t},\Omega^A).
\end{equation}

If $\phi$ solves the equation of motion $(\nabla^2 - m^2) \phi = 0$, then $\phi^A$ will solve it as well. This enables us to define a certain symmetry transformation which will play a prominent role in the rest of this paper. Due to the linearity of the equation of motion for the free scalar field, we may consider the symmetry transformation
\begin{equation}\label{varA}
    \delta \phi = \epsilon \, \phi^A
\end{equation}
where $\epsilon$ is a tiny constant. This infinitesimal variation exponentiates to the finite transformation
\begin{equation}
    \phi' = \cosh(\alpha) \phi + \sinh(\alpha) \phi^A
\end{equation}
for finite parameter $\alpha$.

The variation \eqref{varA} is similar to a variation we previously considered, \eqref{varf}, except that now the linear shift in the field $\phi$ is field-dependent. 

Let us now use the Noether trick to find the corresponding conserved current. We promote $\epsilon$ to a tiny function $\varepsilon(\tx)$. The variation of the euclidean action on the spacetime patch $U \subset S_d$ under variation $\delta \phi(\tx) = \varepsilon(\tx) \phi^A(\tx)$ is
\begin{equation}\label{SvarA}
    \delta S_U[\phi] = \int_U d^d \tx \sqrt{g} \left(  ( \partial^\mu \varepsilon ) J^A_\mu - (\varepsilon \, \phi) ( \nabla^2 \phi^A - m^2 \phi^A)   \right) 
\end{equation}
where we required that $\varepsilon$ vanishes on $\partial U$ and have defined the current
\begin{equation}
    J^A_\mu \equiv \phi^A \pd_\mu \phi.
\end{equation}
Because the antipodal map $X \mapsto -X$ commutes with the dS isometry group, the current $J^A_\mu$ transforms like a proper spin-1 current under said isometry group.

Using \eqref{SvarA}, because $\varepsilon$ is an arbitrary function we know that
\begin{equation}\label{class_cons}
    \nabla^\mu J^{A}_\mu = -\phi ( \nabla^2 - m^2 ) \phi^A = 0
\end{equation}
is satisfied on-shell and thus $\nabla^\mu J^{A}_\mu = 0$ because $\phi^A$ solves the equation of motion.

Notice that $J^{A}_\mu$ is not a  locally-defined current. $J^A_\mu(\tx)$ depends on the values of $\phi$ at both $\tx$ and $\tx^A$. In fact, in \eqref{SvarA}, while we are ostensibly integrating over points in the subset $U \subset S_d$, the expression also depends on the value of $\phi$ at points outside of $U$, in the antipodally-flipped version of $U$ we could call $U^A$.

We're interested in investigating how the classical conservation law \eqref{class_cons} manifests as a path integral Ward identity after quantization. Let us first review how standard Ward identities are derived before showing how the Ward identity for $J^A_\mu$ deviates from the usual story. Say that we have some local symmetry variation $\phi \mapsto \phi + \epsilon 
\, \delta_{\rm var} \phi$ of our euclidean action \eqref{SfreeU} where $\epsilon$ is a tiny constant. Promoting $\epsilon$ to a function $\varepsilon(\tx)$ and considering the variation $\delta \phi = \varepsilon \, \delta_{\rm var} \phi$, the action will vary as

\begin{equation}\label{Jvarusual}
    \delta S_{S_d}[\phi] = \int_{S_d} d^d \tx \sqrt{g} \, \varepsilon \nabla^\mu J_\mu
\end{equation}
where $J_\mu$ is the Noether current of $\delta_{\rm var} \phi$. Then, defining $\phi ' = \phi + \varepsilon \delta_{\rm var} \phi$ and assuming  $\mD \phi' = \mD \phi$ (meaning there is no anomaly) we have
\begin{equation}
    \begin{aligned}
        \int \mD \phi \, e^{-S_{S_d}[\phi]} &= \int \mD \phi' e^{-S_{S_d}[\phi']} \\
        &= \int \mD \phi (1 - \int d^d \tx \sqrt{g} \, \varepsilon \, \nabla^\mu J_\mu) e^{-S_{S_d}[\phi]} 
    \end{aligned}
\end{equation}
and thus
\begin{equation}
    \int \mD \phi \, e^{-S_{S_d}[\phi]} \nabla^\mu J_\mu(\tx) = 0
\end{equation}
for any $\tx \in S_d$. This is a traditional Ward identity.

For our antipodal symmetry variation \eqref{varA}, the story is complicated somewhat. That's because the variation of the action \eqref{SvarA} does not look like the usual \eqref{Jvarusual}, due to the $\phi (\nabla^2 - m^2) \phi^A$ term. Of course, classically, this term vanishes on shell, but in the quantum path integral we are integrating over off-shell fields so it is not a priori clear if it vanishes.

If we attempt to repeat the standard Ward identity derivation using the antipodal variation \eqref{varA}, from  \eqref{SvarA} we see that we'll get\footnote{The antipodal variation has no anomaly because $\det ( \begin{smallmatrix} \cosh(\alpha) & \sinh(\alpha) \\ \sinh(\alpha) & \cosh(\alpha) \end{smallmatrix} ) = 1$.}
\begin{equation}\label{JAward}
    \int \mD \phi \, e^{- S_{S_d}[\phi]} \nabla^\mu J^{A}_\mu (\tx) = -\int \mD \phi \, e^{- S_{S_d}[\phi]} ( \phi (\nabla^2 - m^2) \phi^A) (\tx).
\end{equation}
Note that the RHS is not obviously zero. However, because we know that the two point function
\begin{equation}
    \langle \phi(\tx) \phi(\tx') \rangle = \int \mD \phi \,e^{-S_{S_d}[\phi]} \phi(\tx) \phi(\tx')
\end{equation}
satisfies
\begin{equation}
\begin{aligned}
    \sqrt{g(\tx)}(\nabla_{\tx}^2 - m^2) \langle \phi(\tx) \phi(\tx') \rangle &=- \int \mD \phi \, e^{-S_{S_d}[\phi]} \frac{\delta S_{S_d}}{\delta \phi(\tx)} \phi(\tx') \\
    &= \int \mD \phi \, \frac{\delta }{\delta \phi(\tx)} \left( e^{-S_{S_d}[\phi]} \right)  \phi(\tx ') \\
    &= -\int \mD \phi \, \left( e^{-S_{S_d}[\phi]} \right) \frac{\delta \phi(\tx') }{\delta \phi(\tx)} \\
    &= -\delta^{(4)}(\tx - \tx'),
\end{aligned}
\end{equation}
we know that, because $\tx$ and $\tx^A$ are never coincident, we have
\begin{equation}
    (\nabla_{\tx}^2 - m^2) \langle \phi(\tx) \phi(\tx^A) \rangle = 0.
\end{equation}
Thus the RHS of \eqref{JAward} is indeed 0 for all $\tx$. This implies that $\nabla^\mu J^{A}_\mu = 0$ really does hold as a quantum Ward identity, satisfying
\begin{equation}
    \int \mD \phi \, e^{- S_{S_d}[\phi]} \nabla^\mu J^{A}_\mu (\tx) = 0.
\end{equation}

\begin{figure}
    \centering
    \tikzset{every picture/.style={line width=0.75pt}} 

\begin{tikzpicture}[x=0.75pt,y=0.75pt,yscale=-1,xscale=1]

\draw [line width=1.5]  [dash pattern={on 5.63pt off 4.5pt}]  (309.67,72) .. controls (302.33,79.33) and (272.33,67.33) .. (276,86) ;
\draw   (60.02,57.18) .. controls (60.02,28.56) and (83.22,5.36) .. (111.84,5.36) .. controls (140.46,5.36) and (163.66,28.56) .. (163.66,57.18) .. controls (163.66,85.8) and (140.46,109) .. (111.84,109) .. controls (83.22,109) and (60.02,85.8) .. (60.02,57.18) -- cycle ;
\draw  [draw opacity=0][line width=1.5]  (163.66,57.18) .. controls (163.66,57.18) and (163.66,57.18) .. (163.66,57.18) .. controls (163.66,57.18) and (163.66,57.18) .. (163.66,57.18) .. controls (163.66,65.37) and (140.46,72.01) .. (111.84,72.01) .. controls (83.22,72.01) and (60.02,65.37) .. (60.02,57.18) -- (111.84,57.18) -- cycle ; \draw  [line width=1.5]  (163.66,57.18) .. controls (163.66,57.18) and (163.66,57.18) .. (163.66,57.18) .. controls (163.66,57.18) and (163.66,57.18) .. (163.66,57.18) .. controls (163.66,65.37) and (140.46,72.01) .. (111.84,72.01) .. controls (83.22,72.01) and (60.02,65.37) .. (60.02,57.18) ;  
\draw  [draw opacity=0][dash pattern={on 5.63pt off 4.5pt}][line width=1.5]  (60.02,57.18) .. controls (60.02,57.18) and (60.02,57.18) .. (60.02,57.18) .. controls (60.02,48.99) and (83.22,42.34) .. (111.84,42.34) .. controls (140.46,42.34) and (163.66,48.99) .. (163.66,57.18) -- (111.84,57.18) -- cycle ; \draw  [dash pattern={on 5.63pt off 4.5pt}][line width=1.5]  (60.02,57.18) .. controls (60.02,57.18) and (60.02,57.18) .. (60.02,57.18) .. controls (60.02,48.99) and (83.22,42.34) .. (111.84,42.34) .. controls (140.46,42.34) and (163.66,48.99) .. (163.66,57.18) ;  
\draw  [draw opacity=0][line width=0.75]  (370.99,56.85) .. controls (370.99,56.85) and (370.99,56.85) .. (370.99,56.85) .. controls (370.99,56.85) and (370.99,56.85) .. (370.99,56.85) .. controls (370.99,65.04) and (347.79,71.68) .. (319.17,71.68) .. controls (290.55,71.68) and (267.35,65.04) .. (267.35,56.85) -- (319.17,56.85) -- cycle ; \draw  [color={rgb, 255:red, 179; green, 179; blue, 179 }  ,draw opacity=1 ][line width=0.75]  (370.99,56.85) .. controls (370.99,56.85) and (370.99,56.85) .. (370.99,56.85) .. controls (370.99,56.85) and (370.99,56.85) .. (370.99,56.85) .. controls (370.99,65.04) and (347.79,71.68) .. (319.17,71.68) .. controls (290.55,71.68) and (267.35,65.04) .. (267.35,56.85) ;  
\draw  [draw opacity=0][dash pattern={on 4.5pt off 4.5pt}][line width=0.75]  (267.35,56.85) .. controls (267.35,56.85) and (267.35,56.85) .. (267.35,56.85) .. controls (267.35,48.65) and (290.55,42.01) .. (319.17,42.01) .. controls (347.79,42.01) and (370.99,48.65) .. (370.99,56.85) -- (319.17,56.85) -- cycle ; \draw  [color={rgb, 255:red, 179; green, 179; blue, 179 }  ,draw opacity=1 ][dash pattern={on 4.5pt off 4.5pt}][line width=0.75]  (267.35,56.85) .. controls (267.35,56.85) and (267.35,56.85) .. (267.35,56.85) .. controls (267.35,48.65) and (290.55,42.01) .. (319.17,42.01) .. controls (347.79,42.01) and (370.99,48.65) .. (370.99,56.85) ;  
\draw   (267.35,56.85) .. controls (267.35,28.23) and (290.55,5.02) .. (319.17,5.02) .. controls (347.79,5.02) and (370.99,28.23) .. (370.99,56.85) .. controls (370.99,85.47) and (347.79,108.67) .. (319.17,108.67) .. controls (290.55,108.67) and (267.35,85.47) .. (267.35,56.85) -- cycle ;
\draw [line width=1.5]    (276,86) .. controls (301.67,96) and (309.67,82.67) .. (333,86.67) ;
\draw [line width=1.5]    (333,86.67) .. controls (348.33,90.67) and (353,94) .. (362.33,84.67) ;
\draw [line width=1.5]  [dash pattern={on 5.63pt off 4.5pt}]  (362.33,84.67) .. controls (363.67,65.33) and (335.67,56) .. (309.67,72) ;
\draw [line width=1.5]    (328.99,41.69) .. controls (336.33,34.36) and (366.33,46.36) .. (362.66,27.69) ;
\draw [line width=1.5]  [dash pattern={on 5.63pt off 4.5pt}]  (362.66,27.69) .. controls (336.99,17.69) and (328.99,31.02) .. (305.66,27.02) ;
\draw [line width=1.5]  [dash pattern={on 5.63pt off 4.5pt}]  (305.66,27.02) .. controls (290.33,23.02) and (285.66,19.69) .. (276.33,29.02) ;
\draw [line width=1.5]    (276.33,29.02) .. controls (274.99,48.36) and (302.99,57.69) .. (328.99,41.69) ;

\draw (170.08,55.36) node [anchor=west] [inner sep=0.75pt]    {$E$};
\draw (53.81,54.37) node [anchor=east] [inner sep=0.75pt]    {$Q_{E}^{A}$};
\draw (372.08,90.36) node [anchor=west] [inner sep=0.75pt]    {$\Sigma $};
\draw (269.81,94.7) node [anchor=east] [inner sep=0.75pt]    {$Q_{\Sigma }^{A}$};
\draw (256.41,14.03) node [anchor=west] [inner sep=0.75pt]    {$\Sigma ^{A}$};
\draw (215.85,53.36) node    {$=$};

\end{tikzpicture}
    \caption{\label{qasigma} On the sphere path integral, the surface $\Sigma$ on which the antipodal charge $Q_\Sigma^A$ resides can be deformed, however one must keep in mind that this charge also depends on field values on the surface antipodal to $\Sigma$, which we denote as $\Sigma^A$.}
\end{figure}
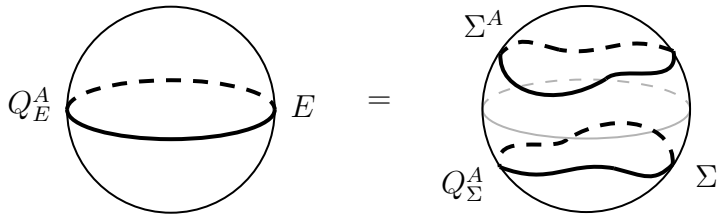

We can now construct a charge $Q_{\Sigma}^A$ by integrating $J^A_\mu$ over a codimension-1 surface $\Sigma$ (and multiplying by $1/2$).
\begin{equation}\label{QAdef}
    Q_{\Sigma}^A \equiv \frac{1}{2} \int_\Sigma d \Sigma^\mu J^A_\mu.
\end{equation}
We refer to this as the ``antipodal charge.'' This charge first appeared in the work of Burges \cite{Burges:1984qm}, who considered it only on the equitorial sliced $\Sigma = E$. The advantage of our current discussion is that we are able to see that, because $J^A_\mu$ is conserved, we can freely deform $\Sigma$ and $Q_\Sigma^A$ will not change. However, it is important to note that because $J^A_\mu(\tx)$ depends on the field at both $\tx$ and $\tx^A$, we know that $Q_\Sigma^A$ depends on field values at both $\Sigma$ and on the antipodally-flipped version of $\Sigma$, which we can call $\Sigma^A$. See figure \ref{qasigma}. For certain surfaces, $\Sigma$ and $\Sigma^A$ are actually the same surface, such as when $\Sigma$ is the equator $E$.

\section{Why the dS-invariant $\alpha$-vacua can be constructed}\label{sec6}

In section \ref{sec3}, we defined the Bunch-Davies state $\ket{0}$ and showed it was dS-invariant (i.e., invariant under the group of dS isometries). We now ask, how can we construct further dS-invariant states?

One possible method of constructing another dS-invariant state would be to act on $\ket{0}$ with some conserved charge, $\hat{Q}$. Here we are imagining that $\hat{Q}$ is defined as an integral over the equator $E$ of some locally-defined conserved current $J_\mu$, where $J_\mu$ itself transforms like a proper vector field under the dS isometry group.

Let us review why $\hat{Q} \ket{0}$ must be dS-invariant. Clearly it is invariant under the $SO(d)$ subgroup of spatial rotations, meaning the non-trivial step is to check that $\hat{L}^{01} \hat{Q} \ket{0} = 0$. We'll accomplish this by proving $[\hat{L}^{01}, \hat{Q}] = 0$. 

First, we insert $Q_E$ on the equator in the sphere path integral. (We can also put some extra arbitrary insertions into the upper or lower half-sphere in the path integral in order to prepare arbitrary bra and ket states.) Then we surround the equator by two dS-isometry generator operator insertions defined just above and below the equator, which are $1 - \epsilon L^{0 1}_{E_{\rm above}}$ and $1 + \epsilon L^{0 1}_{E_{\rm below}}$ respectively, where $\epsilon$ is a tiny angle. Because the current $J_\mu$ which $Q$ is an integral of transforms like a spin-1 vector, $1 - \epsilon L^{01}_{E_{\rm above}}$ and $1 + \epsilon L^{01}_{E_{\rm below}}$ will have the effect of rotating the surface on which $Q$ is defined by the angle $\epsilon$. We denote this new rotated surface as $E'$. See figure \ref{figcommute}. Because $Q$ is conserved, however, we are able to rotate $E'$ back to $E$. This implies $[\hat{L}^{01}, \hat{Q}] = 0$ as an operator equation (as the extra insertions in the upper and lower half-sphere are completely arbitrary), which is what we wanted to show.

\begingroup
\setlength{\abovecaptionskip}{-15pt} 

\begin{figure}
    \centering
    \input{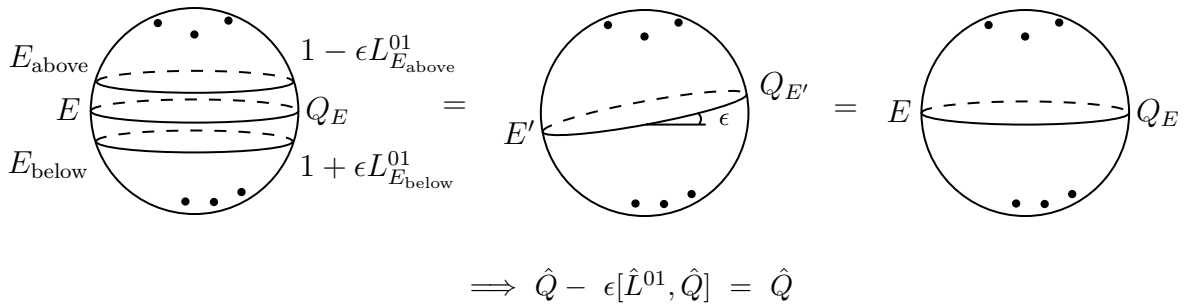}
    \caption{\label{figcommute} The argument that conserved charges commute with the de Sitter isometry generators. Arbitrary insertions are drawn in the upper and lower half-sphere to allow arbitrary bra and ket states to be prepared.}
\end{figure}
\endgroup

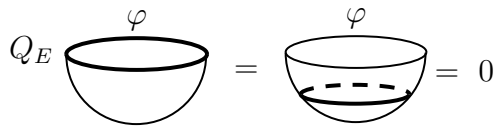
\begin{figure}[t]
    \centering
    \tikzset{every picture/.style={line width=0.75pt}} 

\begin{tikzpicture}[x=0.75pt,y=0.75pt,yscale=-1,xscale=1]

\draw  [draw opacity=0][line width=1.5]  (393,59.52) .. controls (393,59.52) and (393,59.52) .. (393,59.52) .. controls (393,63.93) and (377.27,67.5) .. (357.87,67.5) .. controls (338.47,67.5) and (322.74,63.93) .. (322.74,59.52) -- (357.87,59.52) -- cycle ; \draw  [line width=1.5]  (393,59.52) .. controls (393,59.52) and (393,59.52) .. (393,59.52) .. controls (393,63.93) and (377.27,67.5) .. (357.87,67.5) .. controls (338.47,67.5) and (322.74,63.93) .. (322.74,59.52) ;  
\draw  [draw opacity=0] (392.57,59.84) .. controls (392.57,79.13) and (376.94,94.76) .. (357.66,94.76) .. controls (338.37,94.76) and (322.74,79.13) .. (322.74,59.84) -- (357.66,59.84) -- cycle ; \draw   (392.57,59.84) .. controls (392.57,79.13) and (376.94,94.76) .. (357.66,94.76) .. controls (338.37,94.76) and (322.74,79.13) .. (322.74,59.84) ;  
\draw  [draw opacity=0][line width=1.5]  (322.74,59.52) .. controls (322.74,59.52) and (322.74,59.52) .. (322.74,59.52) .. controls (322.74,55.11) and (338.47,51.53) .. (357.87,51.53) .. controls (377.27,51.53) and (393,55.11) .. (393,59.52) -- (357.87,59.52) -- cycle ; \draw  [line width=1.5]  (322.74,59.52) .. controls (322.74,59.52) and (322.74,59.52) .. (322.74,59.52) .. controls (322.74,55.11) and (338.47,51.53) .. (357.87,51.53) .. controls (377.27,51.53) and (393,55.11) .. (393,59.52) ;  
\draw  [draw opacity=0][line width=0.75]  (502.5,58.52) .. controls (502.5,58.52) and (502.5,58.52) .. (502.5,58.52) .. controls (502.5,58.52) and (502.5,58.52) .. (502.5,58.52) .. controls (502.5,62.93) and (486.77,66.5) .. (467.37,66.5) .. controls (447.97,66.5) and (432.24,62.93) .. (432.24,58.52) -- (467.37,58.52) -- cycle ; \draw  [line width=0.75]  (502.5,58.52) .. controls (502.5,58.52) and (502.5,58.52) .. (502.5,58.52) .. controls (502.5,58.52) and (502.5,58.52) .. (502.5,58.52) .. controls (502.5,62.93) and (486.77,66.5) .. (467.37,66.5) .. controls (447.97,66.5) and (432.24,62.93) .. (432.24,58.52) ;  
\draw  [draw opacity=0] (502.07,58.84) .. controls (502.07,78.13) and (486.44,93.76) .. (467.16,93.76) .. controls (447.87,93.76) and (432.24,78.13) .. (432.24,58.84) -- (467.16,58.84) -- cycle ; \draw   (502.07,58.84) .. controls (502.07,78.13) and (486.44,93.76) .. (467.16,93.76) .. controls (447.87,93.76) and (432.24,78.13) .. (432.24,58.84) ;  
\draw  [draw opacity=0][line width=0.75]  (432.24,58.52) .. controls (432.24,58.52) and (432.24,58.52) .. (432.24,58.52) .. controls (432.24,54.11) and (447.97,50.53) .. (467.37,50.53) .. controls (486.77,50.53) and (502.5,54.11) .. (502.5,58.52) -- (467.37,58.52) -- cycle ; \draw  [line width=0.75]  (432.24,58.52) .. controls (432.24,58.52) and (432.24,58.52) .. (432.24,58.52) .. controls (432.24,54.11) and (447.97,50.53) .. (467.37,50.53) .. controls (486.77,50.53) and (502.5,54.11) .. (502.5,58.52) ;  
\draw  [draw opacity=0][line width=1.5]  (494,79.75) .. controls (494,79.75) and (494,79.75) .. (494,79.75) .. controls (494,82.37) and (481.85,84.5) .. (466.87,84.5) .. controls (451.89,84.5) and (439.74,82.37) .. (439.74,79.75) -- (466.87,79.75) -- cycle ; \draw  [line width=1.5]  (494,79.75) .. controls (494,79.75) and (494,79.75) .. (494,79.75) .. controls (494,82.37) and (481.85,84.5) .. (466.87,84.5) .. controls (451.89,84.5) and (439.74,82.37) .. (439.74,79.75) ;  
\draw  [draw opacity=0][dash pattern={on 5.63pt off 4.5pt}][line width=1.5]  (439.74,79.75) .. controls (439.74,79.75) and (439.74,79.75) .. (439.74,79.75) .. controls (439.74,77.13) and (451.89,75) .. (466.87,75) .. controls (481.85,75) and (494,77.13) .. (494,79.75) -- (466.87,79.75) -- cycle ; \draw  [dash pattern={on 5.63pt off 4.5pt}][line width=1.5]  (439.74,79.75) .. controls (439.74,79.75) and (439.74,79.75) .. (439.74,79.75) .. controls (439.74,77.13) and (451.89,75) .. (466.87,75) .. controls (481.85,75) and (494,77.13) .. (494,79.75) ;  

\draw (357.9,48.99) node [anchor=south] [inner sep=0.75pt]    {$\varphi $};
\draw (317.74,57.52) node [anchor=east] [inner sep=0.75pt]    {$Q_{E}$};
\draw (420.24,68.02) node [anchor=east] [inner sep=0.75pt]    {$=$};
\draw (467.4,47.99) node [anchor=south] [inner sep=0.75pt]    {$\varphi $};
\draw (505.21,68.02) node [anchor=west] [inner sep=0.75pt]    {$=\ 0$};

\end{tikzpicture}
    \caption{\label{fiqQEcontract}Locally-defined conserved charges will annihilate the Bunch-Davies state because the equator $E$ can be contracted.}
\end{figure}

We have now shown that $\hat{Q} \ket{0}$ is dS-invariant. However, there is a problem. Because $E$ is contractible on the half-sphere, the charge $\hat{Q}$ will actually annihilate the Bunch-Davies vacuum and $\hat Q \ket{0} = 0$! See figure \ref{fiqQEcontract}. So, while $\hat{Q} \ket{0}$ is dS-invariant for any locally-defined $\hat{Q}$, it is also zero.

However, our antipodal charge $Q^A$ has a unique property that locally-defined charges do not have: it is \textit{not} contractible in the usual way, and this implies that $\hat{Q}^A \ket{0} \neq 0$.

To see why this is the case, consider the path integral expression for $\hat{Q}^A \ket{0}$:
\begin{equation}
    (\hat{Q}^A \Psi_0) [\varphi] = \frac{1}{2} \int_{\eval{\phi}_E = \, \varphi } \mD \phi \;  e^{- S_{S_{d,-}}[\phi]}  \int_E d E^\mu  J^A_\mu.
\end{equation}

Because the equator $E$ is contractible on the half-sphere and $J^A_\mu$ is conserved, we might still expect that we could contract $E$ and the state would be zero. However, recall that $J^A_\mu(\tx)$ depends on both $\phi(\tx)$ and $\phi(\tx^A)$, and the path integral above \textit{only} integrates over values of $\phi(\tx)$ for $\tx$ contained in the lower half-sphere. This implies that we are \textit{disallowed} from inserting $J^A_\mu(\tx)$ anywhere on the lower half-sphere, aside from the equator itself.

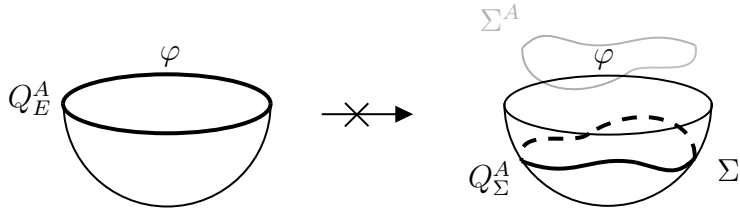
\begin{figure}[t]
    \centering
    \tikzset{every picture/.style={line width=0.75pt}} 

\begin{tikzpicture}[x=0.75pt,y=0.75pt,yscale=-1,xscale=1]

\draw [line width=1.5]  [dash pattern={on 5.63pt off 4.5pt}]  (322.67,72) .. controls (315.33,79.33) and (285.33,67.33) .. (289,86) ;
\draw  [draw opacity=0][line width=1.5]  (163.66,57.18) .. controls (163.66,57.18) and (163.66,57.18) .. (163.66,57.18) .. controls (163.66,57.18) and (163.66,57.18) .. (163.66,57.18) .. controls (163.66,65.37) and (140.46,72.01) .. (111.84,72.01) .. controls (83.22,72.01) and (60.02,65.37) .. (60.02,57.18) -- (111.84,57.18) -- cycle ; \draw  [line width=1.5]  (163.66,57.18) .. controls (163.66,57.18) and (163.66,57.18) .. (163.66,57.18) .. controls (163.66,57.18) and (163.66,57.18) .. (163.66,57.18) .. controls (163.66,65.37) and (140.46,72.01) .. (111.84,72.01) .. controls (83.22,72.01) and (60.02,65.37) .. (60.02,57.18) ;  
\draw  [draw opacity=0][line width=1.5]  (60.02,57.18) .. controls (60.02,57.18) and (60.02,57.18) .. (60.02,57.18) .. controls (60.02,48.99) and (83.22,42.34) .. (111.84,42.34) .. controls (140.46,42.34) and (163.66,48.99) .. (163.66,57.18) -- (111.84,57.18) -- cycle ; \draw  [line width=1.5]  (60.02,57.18) .. controls (60.02,57.18) and (60.02,57.18) .. (60.02,57.18) .. controls (60.02,48.99) and (83.22,42.34) .. (111.84,42.34) .. controls (140.46,42.34) and (163.66,48.99) .. (163.66,57.18) ;  
\draw [line width=1.5]    (289,86) .. controls (314.67,96) and (322.67,82.67) .. (346,86.67) ;
\draw [line width=1.5]    (346,86.67) .. controls (361.33,90.67) and (366,94) .. (375.33,84.67) ;
\draw [line width=1.5]  [dash pattern={on 5.63pt off 4.5pt}]  (375.33,84.67) .. controls (376.67,65.33) and (348.67,56) .. (322.67,72) ;
\draw [color={rgb, 255:red, 191; green, 191; blue, 191 }  ,draw opacity=1 ][line width=0.75]    (341.99,41.69) .. controls (349.33,34.36) and (379.33,46.36) .. (375.66,27.69) ;
\draw [color={rgb, 255:red, 179; green, 179; blue, 179 }  ,draw opacity=1 ][line width=0.75]    (375.66,27.69) .. controls (349.99,17.69) and (341.99,31.02) .. (318.66,27.02) ;
\draw [color={rgb, 255:red, 173; green, 173; blue, 173 }  ,draw opacity=1 ][line width=0.75]    (318.66,27.02) .. controls (303.33,23.02) and (298.66,19.69) .. (289.33,29.02) ;
\draw [color={rgb, 255:red, 173; green, 173; blue, 173 }  ,draw opacity=1 ][line width=0.75]    (289.33,29.02) .. controls (287.99,48.36) and (315.99,57.69) .. (341.99,41.69) ;
\draw  [draw opacity=0] (163.66,57.18) .. controls (163.66,57.18) and (163.66,57.18) .. (163.66,57.18) .. controls (163.66,85.71) and (140.53,108.84) .. (112,108.84) .. controls (83.46,108.84) and (60.33,85.71) .. (60.33,57.18) .. controls (60.33,57.18) and (60.33,57.18) .. (60.33,57.18) -- (112,57.18) -- cycle ; \draw   (163.66,57.18) .. controls (163.66,57.18) and (163.66,57.18) .. (163.66,57.18) .. controls (163.66,85.71) and (140.53,108.84) .. (112,108.84) .. controls (83.46,108.84) and (60.33,85.71) .. (60.33,57.18) .. controls (60.33,57.18) and (60.33,57.18) .. (60.33,57.18) ;  
\draw  [draw opacity=0] (383.68,56.85) .. controls (383.68,56.85) and (383.68,56.85) .. (383.68,56.85) .. controls (383.68,85.38) and (360.55,108.51) .. (332.02,108.51) .. controls (303.48,108.51) and (280.35,85.38) .. (280.35,56.85) .. controls (280.35,56.85) and (280.35,56.85) .. (280.35,56.85) -- (332.02,56.85) -- cycle ; \draw   (383.68,56.85) .. controls (383.68,56.85) and (383.68,56.85) .. (383.68,56.85) .. controls (383.68,85.38) and (360.55,108.51) .. (332.02,108.51) .. controls (303.48,108.51) and (280.35,85.38) .. (280.35,56.85) .. controls (280.35,56.85) and (280.35,56.85) .. (280.35,56.85) ;  
\draw  [draw opacity=0][line width=0.75]  (384,57.85) .. controls (384,57.85) and (384,57.85) .. (384,57.85) .. controls (384,57.85) and (384,57.85) .. (384,57.85) .. controls (384,66.04) and (360.79,72.68) .. (332.17,72.68) .. controls (303.55,72.68) and (280.35,66.04) .. (280.35,57.85) -- (332.17,57.85) -- cycle ; \draw  [line width=0.75]  (384,57.85) .. controls (384,57.85) and (384,57.85) .. (384,57.85) .. controls (384,57.85) and (384,57.85) .. (384,57.85) .. controls (384,66.04) and (360.79,72.68) .. (332.17,72.68) .. controls (303.55,72.68) and (280.35,66.04) .. (280.35,57.85) ;  
\draw  [draw opacity=0][line width=0.75]  (280.35,57.85) .. controls (280.35,57.85) and (280.35,57.85) .. (280.35,57.85) .. controls (280.35,49.65) and (303.55,43.01) .. (332.17,43.01) .. controls (360.79,43.01) and (384,49.65) .. (384,57.85) -- (332.17,57.85) -- cycle ; \draw  [line width=0.75]  (280.35,57.85) .. controls (280.35,57.85) and (280.35,57.85) .. (280.35,57.85) .. controls (280.35,49.65) and (303.55,43.01) .. (332.17,43.01) .. controls (360.79,43.01) and (384,49.65) .. (384,57.85) ;  
\draw    (189.17,62.41) -- (231.83,62.41) ;
\draw [shift={(234.83,62.41)}, rotate = 180] [fill={rgb, 255:red, 0; green, 0; blue, 0 }  ][line width=0.08]  [draw opacity=0] (8.93,-4.29) -- (0,0) -- (8.93,4.29) -- cycle    ;
\draw    (200.83,56.75) -- (212.5,68.41) ;
\draw    (212.42,56.91) -- (200.75,68.58) ;

\draw (55.89,54.37) node [anchor=east] [inner sep=0.75pt]    {$Q_{E}^{A}$};
\draw (385.08,90.36) node [anchor=west] [inner sep=0.75pt]    {$\Sigma $};
\draw (285.3,94.7) node [anchor=east] [inner sep=0.75pt]    {$Q_{\Sigma }^{A}$};
\draw (290,14.03) node [anchor=east] [inner sep=0.75pt]  [color={rgb, 255:red, 179; green, 179; blue, 179 }  ,opacity=1 ]  {$\Sigma ^{A}$};
\draw (113.57,39.7) node [anchor=south] [inner sep=0.75pt]    {$\varphi $};
\draw (329.9,41.03) node [anchor=south] [inner sep=0.75pt]    {$\varphi $};

\end{tikzpicture}
    \caption{\label{figcannotcontract} In the half-sphere path integral, the antipodal charge integrated over the equator cannot be contracted to zero, because the points on the lower half-sphere have their antipodes in the upper half-sphere, and the field values there are not being integrated over in the path integral.}
\end{figure}

Because the antipodal mirror of $E$ would venture into the upper half-sphere as we contract $E$ in the lower half-sphere, we are therefore not allowed to contract the equator after acting $Q_E^A$ on the half-sphere, and this is why $\hat{Q}^A \ket{0} \neq 0$. See figure \ref{figcannotcontract}. This is the fact that lets us construct an infinite number of dS-invariant states.

Explicitly, the charge $\hat{Q}^A$ on the $t=0$ slice is equal to
\begin{equation}\label{QAhat}
    \hat{Q}^A = \int d^{d-1} \Omega  \;  \hat{\phi}(0,\Omega) \hat{\pi}(0,\Omega^A)
\end{equation}
where $d^{d-1} \Omega$ is the measure on the unit $S_{d-1}$. The commutator of this charge with the field operator and momentum operator is\footnote{Note that we defined the charge $Q^A$ with an extra factor of $1/2$ in \eqref{QAdef}. And yet, in \eqref{QAcomm} we see that $\hat{Q}^A$ does generate the original antipodal symmetry \eqref{varA} from which our Noether charge was constructed. It is strange that the charge requires an ad hoc multiplication by $1/2$ in order to generate the correct symmetry. This runs counter to the usual state of affairs, where the Noether charge generates the original symmetry under the commutator exactly.}
\begin{equation}\label{QAcomm}
    [\hat{\phi}(0,\Omega), \hat{Q}^A] = i \hat{\phi}(0,\Omega^A), \hspace{1 cm} [\hat{\pi}(0,\Omega), \hat{Q}^A] = -i \hat{\pi}(0,\Omega^A).
\end{equation}

We can exponentiate this charge to get a 1-parameter family of dS-invariant states
\begin{equation}
    \ket{\alpha} \equiv e^{i \alpha \hat{Q}^A} \ket{0}.
\end{equation}
These are the $\alpha$-vacua. They are not annihilated by the euclidean modes, but rather by a Bogoliubov transformation of the euclidean modes. From the relation
\begin{equation}
    e^{i \alpha \hat{Q}^A} \hat{a}(f) e^{-i \alpha \hat{Q}^A} = \cosh(\alpha) \hat{a}(f) - \sinh(\alpha) \hat{a}(f^A)
\end{equation}
we know that the $\alpha$-vacua satisfy
\begin{equation}\label{alpha_annihilate}
    \Big( \cosh(\alpha) \hat{a}(f_e) - \sinh (\alpha) \hat{a}(f_e^A) \Big) \ket{\alpha} = 0
\end{equation}
where $f_e$ is any euclidean mode.

\section{Normalizability of $\alpha$-vacua wavefunctionals}\label{sec7}

In principle, $\alpha$ can be a complex number if one does not mind having a non CPT-invariant vacuum. However, $\ket{\alpha}$ won't be normalizable for all $\alpha \in \mathbb{C}$. In this section we solve for the allowed range of $\alpha$ for which $\ket{\alpha}$ is normalizable, replicating the findings of \cite{Chernikov:1968zm,Burges:1984qm}.

The wavefunctionals of the $\alpha$-vacua $\Psi^E_\alpha[\varphi] = \braket{\varphi}{\alpha}$ on the equator are related to the wavefunctional of the Bunch-Davies state $\Psi^E_0[\varphi]$ via
\begin{equation}\label{psialpha}
    \Psi^E_\alpha[\varphi] = \Psi^E_0\left[\cosh(\alpha) \varphi + \sinh(\alpha) \varphi^A\right].
\end{equation}

Because $\Psi^E_0$ has short distance singularities, i.e., a diverging amount of correlation between nearby field variables, the above equation tells us that $\Psi^E_\alpha$ has both short-distance and ``antipodal-short-distance'' singularities, where there is a high-degree of correlation between field values at antipodal points.

First, we define a basis of spherical harmonics on the $d-1$ sphere $Y_{\ell m}(\Omega)$, where $\ell = 0, 1, 2, \ldots$ and $m$ is a generalized multi-index. Let us assume all the spherical harmonics are real
\begin{equation}
    Y_{\ell m}(\Omega)^* = Y_{\ell m}(\Omega)
\end{equation}
and normalized such that
\begin{equation}
    \int d^{d-1} \Omega \, Y_{\ell m}(\Omega) Y_{\ell' m'}(\Omega) = \delta_{\ell \ell'} \delta_{mm'}.
\end{equation}
Under the antipodal map, the spherical harmonics satisfy
\begin{equation}\label{YA}
    Y_{\ell m}(\Omega^A) = (-1)^\ell Y_{\ell m}(\Omega).
\end{equation}

Let us expand the test function $\varphi$ on the equatorial $d-1$ sphere into a basis of spherical harmonics as
\begin{equation}
    \varphi(\Omega) = \sum_{\ell,m} c_{\ell m} Y_{\ell m}(\Omega).
\end{equation}
The norm of $\ket{\alpha}$ on the $t = 0$ time slice is given by
\begin{equation}
    \braket{\alpha} = \int \prod_{\ell,m} d c_{\ell m} \Psi^{E*}_\alpha[\varphi] \Psi^E_\alpha[\varphi].
\end{equation}

The Bunch-Davies state on the equator $E$ is a gaussian wavefunctional
\begin{equation}\label{hhwavefunction}
    \Psi^E_0[\varphi] = \mathcal{N}^E_0 \exp(-\sum_{\ell,m} B^{E}_{0,\ell} (c_{\ell m})^2)
\end{equation}
where $\mathcal{N}^E_0$ is a normalization constant and $B^{E}_{0,\ell}$ are a collection of positive real coefficients. We won't need the exact expressions for $B^{E}_{0,\ell}$, but they can be found in \cite{Burges:1984qm} for dimensions $d=2,3,4$.

Let us now write down the wavefunctional for the $\alpha$-vacuum. Using \eqref{YA}, we have the equation
\begin{equation}
    \cosh(\alpha) Y_{\ell m}(\Omega) + \sinh(\alpha) Y_{\ell m}(\Omega^A) = e^{(-1)^\ell \alpha} Y_{\ell m}(\Omega)
\end{equation}
and by \eqref{psialpha}, we then have
\begin{equation}\label{alphawavefunction}
    \Psi^E_{\alpha}[\varphi] = \mathcal{N}^E_0 \exp( - \sum_{\ell,m} e^{(-1)^\ell 2 \alpha} B^{E}_{0,\ell} (c_{\ell m})^2 ).
\end{equation}
The $\alpha$-vacuum is therefore normalizable if and only if $\Re(e^{2 \alpha}) > 0$ and $\Re(e^{-2 \alpha}) > 0$, which is equivalent to just
\begin{equation}\label{Re}
    \Re(e^{2 \alpha}) > 0.
\end{equation}

As a final observation, from the expression of the wavefunctional \eqref{alphawavefunction}, we can also see that the $\alpha$ which parameterize $\ket{\alpha}$ are periodic up to shifts in $\pi i$.
\begin{equation}
    \ket{\alpha + \pi i} = \ket{\alpha}.
\end{equation}

\section{In and Out vacua}\label{sec8}

\subsection{Definition of in/out states}

The ``in'' and ``out'' vacua in de Sitter space are dS-invariant vacuum states in which the quantum fields obey definite exponential fall-offs at $t = \pm \infty$. It turns out that these ``in'' and ``out'' vacua are actually $\alpha$-vacua for special values of $\alpha$. In this section we review the definition the in/out vacua and compute these values of $\alpha$ using a simple argument.

The classical equation of motion for a free scalar field in de Sitter space is
\begin{equation}\label{lorentzianeom}
    (\nabla^2 - m^2) \phi = \Big( - \partial_t^2 - (d-1) \tanh t \, \partial_t + \frac{\nabla^2_{S_{d-1}}}{\cosh^2t} - m^2 \Big) \phi = 0.
\end{equation}
At $t \to \pm \infty$, solutions to this equation have the form
\begin{equation}
\begin{aligned}
    \lim_{t \to \infty} \phi(t,\Omega) &=   A(\Omega) e^{-h_+ t}  + B(\Omega)  e^{-h_- t} + \ldots, \\
    \lim_{t \to -\infty} \phi(t,\Omega) &=   C(\Omega) e^{+h_+ t} + D(\Omega) e^{+h_- t} + \ldots,
\end{aligned}
\end{equation}
where $A(\Omega)$, $B(\Omega)$, $C(\Omega)$, $D(\Omega)$ are arbitrary sphere functions, the ``$\ldots$'' terms are terms of order $\mathcal{O}(e^{-(h_+ + 2)|t|})$ and $\mathcal{O}(e^{-(h_- + 2)|t|})$, and $h_+$ and $h_-$ satisfy
\begin{equation}
    h_\pm^2 - (d-1) h_\pm + m^2 = 0
\end{equation}
and are given by
\begin{equation}
    h_\pm = \frac{d-1}{2} \pm \sqrt{\left(\frac{d-1}{2}\right)^2 - m^2}.
\end{equation}

These solutions behave qualitatively differently depending on if $m$ is bigger or smaller than $(d-1)/2$. For a ``heavy scalar'' with $m > (d-1)/2$, $h_+$ and $h_-$ are complex numbers and conjugates of each other. For a ``light scalar'' with $0 < m < (d-1)/2$, $h_\pm$ are two positive real numbers.

Say you have some solutions to the wave equation \eqref{lorentzianeom} called $f^{\rm out}_{+}$, $f^{\rm out}_{-}$, $f^{\rm in}_{+}$, $f^{\rm in}_{-}$, which at $t = \pm \infty$ satisfy
\begin{equation}
    \begin{aligned}
        &f^{\rm out}_+(t,\Omega) = A(\Omega) e^{-h_+ t} \hspace{0.5cm}&& f^{\rm out}_-(t,\Omega) = B(\Omega) e^{-h_- t}  &&\hspace{0.5 cm} \text{ as } t \to \infty\\
        &\hspace{0.18cm}f^{\rm in}_+(t,\Omega) = C(\Omega)e^{+h_+ t}  \hspace{0.5cm}&&\hspace{0.18cm}f^{\rm in}_-(t,\Omega) = D(\Omega)e^{+h_- t} && \hspace{0.5 cm} \text{ as } t \to -\infty
    \end{aligned}
\end{equation}
for some arbitrary functions $A$, $B$, $C$, $D$. The in/out vacua are the states which satisfy
\begin{equation}\label{inoutdef}
\begin{aligned}
    \hat{a}(f^{\rm out}_-) \ket{{\rm out},+} = 0, & \hspace{1 cm} \hat{a}(f^{\rm in}_-) \ket{{\rm in},+} = 0,  \\
    \hat{a}(f^{\rm out}_+) \ket{{\rm out},-} = 0, & \hspace{1 cm} \hat{a}(f^{\rm in}_+) \ket{{\rm in},-} = 0,
\end{aligned}
\end{equation}
for all possible $f^{\rm out/in}_{\pm}$. We alert the reader that not all of these states will be normalizable, as we'll soon see.

\subsection{Calculating $\alpha^{\rm out/in}_\pm$ using a simple argument}

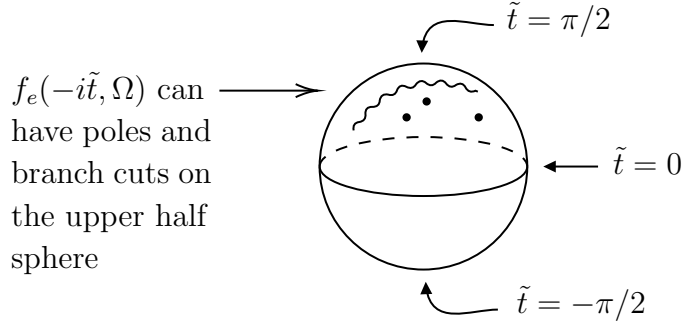
\begin{figure}
    \centering
    \tikzset{every picture/.style={line width=0.75pt}} 

\begin{tikzpicture}[x=0.75pt,y=0.75pt,yscale=-1,xscale=1]

\draw   (249.02,106.18) .. controls (249.02,77.56) and (272.22,54.36) .. (300.84,54.36) .. controls (329.46,54.36) and (352.66,77.56) .. (352.66,106.18) .. controls (352.66,134.8) and (329.46,158) .. (300.84,158) .. controls (272.22,158) and (249.02,134.8) .. (249.02,106.18) -- cycle ;
\draw  [draw opacity=0][line width=0.75]  (352.66,106.18) .. controls (352.66,106.18) and (352.66,106.18) .. (352.66,106.18) .. controls (352.66,106.18) and (352.66,106.18) .. (352.66,106.18) .. controls (352.66,114.37) and (329.46,121.01) .. (300.84,121.01) .. controls (272.22,121.01) and (249.02,114.37) .. (249.02,106.18) -- (300.84,106.18) -- cycle ; \draw  [line width=0.75]  (352.66,106.18) .. controls (352.66,106.18) and (352.66,106.18) .. (352.66,106.18) .. controls (352.66,106.18) and (352.66,106.18) .. (352.66,106.18) .. controls (352.66,114.37) and (329.46,121.01) .. (300.84,121.01) .. controls (272.22,121.01) and (249.02,114.37) .. (249.02,106.18) ;  
\draw  [draw opacity=0][dash pattern={on 4.5pt off 4.5pt}][line width=0.75]  (249.02,106.18) .. controls (249.02,97.99) and (272.22,91.34) .. (300.84,91.34) .. controls (329.46,91.34) and (352.66,97.99) .. (352.66,106.18) -- (300.84,106.18) -- cycle ; \draw  [dash pattern={on 4.5pt off 4.5pt}][line width=0.75]  (249.02,106.18) .. controls (249.02,97.99) and (272.22,91.34) .. (300.84,91.34) .. controls (329.46,91.34) and (352.66,97.99) .. (352.66,106.18) ;  
\draw    (336,35.01) .. controls (322.56,35.01) and (302.67,25.79) .. (300.21,46.29) ;
\draw [shift={(300,49.01)}, rotate = 272.39] [fill={rgb, 255:red, 0; green, 0; blue, 0 }  ][line width=0.08]  [draw opacity=0] (6.25,-3) -- (0,0) -- (6.25,3) -- cycle    ;
\draw    (388,106.18) -- (363.66,106.18) ;
\draw [shift={(360.66,106.18)}, rotate = 360] [fill={rgb, 255:red, 0; green, 0; blue, 0 }  ][line width=0.08]  [draw opacity=0] (6.25,-3) -- (0,0) -- (6.25,3) -- cycle    ;
\draw    (338,178.13) .. controls (324.56,178.13) and (304.67,187.35) .. (302.21,166.85) ;
\draw [shift={(302,164.13)}, rotate = 87.61] [fill={rgb, 255:red, 0; green, 0; blue, 0 }  ][line width=0.08]  [draw opacity=0] (6.25,-3) -- (0,0) -- (6.25,3) -- cycle    ;
\draw    (266,88.01) .. controls (265.58,85.52) and (266.49,84.12) .. (268.73,83.79) .. controls (271.08,83.41) and (272.14,82.05) .. (271.89,79.72) .. controls (271.78,77.35) and (272.89,76.16) .. (275.23,76.17) .. controls (277.66,76.22) and (278.93,75.13) .. (279.03,72.9) .. controls (279.5,70.5) and (280.92,69.53) .. (283.29,70) .. controls (285.51,70.67) and (286.98,69.91) .. (287.69,67.72) .. controls (288.6,65.55) and (290.11,64.98) .. (292.22,66.02) .. controls (294.36,67.15) and (296,66.74) .. (297.15,64.79) .. controls (298.72,62.86) and (300.38,62.64) .. (302.14,64.14) .. controls (303.95,65.72) and (305.62,65.68) .. (307.15,64.02) .. controls (309.05,62.45) and (310.71,62.57) .. (312.13,64.38) .. controls (313.6,66.28) and (315.24,66.55) .. (317.03,65.19) .. controls (319.16,63.96) and (320.76,64.36) .. (321.84,66.38) .. controls (322.98,68.48) and (324.63,69.03) .. (326.79,68.04) -- (328,68.51) ;
\draw  [fill={rgb, 255:red, 0; green, 0; blue, 0 }  ,fill opacity=1 ] (326.87,81.44) .. controls (326.87,80.58) and (327.57,79.87) .. (328.43,79.87) .. controls (329.3,79.87) and (330,80.58) .. (330,81.44) .. controls (330,82.31) and (329.3,83.01) .. (328.43,83.01) .. controls (327.57,83.01) and (326.87,82.31) .. (326.87,81.44) -- cycle ;
\draw  [fill={rgb, 255:red, 0; green, 0; blue, 0 }  ,fill opacity=1 ] (290.87,81.44) .. controls (290.87,80.58) and (291.57,79.87) .. (292.43,79.87) .. controls (293.3,79.87) and (294,80.58) .. (294,81.44) .. controls (294,82.31) and (293.3,83.01) .. (292.43,83.01) .. controls (291.57,83.01) and (290.87,82.31) .. (290.87,81.44) -- cycle ;
\draw  [fill={rgb, 255:red, 0; green, 0; blue, 0 }  ,fill opacity=1 ] (300.87,73.44) .. controls (300.87,72.58) and (301.57,71.87) .. (302.43,71.87) .. controls (303.3,71.87) and (304,72.58) .. (304,73.44) .. controls (304,74.31) and (303.3,75.01) .. (302.43,75.01) .. controls (301.57,75.01) and (300.87,74.31) .. (300.87,73.44) -- cycle ;
\draw    (199,68) -- (246,68) ;
\draw [shift={(248,68)}, rotate = 180] [color={rgb, 255:red, 0; green, 0; blue, 0 }  ][line width=0.75]    (10.93,-3.29) .. controls (6.95,-1.4) and (3.31,-0.3) .. (0,0) .. controls (3.31,0.3) and (6.95,1.4) .. (10.93,3.29)   ;

\draw (341,32.2) node [anchor=west] [inner sep=0.75pt]    {$\tilde{t} =\pi /2$};
\draw (346,175.2) node [anchor=west] [inner sep=0.75pt]    {$\tilde{t} =-\pi /2$};
\draw (394,103.2) node [anchor=west] [inner sep=0.75pt]    {$\tilde{t} =0$};
\draw (196.59,57) node [anchor=north east] [inner sep=0.75pt]   [align=left] {\begin{minipage}[lt]{75.01pt}\setlength\topsep{0pt}
$\displaystyle f_{e}( -i\tilde{t} ,\Omega )$ can 
\begin{center}
have poles and 
\end{center}
branch cuts on \\the upper half\\sphere
\end{minipage}};

\end{tikzpicture}
    \caption{\label{figf} The analytic structure of a euclidean mode $f_e$ on the euclidean sphere.}
\end{figure}

Consider a function $f_e(t,\Omega)$ which is a euclidean mode solution to \eqref{lorentzianeom}, meaning it is non-singular when analytically continued to the lower half euclidean sphere. On the upper half-sphere it may have poles and branch cuts, as depicted in figure \ref{figf}. 

At $t = \pm \infty$, $f_e(t,\Omega)$ must behave as
\begin{equation}\label{f_falloff}
\begin{aligned}
    \lim_{t \to \infty} f_e(t,\Omega) &=   A(\Omega) e^{-h_+ t} + B(\Omega) e^{-h_- t} + \ldots , \\
    \lim_{t \to -\infty} f_e(t,\Omega) &=   C(\Omega) e^{+h_+ t} + D(\Omega) e^{+h_- t} + \ldots,
\end{aligned}
\end{equation}
for some functions $A, B, C, D$, just like any solution to \eqref{lorentzianeom} must.

Let us now consider the behavior of the function $f_e(t + \pi i, \Omega)$ (which also solves \eqref{lorentzianeom}) at $t = \pm \infty$. It behaves as
\begin{equation}\label{fA_falloff}
\begin{aligned}
    \lim_{t \to \infty} f_e(t + \pi i,\Omega) &=   e^{-\pi i h_+}A(\Omega) e^{-h_+ t} + e^{-\pi i h_-} B(\Omega) e^{-h_- t} + \ldots , \\
    \lim_{t \to -\infty} f_e(t + \pi i,\Omega) &=   e^{+\pi i h_+}C(\Omega) e^{+h_+ t}  + e^{+\pi i h_+}D(\Omega) e^{+h_- t} + \ldots .
\end{aligned}
\end{equation}
In the embedding space parameterization of de Sitter $X^M(t,\Omega)$ \eqref{XM}, note that
\begin{equation}
    X^M(t + \pi i,\Omega) = - X^M(t,\Omega)
\end{equation}
so $t \mapsto t + \pi i$ acts the same as the antipodal map.

\begin{figure}
    \centering
    \tikzset{every picture/.style={line width=0.75pt}} 

\begin{tikzpicture}[x=0.75pt,y=0.75pt,yscale=-1,xscale=1]

\draw   (33.93,105.84) .. controls (33.93,86.61) and (49.52,71.02) .. (68.75,71.02) .. controls (87.98,71.02) and (103.57,86.61) .. (103.57,105.84) .. controls (103.57,125.07) and (87.98,140.66) .. (68.75,140.66) .. controls (49.52,140.66) and (33.93,125.07) .. (33.93,105.84) -- cycle ;
\draw  [draw opacity=0][line width=0.75]  (103.57,105.84) .. controls (103.57,111.35) and (87.98,115.81) .. (68.75,115.81) .. controls (49.52,115.81) and (33.93,111.35) .. (33.93,105.84) -- (68.75,105.84) -- cycle ; \draw  [line width=0.75]  (103.57,105.84) .. controls (103.57,111.35) and (87.98,115.81) .. (68.75,115.81) .. controls (49.52,115.81) and (33.93,111.35) .. (33.93,105.84) ;  
\draw  [draw opacity=0][dash pattern={on 4.5pt off 4.5pt}][line width=0.75]  (33.93,105.84) .. controls (33.93,100.34) and (49.52,95.88) .. (68.75,95.88) .. controls (87.98,95.88) and (103.57,100.34) .. (103.57,105.84) -- (68.75,105.84) -- cycle ; \draw  [dash pattern={on 4.5pt off 4.5pt}][line width=0.75]  (33.93,105.84) .. controls (33.93,100.34) and (49.52,95.88) .. (68.75,95.88) .. controls (87.98,95.88) and (103.57,100.34) .. (103.57,105.84) ;  
\draw   (140.33,105.64) .. controls (140.33,86.41) and (155.92,70.82) .. (175.15,70.82) .. controls (194.38,70.82) and (209.97,86.41) .. (209.97,105.64) .. controls (209.97,124.87) and (194.38,140.46) .. (175.15,140.46) .. controls (155.92,140.46) and (140.33,124.87) .. (140.33,105.64) -- cycle ;
\draw  [draw opacity=0][line width=0.75]  (197.46,130.48) .. controls (197.46,130.48) and (197.46,130.48) .. (197.46,130.48) .. controls (197.46,134.01) and (187.44,136.88) .. (175.09,136.88) .. controls (162.74,136.88) and (152.73,134.01) .. (152.73,130.48) -- (175.09,130.48) -- cycle ; \draw  [line width=0.75]  (197.46,130.48) .. controls (197.46,130.48) and (197.46,130.48) .. (197.46,130.48) .. controls (197.46,134.01) and (187.44,136.88) .. (175.09,136.88) .. controls (162.74,136.88) and (152.73,134.01) .. (152.73,130.48) ;  
\draw  [draw opacity=0][dash pattern={on 4.5pt off 4.5pt}][line width=0.75]  (152.73,130.48) .. controls (152.73,126.94) and (162.74,124.08) .. (175.09,124.08) .. controls (187.44,124.08) and (197.46,126.94) .. (197.46,130.48) -- (175.09,130.48) -- cycle ; \draw  [dash pattern={on 4.5pt off 4.5pt}][line width=0.75]  (152.73,130.48) .. controls (152.73,126.94) and (162.74,124.08) .. (175.09,124.08) .. controls (187.44,124.08) and (197.46,126.94) .. (197.46,130.48) ;  
\draw   (245.73,106.24) .. controls (245.73,87.01) and (261.32,71.42) .. (280.55,71.42) .. controls (299.78,71.42) and (315.37,87.01) .. (315.37,106.24) .. controls (315.37,125.47) and (299.78,141.06) .. (280.55,141.06) .. controls (261.32,141.06) and (245.73,125.47) .. (245.73,106.24) -- cycle ;
\draw  [fill={rgb, 255:red, 0; green, 0; blue, 0 }  ,fill opacity=1 ] (280.55,141.06) .. controls (280.55,141.82) and (279.94,142.44) .. (279.18,142.44) .. controls (278.42,142.44) and (277.81,141.82) .. (277.81,141.06) .. controls (277.81,140.31) and (278.42,139.69) .. (279.18,139.69) .. controls (279.94,139.69) and (280.55,140.31) .. (280.55,141.06) -- cycle ;
\draw   (351.33,105.64) .. controls (351.33,86.41) and (366.92,70.82) .. (386.15,70.82) .. controls (405.38,70.82) and (420.97,86.41) .. (420.97,105.64) .. controls (420.97,124.87) and (405.38,140.46) .. (386.15,140.46) .. controls (366.92,140.46) and (351.33,124.87) .. (351.33,105.64) -- cycle ;
\draw  [draw opacity=0][line width=0.75]  (408.46,130.48) .. controls (408.46,130.48) and (408.46,130.48) .. (408.46,130.48) .. controls (408.46,134.01) and (398.44,136.88) .. (386.09,136.88) .. controls (373.74,136.88) and (363.73,134.01) .. (363.73,130.48) -- (386.09,130.48) -- cycle ; \draw  [line width=0.75]  (408.46,130.48) .. controls (408.46,130.48) and (408.46,130.48) .. (408.46,130.48) .. controls (408.46,134.01) and (398.44,136.88) .. (386.09,136.88) .. controls (373.74,136.88) and (363.73,134.01) .. (363.73,130.48) ;  
\draw  [draw opacity=0][dash pattern={on 4.5pt off 4.5pt}][line width=0.75]  (363.73,130.48) .. controls (363.73,126.94) and (373.74,124.08) .. (386.09,124.08) .. controls (398.44,124.08) and (408.46,126.94) .. (408.46,130.48) -- (386.09,130.48) -- cycle ; \draw  [dash pattern={on 4.5pt off 4.5pt}][line width=0.75]  (363.73,130.48) .. controls (363.73,126.94) and (373.74,124.08) .. (386.09,124.08) .. controls (398.44,124.08) and (408.46,126.94) .. (408.46,130.48) ;  
\draw   (456.13,105.24) .. controls (456.13,86.01) and (471.72,70.42) .. (490.95,70.42) .. controls (510.18,70.42) and (525.77,86.01) .. (525.77,105.24) .. controls (525.77,124.47) and (510.18,140.06) .. (490.95,140.06) .. controls (471.72,140.06) and (456.13,124.47) .. (456.13,105.24) -- cycle ;
\draw  [draw opacity=0][line width=0.75]  (525.77,105.24) .. controls (525.77,110.75) and (510.18,115.21) .. (490.95,115.21) .. controls (471.72,115.21) and (456.13,110.75) .. (456.13,105.24) -- (490.95,105.24) -- cycle ; \draw  [line width=0.75]  (525.77,105.24) .. controls (525.77,110.75) and (510.18,115.21) .. (490.95,115.21) .. controls (471.72,115.21) and (456.13,110.75) .. (456.13,105.24) ;  
\draw  [draw opacity=0][dash pattern={on 4.5pt off 4.5pt}][line width=0.75]  (456.13,105.24) .. controls (456.13,99.74) and (471.72,95.28) .. (490.95,95.28) .. controls (510.18,95.28) and (525.77,99.74) .. (525.77,105.24) -- (490.95,105.24) -- cycle ; \draw  [dash pattern={on 4.5pt off 4.5pt}][line width=0.75]  (456.13,105.24) .. controls (456.13,99.74) and (471.72,95.28) .. (490.95,95.28) .. controls (510.18,95.28) and (525.77,99.74) .. (525.77,105.24) ;  
\draw    (111.57,105.84) -- (129.17,105.84) ;
\draw [shift={(132.17,105.84)}, rotate = 180] [fill={rgb, 255:red, 0; green, 0; blue, 0 }  ][line width=0.08]  [draw opacity=0] (10.72,-5.15) -- (0,0) -- (10.72,5.15) -- (7.12,0) -- cycle    ;
\draw    (218.97,105.64) -- (236.57,105.64) ;
\draw [shift={(239.57,105.64)}, rotate = 180] [fill={rgb, 255:red, 0; green, 0; blue, 0 }  ][line width=0.08]  [draw opacity=0] (10.72,-5.15) -- (0,0) -- (10.72,5.15) -- (7.12,0) -- cycle    ;
\draw    (323.37,106.24) -- (340.97,106.24) ;
\draw [shift={(343.97,106.24)}, rotate = 180] [fill={rgb, 255:red, 0; green, 0; blue, 0 }  ][line width=0.08]  [draw opacity=0] (10.72,-5.15) -- (0,0) -- (10.72,5.15) -- (7.12,0) -- cycle    ;
\draw    (426.97,105.64) -- (444.57,105.64) ;
\draw [shift={(447.57,105.64)}, rotate = 180] [fill={rgb, 255:red, 0; green, 0; blue, 0 }  ][line width=0.08]  [draw opacity=0] (10.72,-5.15) -- (0,0) -- (10.72,5.15) -- (7.12,0) -- cycle    ;
\draw    (45,161.35) .. controls (29.56,154.91) and (9.15,132.34) .. (26.91,113.39) ;
\draw [shift={(29,111.35)}, rotate = 137.82] [fill={rgb, 255:red, 0; green, 0; blue, 0 }  ][line width=0.08]  [draw opacity=0] (8.04,-3.86) -- (0,0) -- (8.04,3.86) -- (5.34,0) -- cycle    ;
\draw    (417,161.68) .. controls (425.02,153.66) and (425.62,145.65) .. (417.5,137.37) ;
\draw [shift={(415.36,135.35)}, rotate = 41.12] [fill={rgb, 255:red, 0; green, 0; blue, 0 }  ][line width=0.08]  [draw opacity=0] (8.04,-3.86) -- (0,0) -- (8.04,3.86) -- (5.34,0) -- cycle    ;
\draw    (517.67,162.35) .. controls (533.75,152.05) and (549.55,129.65) .. (531.46,110.72) ;
\draw [shift={(529.36,108.68)}, rotate = 42.18] [fill={rgb, 255:red, 0; green, 0; blue, 0 }  ][line width=0.08]  [draw opacity=0] (8.04,-3.86) -- (0,0) -- (8.04,3.86) -- (5.34,0) -- cycle    ;
\draw    (147,160.68) .. controls (145.32,159) and (143.97,157.32) .. (142.95,155.64) .. controls (139.1,149.27) and (140.09,142.87) .. (146.56,136.3) ;
\draw [shift={(148.64,134.35)}, rotate = 138.88] [fill={rgb, 255:red, 0; green, 0; blue, 0 }  ][line width=0.08]  [draw opacity=0] (8.04,-3.86) -- (0,0) -- (8.04,3.86) -- (5.34,0) -- cycle    ;
\draw    (280,159.33) -- (280,149.01) ;
\draw [shift={(280,146.01)}, rotate = 90] [fill={rgb, 255:red, 0; green, 0; blue, 0 }  ][line width=0.08]  [draw opacity=0] (8.04,-3.86) -- (0,0) -- (8.04,3.86) -- (5.34,0) -- cycle    ;

\draw (69.44,159.93) node [anchor=north] [inner sep=0.75pt]    {$\tilde{t} =0$};
\draw (175.04,159.43) node [anchor=north] [inner sep=0.75pt]    {$\tilde{t} =-\tfrac{\pi }{4}$};
\draw (279.51,159.96) node [anchor=north] [inner sep=0.75pt]    {$\tilde{t} =-\tfrac{\pi }{2}$};
\draw (387.56,160.03) node [anchor=north] [inner sep=0.75pt]    {$\tilde{t} =-\tfrac{3\pi }{4}$};
\draw (490.96,159.8) node [anchor=north] [inner sep=0.75pt]    {$\tilde{t} =-\pi $};

\end{tikzpicture}
    \caption{\label{fig5spheres} As the surface $\tilde{t} = 0$ is smoothly interpolated to $\tilde{t} = - \pi$, the surface passes through the south pole and comes out antipodally flipped.}
\end{figure}
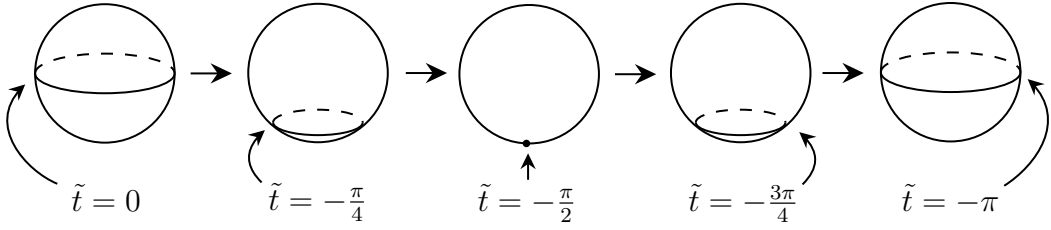

At a first glance, it seems as though $t \mapsto t - \pi i$ would also correspond to the antipodal map, however there is a key difference between $t + \pi i$ and $t - \pi i$. Recalling our notation $t = -i \tilde{t}$, consider the how the surface $\tilde{t} = w$ evolves on the sphere as $w$ is smoothly interpolated from $0$ to $-\pi$. The surface $\tilde{t} = 0$ is the $S_{d-1}$ equator, and then as $w$ decreases, the $S_{d-1}$ moves downward on the euclidean $S_d$ sphere, shrinks to a point at the south pole when $w = - \pi /2$, and then moves back up the sphere until it reaches the equator again at $\tilde{t} = - \pi$, but flipped antipodally. This is shown in figure \ref{fig5spheres}. Therefore, $f_e(\pi i,\Omega) = f_e(0,\Omega^A)$. Likewise, one can show that a point just above the equator  $(\epsilon,\Omega)$ gets mapped to an antipodal point just under the equator $(- \epsilon,\Omega^A)$ under $\tilde{t} \mapsto \tilde{t} - \pi$, implying $(\partial_t f_e)(\pi i,\Omega) = -(\partial_t f_e)(0,\Omega^A)$. Because $f_e(t+\pi i,\Omega)$ is guaranteed to also solve the second order wave equation \eqref{lorentzianeom}, these properties imply that all euclidean modes $f_e$ satisfy
\begin{equation}\label{ftA}
    f_e(t + \pi i, \Omega) = f_e^A(t,\Omega) = f_e(- t, \Omega^A)
\end{equation}
for all $t,\Omega$. The same could not necessarily be said for $f_e(t - \pi i, \Omega)$, because any path that travels through the upper half-sphere may encounter poles or branch cuts which cannot be safely navigated through.

If we use \eqref{f_falloff}, \eqref{fA_falloff}, and \eqref{ftA}, we can engineer in/out modes with the falloffs $h_\pm$ at $t = \pm \infty$ via the linear combinations
\begin{equation}\label{feout}
    \begin{aligned}
        &f_e(x) - e^{+ \pi i h_+}f_e(x^A) \sim e^{-h_- t} \hspace{1 cm} \text{at } t \to +\infty \\
        &f_e(x) - e^{+ \pi i h_-}f_e(x^A) \sim e^{-h_+ t} \hspace{1 cm} \text{at } t \to +\infty \\
        &f_e(x) - e^{-\pi i h_+}f_e(x^A) \sim e^{+h_- t} \hspace{1 cm} \text{at } t \to -\infty \\
        &f_e(x) - e^{-\pi i h_-}f_e(x^A) \sim e^{+h_+ t} \hspace{1 cm} \text{at } t \to -\infty
    \end{aligned}
\end{equation}
where we remind the reader that $f_e$ could be any euclidean mode. Comparing the above equation with the definition of the alpha vacua \eqref{alpha_annihilate} and the definition of the in/out vacua \eqref{inoutdef}, we can immediately see that
\begin{equation}
\begin{aligned}
    \ket{{\rm out},+} = \ket*{\alpha^{\rm out}_+}, &\hspace{0.5 cm} \ket{{\rm in},+} = \ket*{\alpha^{\rm in}_+}, \\
    \ket{{\rm out},-} = \ket*{\alpha^{\rm out}_-}, &\hspace{0.5 cm} \ket{{\rm in},-} = \ket*{\alpha^{\rm in}_-},
\end{aligned}
\end{equation}
where we define $\alpha^{\rm out/in}_\pm$ by
\begin{equation}\label{alpha_inout}
    \begin{aligned}
        \tanh \alpha^{\rm out}_+ &= e^{+ \pi i h_+}  ,   \\
        \tanh \alpha^{\rm out}_- &= e^{+ \pi i h_-} ,\\
        \tanh \alpha^{\rm in}_+ &= e^{- \pi i h_+}  ,  \\
        \tanh \alpha^{\rm in}_- &= e^{- \pi i h_-} .
    \end{aligned}
\end{equation}

We have now shown that the in/out vacua are $\alpha$-vacua and have solved for the corresponding $\alpha^{\rm out/in}_\pm$'s for all $d$ and $m$. The properties of the in/out vacua change qualitatively depending on whether $d$ is even or odd and whether the mass $m$ is heavy or light. Let us now go over all four cases separately. For heavy scalars we'll define
\begin{equation}
    h_\pm = \frac{d-1}{2} \pm i \mu, \hspace{1 cm} \mu = \sqrt{m^2 - \left( \frac{d-1}{2} \right)^2 } > 0,
\end{equation}
and for light scalars we'll define
\begin{equation}
    h_\pm = \frac{d-1}{2} \pm \nu, \hspace{1 cm} \nu = \sqrt{\left( \frac{d-1}{2} \right)^2 -m^2} > 0.
\end{equation}

\subsubsection*{Heavy scalar, $d$ odd:}
The solutions to \eqref{alpha_inout} in this case are
\begin{equation}
    \begin{aligned}
        \alpha^{\rm out}_+ &= (-1)^{\frac{d-1}{2}} \frac{1}{2} \log(\frac{e^{ \pi \mu}+1}{e^{ \pi \mu} - 1}) , \\
        \alpha^{\rm out}_- &= (-1)^{\frac{d-1}{2}} \frac{1}{2} \log(\frac{e^{\pi \mu} + 1}{e^{\pi \mu} - 1}) + \frac{i \pi}{2},\\
        \alpha^{\rm in}_+ &= (-1)^{\frac{d-1}{2}} \frac{1}{2} \log(\frac{e^{\pi \mu} + 1}{e^{\pi \mu} - 1}) + \frac{i \pi}{2},\\
        \alpha^{\rm in}_- &= (-1)^{\frac{d-1}{2}} \frac{1}{2} \log(\frac{e^{ \pi \mu}+1}{e^{ \pi \mu} - 1}) .
    \end{aligned}
\end{equation}
Notice that
\begin{equation}
    \Re(e^{2 \alpha^{\rm out}_+}) = \Re(e^{2 \alpha^{\rm in}_-}) > 0,
    \hspace{1 cm}
    \Re(e^{2 \alpha^{\rm out}_-}) = \Re(e^{2 \alpha^{\rm in}_+}) < 0 ,
\end{equation}
so only $\ket{{\rm out},+}$ and $\ket{{\rm in},-}$ are normalizable. $\ket{{\rm out},-}$ and $\ket{{\rm in},+}$ are not normalizable. Furthermore, because $e^{\pi i h_+} = e^{- \pi i h_-}$ in odd dimensions,
\begin{equation}\label{outequalsin}
    \alpha^{\rm out}_+ = \alpha^{\rm in}_- 
\end{equation}
which implies
\begin{equation}
    \ket{{\rm in},-} = \ket{{\rm out},+}. 
\end{equation}
We have now given a simple proof of the famous fact that the ``in'' vacuum equals the ``out'' vacuum for a heavy field in odd spacetime dimension. This fact was originally discovered by Bousso, Maloney, and Strominger in \cite{Bousso:2001mw}. Later, in a beautiful work by Lagogiannis, Maloney, and Wang \cite{Lagogiannis:2011st}, it was shown that the wave equation in odd dimensional de Sitter space can be recast into a problem in 1D supersymmetric quantum mechanics, and in that setting this fact can also be easily derived.

\subsubsection*{Heavy scalar, $d$ even:}
The solutions to \eqref{alpha_inout} here are
\begin{equation}
    \begin{aligned}
        \alpha^{\rm out}_\pm &=  (-1)^{\frac{d}{2}} \frac{1}{2} \log(\frac{1 - i e^{ \mp \pi \mu}}{1 + i e^{ \mp \pi \mu} }),\\
    \alpha^{\rm in}_\pm &= (-1)^{\frac{d}{2}} \frac{1}{2} \log(\frac{1 + i e^{ \pm \pi \mu}}{1 - 
 i e^{ \pm \pi \mu} }) .
    \end{aligned}
\end{equation}
Note that $\alpha^{\rm in/out}_\pm \in i \mathbb{R}$. Furthermore,
\begin{equation}
    \Re(e^{2 \alpha^{\rm out}_+}) = \Re(e^{2 \alpha^{\rm in}_-}) > 0 \hspace{1 cm} 
    \Re(e^{2 \alpha^{\rm out}_-}) = \Re(e^{2 \alpha^{\rm in}_+}) < 0 
\end{equation}
so again only $\ket{{\rm out},+}$ and $\ket{{\rm in},-}$ are normalizable. However now
\begin{equation}
        \alpha^{\rm out}_+ = (\alpha^{\rm in}_-)^*
\end{equation}
and therefore $\ket{{\rm out},+}$ and $\ket{{\rm in},-}$ are not the same state like they were in odd dimensions.

\subsubsection*{Light scalar, $d$ odd:}
Light scalars are fundamentally different from heavy scalars. The solutions to \eqref{alpha_inout} become
\begin{equation}
    \begin{aligned}
        \alpha^{\rm out}_\pm &= (-1)^{\frac{d-1}{2}} \frac{1}{2} \log(\cot \frac{\pi \nu}{2} ) \pm (-1)^{\frac{d-1}{2}} \frac{i \pi}{4} \, , \\
        \alpha^{\rm in}_\pm &= (-1)^{\frac{d-1}{2}} \frac{1}{2} \log( \cot \frac{\pi \nu}{2} ) \mp (-1)^{\frac{d-1}{2}} \frac{i \pi}{4} \, .
    \end{aligned}
\end{equation}
These equations imply
\begin{equation}
    \Re( e^{2 \alpha^{\rm out/in}_\pm}) = 0
\end{equation}
so strictly speaking, none of the states $\ket{{\rm out/in},\pm}$ are normalizable, but all of them are right at the edge of the normalizability condition \eqref{Re}. Notice that
\begin{equation}
    \alpha^{\rm out}_{\pm} = \alpha^{\rm in}_{\mp}
\end{equation}
which implies
\begin{equation}
    \begin{aligned}
        \ket{{\rm in},+} &= \ket{{\rm out},-},\\
        \ket{{\rm in},-} &= \ket{{\rm out},+},
    \end{aligned}
\end{equation}
for light scalars in odd dimensions. One consequence of the above equation is that a classical field in de Sitter which goes as $e^{h_\pm t}$ in the far past evolves into a field which goes as $e^{-h_\mp t}$ in the far future, meaning the field ``flips'' the rate at which it decays at the two boundaries from $h_\pm$ to $h_\mp$.

There is however a pathological case we will briefly mention. If $\nu \in \mathbb{Z}$ then $e^{\pi i h_+} = e^{\pi i h_-}$, meaning the $e^{- h_+ t}$ and $e^{- h_- t}$ modes at $\mIp$ and the $e^{ h_+ t}$ and $e^{ h_- t}$ modes at $\mIm$ cannot be separated out from the euclidean mode using the $t \mapsto t + i \pi$ shift as we did in \eqref{fA_falloff}. This implies that when $\nu \in \mathbb{Z}$, the in/out vacua are not actually $\alpha$-vacua and the entire analysis we've given breaks down.

\subsubsection*{Light scalar, $d$ even:}
The solutions to \eqref{alpha_inout} are
\begin{equation}
    \begin{aligned}
        \alpha^{\rm out}_\pm &= (-1)^{\frac{d}{2}} \frac{1}{2} \log (\frac{1 - i e^{\pm \pi i \nu}}{1 + i e^{\pm \pi i \nu}}), \\
        \alpha^{\rm in}_\pm &= (-1)^{\frac{d}{2}} \frac{1}{2} \log (\frac{1 + i e^{\mp \pi i \nu}}{1 - i e^{\mp \pi i \nu}}).
    \end{aligned}
\end{equation}
Once again,
\begin{equation}
    \Re( e^{2 \alpha^{\rm out/in}_\pm}) = 0
\end{equation}
so all of the in/out vacua are on the edge of normalizability. Furthermore,
\begin{equation}
    \alpha^{\rm out}_{\pm} = (\alpha^{\rm in}_{\mp})^*
\end{equation}
so generically we don't have an equality between in/out vacua like we had in odd dimensions. However, for special values of the mass where $\nu \in \mathbb{Z}+ \frac{1}{2}$, we have $e^{\pi i h_-} = e^{- \pi i h_-}$ and $e^{\pi i h_+} = e^{-\pi i h_+}$, implying
\begin{equation}
    \ket{{\rm out},+} = \ket{{\rm in},+}, \hspace{1 cm} \ket{{\rm out},-} = \ket{{\rm in},-}  \hspace{0.5 cm} \text{ if } \nu \in \mathbb{Z} + \frac{1}{2}.
\end{equation}
For these values of $\nu$, a free field which at $t = -\infty$ comes in as $e^{h_\pm t}$ leaves at $t = +\infty$ as $e^{-h_\pm t}$, meaning there is no ``flipping'' like we saw with light fields in odd dimensions.

There are $d-1$ possible half-integer values for $\nu$, which are
\begin{equation}
    \nu \in \left\{ \frac{1}{2}, \frac{3}{2}, \ldots, \frac{d-1}{2}\right\}.
\end{equation}
The conformally coupled scalar $m^2 = d(d-2)/4$ corresponds to $\nu = 1/2$. We note that the $\nu = (d-1)/2$ case, corresponding to a massless field $m = 0$, requires a bit more care in its analysis due to the zero mode. We are not providing that more careful analysis here. We also point out that the in/out vacua in the special case of the $d = 4$ conformally coupled scalar were studied in the work \cite{Ng:2012xp}.

Finally, just like when $d$ is odd, for the pathological $\nu \in \mathbb{Z}$ case the in/out vacua are not actually $\alpha$-vacua.

\section{Wavefunctionals of $\alpha$-vacua at $\mathcal{I}^+$ (for odd $d$ and $m > \tfrac{d-1}{2}$)}\label{sec9}

In this section we will study the $\alpha$-vacua wavefunctionals at timelike infinity $\mIp$ using the path integral. We will restrict our attention to odd spacetime dimension $d$ and heavy mass $m > (d-1)/2$. While we could do the $d$ even or light mass case if we wished, we are mainly interested in studying the basic features of this wavefunctional and therefore adopt these simplifications.

In subsection \ref{sec91} we solve for the Bunch-Davies wavefunctional at $\mIp$. In subsection \ref{sec92} we solve for the $\alpha$-vacuum wavefunctional at $\mIp$. In subsection \ref{sec93} we show how, for $|\alpha| \ll 1$, the $\alpha$-vacuum wavefunctional at $\mIp$ can also be prepared by moving the antipodal charge from $E$ to $\mIp$/$\mIm$ in the path integral. In subsection \ref{sec94} we explain why operator ordering problems prevent us from giving a similar analysis for the case of finite $\alpha$.

We define $\mIp$ and $\mIm$ as the constant $t$ surfaces
\begin{equation}
    \mIp = \{ (t = T,\Omega) \text{ surface} \}, \hspace{0.5 cm}\mIm = \{( t = -T,\Omega) \text{ surface} \}, \hspace{0.5 cm} \text{ for } T \to \infty.
\end{equation}

\subsection{Bunch-Davies wavefunctional at $\mIp$}\label{sec91}

The Bunch-Davies wavefunctional at $\mIp$, which we will denote as $\Psi_0^{\mIp}[\varphi]$, can be prepared by a path integral over the euclidean half-sphere followed by a lorentzian path integral from the equator $E$ to $\mIp$.
\begin{equation}
    \vcenter{\hbox{
    \tikzset{every picture/.style={line width=0.75pt}} 

\begin{tikzpicture}[x=0.75pt,y=0.75pt,yscale=-1,xscale=1]

\draw    (346.17,55.83) .. controls (331.13,68.48) and (332.45,88.16) .. (331.85,87.68) ;
\draw  [draw opacity=0][line width=0.75]  (346.17,54.54) .. controls (346.17,54.54) and (346.17,54.54) .. (346.17,54.54) .. controls (346.17,58.59) and (332.67,61.87) .. (316.03,61.87) .. controls (299.38,61.87) and (285.88,58.59) .. (285.88,54.54) -- (316.03,54.54) -- cycle ; \draw  [line width=0.75]  (346.17,54.54) .. controls (346.17,54.54) and (346.17,54.54) .. (346.17,54.54) .. controls (346.17,58.59) and (332.67,61.87) .. (316.03,61.87) .. controls (299.38,61.87) and (285.88,58.59) .. (285.88,54.54) ;  
\draw  [draw opacity=0][line width=0.75]  (285.88,54.54) .. controls (285.88,51.12) and (299.38,48.35) .. (316.03,48.35) .. controls (332.67,48.35) and (346.17,51.12) .. (346.17,54.54) -- (316.03,54.54) -- cycle ; \draw  [line width=0.75]  (285.88,54.54) .. controls (285.88,51.12) and (299.38,48.35) .. (316.03,48.35) .. controls (332.67,48.35) and (346.17,51.12) .. (346.17,54.54) ;  
\draw  [draw opacity=0][line width=0.75]  (331.85,87.68) .. controls (331.85,87.68) and (331.85,87.68) .. (331.85,87.68) .. controls (331.85,89.78) and (324.84,91.48) .. (316.2,91.48) .. controls (307.55,91.48) and (300.55,89.78) .. (300.55,87.68) -- (316.2,87.68) -- cycle ; \draw  [line width=0.75]  (331.85,87.68) .. controls (331.85,87.68) and (331.85,87.68) .. (331.85,87.68) .. controls (331.85,89.78) and (324.84,91.48) .. (316.2,91.48) .. controls (307.55,91.48) and (300.55,89.78) .. (300.55,87.68) ;  
\draw  [draw opacity=0][dash pattern={on 3pt off 2.25pt}][line width=0.75]  (300.55,87.68) .. controls (300.55,85.9) and (307.55,84.46) .. (316.2,84.46) .. controls (324.84,84.46) and (331.85,85.9) .. (331.85,87.68) -- (316.2,87.68) -- cycle ; \draw  [dash pattern={on 3pt off 2.25pt}][line width=0.75]  (300.55,87.68) .. controls (300.55,85.9) and (307.55,84.46) .. (316.2,84.46) .. controls (324.84,84.46) and (331.85,85.9) .. (331.85,87.68) ;  
\draw  [draw opacity=0] (332.05,87.68) .. controls (332.05,96.37) and (324.99,103.42) .. (316.3,103.42) .. controls (307.6,103.42) and (300.55,96.37) .. (300.55,87.68) -- (316.3,87.68) -- cycle ; \draw   (332.05,87.68) .. controls (332.05,96.37) and (324.99,103.42) .. (316.3,103.42) .. controls (307.6,103.42) and (300.55,96.37) .. (300.55,87.68) ;  
\draw    (286.22,55.83) .. controls (301.27,68.48) and (299.95,88.16) .. (300.55,87.68) ;

\draw (318.44,45.7) node [anchor=south] [inner sep=0.75pt]    {$\varphi $};
\draw (285.34,73.17) node [anchor=east] [inner sep=0.75pt]    {$\Psi _{0}^{\mIp}[ \varphi ] =$};

\end{tikzpicture}
    }}
\end{equation}

If one has a euclidean mode solution $f_e$, then the annihilation operator $\hat{a}(f_e)$ will annihilate the Bunch-Davies vacuum at $\mIp$ because $\mIp$ is contractible using the euclidean section.
\begin{equation}
    \vcenter{\hbox{
    \tikzset{every picture/.style={line width=0.75pt}} 

\begin{tikzpicture}[x=0.75pt,y=0.75pt,yscale=-1,xscale=1]

\draw    (241.67,69.83) .. controls (226.63,82.48) and (227.95,102.16) .. (227.35,101.68) ;
\draw  [draw opacity=0][line width=1.5]  (241.67,68.54) .. controls (241.67,68.54) and (241.67,68.54) .. (241.67,68.54) .. controls (241.67,72.59) and (228.17,75.87) .. (211.53,75.87) .. controls (194.88,75.87) and (181.38,72.59) .. (181.38,68.54) -- (211.53,68.54) -- cycle ; \draw  [line width=1.5]  (241.67,68.54) .. controls (241.67,68.54) and (241.67,68.54) .. (241.67,68.54) .. controls (241.67,72.59) and (228.17,75.87) .. (211.53,75.87) .. controls (194.88,75.87) and (181.38,72.59) .. (181.38,68.54) ;  
\draw  [draw opacity=0][line width=1.5]  (181.38,68.54) .. controls (181.38,65.12) and (194.88,62.35) .. (211.53,62.35) .. controls (228.17,62.35) and (241.67,65.12) .. (241.67,68.54) -- (211.53,68.54) -- cycle ; \draw  [line width=1.5]  (181.38,68.54) .. controls (181.38,65.12) and (194.88,62.35) .. (211.53,62.35) .. controls (228.17,62.35) and (241.67,65.12) .. (241.67,68.54) ;  
\draw  [draw opacity=0][line width=0.75]  (227.35,101.68) .. controls (227.35,101.68) and (227.35,101.68) .. (227.35,101.68) .. controls (227.35,103.78) and (220.34,105.48) .. (211.7,105.48) .. controls (203.05,105.48) and (196.05,103.78) .. (196.05,101.68) -- (211.7,101.68) -- cycle ; \draw  [line width=0.75]  (227.35,101.68) .. controls (227.35,101.68) and (227.35,101.68) .. (227.35,101.68) .. controls (227.35,103.78) and (220.34,105.48) .. (211.7,105.48) .. controls (203.05,105.48) and (196.05,103.78) .. (196.05,101.68) ;  
\draw  [draw opacity=0][dash pattern={on 3pt off 2.25pt}][line width=0.75]  (196.05,101.68) .. controls (196.05,99.9) and (203.05,98.46) .. (211.7,98.46) .. controls (220.34,98.46) and (227.35,99.9) .. (227.35,101.68) -- (211.7,101.68) -- cycle ; \draw  [dash pattern={on 3pt off 2.25pt}][line width=0.75]  (196.05,101.68) .. controls (196.05,99.9) and (203.05,98.46) .. (211.7,98.46) .. controls (220.34,98.46) and (227.35,99.9) .. (227.35,101.68) ;  
\draw  [draw opacity=0] (227.55,101.68) .. controls (227.55,110.37) and (220.49,117.42) .. (211.8,117.42) .. controls (203.1,117.42) and (196.05,110.37) .. (196.05,101.68) -- (211.8,101.68) -- cycle ; \draw   (227.55,101.68) .. controls (227.55,110.37) and (220.49,117.42) .. (211.8,117.42) .. controls (203.1,117.42) and (196.05,110.37) .. (196.05,101.68) ;  
\draw    (181.72,69.83) .. controls (196.77,82.48) and (195.45,102.16) .. (196.05,101.68) ;
\draw    (340.17,69.83) .. controls (325.13,82.48) and (326.45,102.16) .. (325.85,101.68) ;
\draw  [draw opacity=0][line width=0.75]  (340.17,68.54) .. controls (340.17,68.54) and (340.17,68.54) .. (340.17,68.54) .. controls (340.17,72.59) and (326.67,75.87) .. (310.03,75.87) .. controls (293.38,75.87) and (279.88,72.59) .. (279.88,68.54) -- (310.03,68.54) -- cycle ; \draw  [line width=0.75]  (340.17,68.54) .. controls (340.17,68.54) and (340.17,68.54) .. (340.17,68.54) .. controls (340.17,72.59) and (326.67,75.87) .. (310.03,75.87) .. controls (293.38,75.87) and (279.88,72.59) .. (279.88,68.54) ;  
\draw  [draw opacity=0][line width=0.75]  (279.88,68.54) .. controls (279.88,65.12) and (293.38,62.35) .. (310.03,62.35) .. controls (326.67,62.35) and (340.17,65.12) .. (340.17,68.54) -- (310.03,68.54) -- cycle ; \draw  [line width=0.75]  (279.88,68.54) .. controls (279.88,65.12) and (293.38,62.35) .. (310.03,62.35) .. controls (326.67,62.35) and (340.17,65.12) .. (340.17,68.54) ;  
\draw  [draw opacity=0][line width=0.75]  (325.85,101.68) .. controls (325.85,101.68) and (325.85,101.68) .. (325.85,101.68) .. controls (325.85,103.78) and (318.84,105.48) .. (310.2,105.48) .. controls (301.55,105.48) and (294.55,103.78) .. (294.55,101.68) -- (310.2,101.68) -- cycle ; \draw  [line width=0.75]  (325.85,101.68) .. controls (325.85,101.68) and (325.85,101.68) .. (325.85,101.68) .. controls (325.85,103.78) and (318.84,105.48) .. (310.2,105.48) .. controls (301.55,105.48) and (294.55,103.78) .. (294.55,101.68) ;  
\draw  [draw opacity=0][dash pattern={on 3pt off 2.25pt}][line width=0.75]  (294.55,101.68) .. controls (294.55,99.9) and (301.55,98.46) .. (310.2,98.46) .. controls (318.84,98.46) and (325.85,99.9) .. (325.85,101.68) -- (310.2,101.68) -- cycle ; \draw  [dash pattern={on 3pt off 2.25pt}][line width=0.75]  (294.55,101.68) .. controls (294.55,99.9) and (301.55,98.46) .. (310.2,98.46) .. controls (318.84,98.46) and (325.85,99.9) .. (325.85,101.68) ;  
\draw  [draw opacity=0] (326.05,101.68) .. controls (326.05,110.37) and (318.99,117.42) .. (310.3,117.42) .. controls (301.6,117.42) and (294.55,110.37) .. (294.55,101.68) -- (310.3,101.68) -- cycle ; \draw   (326.05,101.68) .. controls (326.05,110.37) and (318.99,117.42) .. (310.3,117.42) .. controls (301.6,117.42) and (294.55,110.37) .. (294.55,101.68) ;  
\draw    (280.22,69.83) .. controls (295.27,82.48) and (293.95,102.16) .. (294.55,101.68) ;
\draw  [draw opacity=0][line width=1.5]  (322,111.68) .. controls (322,111.68) and (322,111.68) .. (322,111.68) .. controls (322,113.23) and (316.82,114.49) .. (310.44,114.49) .. controls (304.06,114.49) and (298.88,113.23) .. (298.88,111.68) -- (310.44,111.68) -- cycle ; \draw  [line width=1.5]  (322,111.68) .. controls (322,111.68) and (322,111.68) .. (322,111.68) .. controls (322,113.23) and (316.82,114.49) .. (310.44,114.49) .. controls (304.06,114.49) and (298.88,113.23) .. (298.88,111.68) ;  
\draw  [draw opacity=0][dash pattern={on 3pt off 2.25pt}][line width=1.5]  (298.88,111.68) .. controls (298.88,111.68) and (298.88,111.68) .. (298.88,111.68) .. controls (298.88,110.37) and (304.06,109.31) .. (310.44,109.31) .. controls (316.82,109.31) and (322,110.37) .. (322,111.68) -- (310.44,111.68) -- cycle ; \draw  [dash pattern={on 3pt off 2.25pt}][line width=1.5]  (298.88,111.68) .. controls (298.88,111.68) and (298.88,111.68) .. (298.88,111.68) .. controls (298.88,110.37) and (304.06,109.31) .. (310.44,109.31) .. controls (316.82,109.31) and (322,110.37) .. (322,111.68) ;  

\draw (213.94,58.7) node [anchor=south] [inner sep=0.75pt]    {$\varphi $};
\draw (173.3,69.19) node [anchor=east] [inner sep=0.75pt]    {$a_{\mIp}( f_{e})$};
\draw (312.44,58.7) node [anchor=south] [inner sep=0.75pt]    {$\varphi $};
\draw (260.77,88.7) node    {$=$};
\draw (362.27,88.7) node    {$=0$};

\end{tikzpicture}
    }}
\end{equation}

\noindent We will now use this fact to solve for the wavefunctional. It is a gaussian functional
\begin{equation}
    \Psi^\mIp_0[\varphi]  = \mathcal{N}^\mIp_0\exp( - \sum_{\ell,m} B^{\mIp}_{0,\ell} c_{\ell m}^2 )
\end{equation}
for some normalization constant $\mathcal{N}^\mIp_0$ and coefficients $B^{\mIp}_{0,\ell}$. We are only interested in solving for $B^{\mIp}_{0,\ell}$.

Consider a euclidean mode solution of the form
\begin{equation}
    f_{e,\ell m}(t,\Omega) = y_{e, \ell}(t) Y_{\ell m} (\Omega).
\end{equation}
The equation
\begin{equation}
    \left( \hat{a}(f_{e,\ell m}) \Psi_0^\mIp \right) [\varphi] = 0
\end{equation}
becomes
\begin{equation}
    \Big( \frac{y_{e,\ell}(T)}{i \cosh^{d-1} T}   \pdv{c_{\ell m}} - \dot y_{e,\ell}(T)c_{\ell m} \Big) \exp( - \sum_{\ell,m} B^{\mIp}_{0,\ell} c_{\ell m}^2 ) = 0
\end{equation}
which implies, as $T$ becomes large,
\begin{equation}\label{BIp}
    B^{\mIp}_{0,\ell} = \frac{e^{(d-1)T}}{i 2^d } \frac{\dot y_{e,\ell}(T)}{y_{e,\ell}(T)}.
\end{equation}
We therefore need to solve for the large $T$ behavior of $y_{e,\ell}$ in order to solve for $B^{\mIp}_{0,\ell}$. 

This can be done using the following trick: earlier, we noted that a certain linear combination of euclidean and antipodally flipped euclidean modes sum together to produce an ``out'' mode. See equation \eqref{feout}. Therefore, if we know how the ``out'' mode behaves at $t = T$ and $t = -T$, we can solve for the behavior of the euclidean modes at $t = T$.

We will consider an ``out $-$'' mode which we'll decompose as
\begin{equation}\label{fo}
    f_{o,\ell m}(t,\Omega) = y_{o,\ell}(t) Y_{\ell m}(\Omega).
\end{equation}
As we discussed earlier,\footnote{see for instance \eqref{outequalsin}} solutions to the Klein-Gordon equation which behave like $e^{-h_- t}$ at $\mIp$ will behave like $e^{h_+ t}$ at $\mIm$ when $d$ is odd. However, there is also a relative phase that the solution will obtain between $\mIp$ and $\mIm$. This phase was solved for in \cite{Bousso:2001mw,Lagogiannis:2011st}.

Near $\mIp$ and $\mIm$, this ``out $-$'' mode behaves like
\begin{equation}\label{yout}
\begin{aligned}
    \lim_{t \to \infty} y_{o,\ell}(t) &= e^{+i \theta_\ell} e^{-h_- t} \\
    \lim_{t \to -\infty} y_{o,\ell}(t) &= e^{- i \theta_\ell} e^{+h_+ t}
\end{aligned}
\end{equation}
and the phase $\theta_\ell$ is
\begin{equation}
    e^{2 i \theta_\ell} = \prod_{n = 1}^{\ell + (d-3)/2 } \frac{\mu + i n}{\mu - i n}.
\end{equation}

We now apply \eqref{feout} and write
\begin{equation}
    f_{e,\ell m}(t,\Omega) - e^{ \pi i  h_+} f^A_{e,\ell m}(t,\Omega) \propto f_{o,\ell m}(t,\Omega) .
\end{equation}
This can be re-expressed as the matrix equation
\begin{equation}\label{911}
    \begin{pmatrix}
        1 & - e^{\pi i h_+} \\ - e^{\pi i h_+} & 1 \end{pmatrix} \begin{pmatrix} f_{e,\ell m}(t,\Omega) \\ f_{e,\ell m}^A(t,\Omega) \end{pmatrix} \propto \begin{pmatrix} f_{o,\ell m}(t,\Omega) \\ f^A_{o,\ell m}(t,\Omega) \end{pmatrix}.
\end{equation}
Note that we have not yet fixed the overall proportionality constant for the function $f_{e,\ell m}$. We now invert \eqref{911} and arbitrarily fix this constant with
\begin{equation}\label{828}
     \begin{pmatrix} f_{e,\ell m}(t,\Omega) \\ f^A_{e,\ell m}(t,\Omega) \end{pmatrix} = \begin{pmatrix}
        1 &  e^{\pi i h_+} \\  e^{\pi i h_+} & 1 \end{pmatrix} \begin{pmatrix} f_{o,\ell m}(t,\Omega) \\ f^A_{o,\ell m}(t,\Omega) \end{pmatrix} .
\end{equation}

Using the fact that $Y_{\ell m}(\Omega^A) = (-1)^\ell Y_{\ell m}(\Omega)$, the above equation implies the euclidean mode behaves at large times as
\begin{equation}
\begin{aligned}
    \lim_{t \to \infty} y_{e,\ell}(t) &= e^{i \theta_\ell - h_- t} + (-1)^\ell e^{\pi i h_+} e^{-i \theta_\ell - h_+ t}, \\
    \lim_{t \to- \infty} y_{e,\ell}(t)&=  e^{-i \theta_\ell + h_+ t} + (-1)^\ell e^{\pi i h_+} e^{i \theta_\ell + h_- t}.
\end{aligned}
\end{equation}
Plugging this into \eqref{BIp} we arrive at
\begin{equation}\label{Bip0}
\begin{aligned}
    B^{\mIp}_{0,\ell}  &= \frac{i e^{(d-1)T}}{2^d} \frac{h_- e^{i \theta_\ell - h_- T} + h_+ (-1)^\ell e^{\pi i h_+} e^{-i \theta_\ell - h_+ T}}{e^{i \theta_\ell - h_- T} + (-1)^\ell e^{\pi i h_+} e^{-i \theta_\ell - h_+ T}}\\
    &= \frac{i e^{(d-1)T}}{2^{d}} \left( \frac{d - 1}{2} + (-1)^{\ell}  \mu  \tan^{(-1)^\ell}( \theta_\ell + \mu T - \pi h_+ /2) \right).
\end{aligned}
\end{equation}
Now that we have solved for $B^{\mIp}_{0,\ell}$, we have solved for $\Psi^{\mIp}_{0}[\varphi]$. Notice that $B^{\mIp}_{0,\ell}$ contains an overall factor of $e^{(d-1)T}$, which grows as $T \to \infty$. This is the well-known infrared divergence of the de Sitter wavefunctional.

\subsection{$\alpha$-vacuum wavefunctional at $\mIp$}\label{sec92}

We write the wavefunctional of the $\alpha$-vacuum at $\mIp$ as
\begin{equation}
    \Psi^\mIp_\alpha[\varphi]  = \mathcal{N}^\mIp_\alpha\exp( - \sum_{\ell,m} B^{\mIp}_{\alpha,\ell} c_{\ell m}^2 ).
\end{equation}

If one defines the $\alpha$-rotated euclidean mode
\begin{equation}
    f_{\alpha,\ell m}(t,\Omega) = y_{\alpha,\ell }(t) Y_{\ell m}(\Omega)
\end{equation}
by
\begin{equation}
    f_{\alpha,\ell m}(t,\Omega) = \cosh (\alpha) f_{e, \ell m}(t,\Omega) - \sinh (\alpha) f_{e,\ell m}^A (t,\Omega)
\end{equation}
then one has
\begin{equation}
    y_{\alpha,\ell }(t) = \cosh ( \alpha) \,  y_{e,\ell}(t) - \sinh (\alpha) \, (-1)^\ell y_{e,\ell}(-t).
\end{equation}
The equation
\begin{equation}
    \left( \hat{a}(f_{\alpha, \ell m}) \Psi^{\mIp}_\alpha \right) [\varphi] = 0
\end{equation}
becomes
\begin{equation}
    \Big( \frac{y_{\alpha,\ell}(T)}{i \cosh^{d-1} T}   \pdv{c_{\ell m}} - \dot y_{\alpha,\ell}(T)c_{\ell m} \Big) \exp( - \sum_{\ell,m} B^{\mIp}_{\alpha,\ell} c_{\ell m}^2 ) = 0,
\end{equation}
which then implies
\begin{equation}\label{BIpalpha}
    B^{\mIp}_{\alpha,\ell} = \frac{e^{(d-1)T}}{i 2^d } \frac{\dot y_{\alpha,\ell}(T)}{y_{\alpha,\ell}(T)}.
\end{equation}

Because $y_{\alpha,\ell}(T)$ can be expressed given quantities we have already solved for earlier, we have now solved for $B^{\mIp}_{\alpha,\ell}$. The full expression with all quantities plugged in is not particularly enlightening.

The expression simplifies in the case of the ``out $+$'' vacuum to
\begin{equation}\label{Boutell}
    B^{\mIp}_{\alpha^{\rm out}_{+},\ell} = \frac{e^{(d-1)T}}{i 2^d } \frac{\dot y_{o,\ell}(T)}{y_{o,\ell}(T)} =  \frac{i}{2^d} e^{(d-1)T} h_-.
\end{equation}
Note that $B^{\mIp}_{\alpha^{\rm out}_{+},\ell}$ is $\ell$-independent. The ``out $+$'' wavefunctional is
\begin{equation}
    \Psi_{\alpha^{\rm out}_{+}}^{\mIp}[\varphi] = \mathcal{N}^{\mIp}_{\alpha^{\rm out}_+} \exp( -  \frac{i}{2^d} e^{(d-1)T} h_- \int d^{d-1}\Omega \, \varphi(\Omega)^2 ).
\end{equation}
We can see that there is no entanglement in this state between any two points on $\mIp$.

\subsection{Path integral calculation of $\Psi^{\mIp}_\alpha[\varphi]$ for $|\alpha| \ll 1$}\label{sec93}

In this section we will use some path integral tricks to compute the wavefunctional of the $\alpha$-vacuum at $\mIp$ for small $\alpha$. In order to convince the reader that $|\alpha| \ll 1$ is an interesting regime to consider, we note that the ``in'' and ``out'' vacua are $\alpha$-vacua with $|\alpha| \ll 1$ in the large mass limit with $m \Lambda^{-1/2} \gg 1$, where $\Lambda$ is the cosmological constant:
\begin{equation}
    m \to \infty \;\; \text{implies} \;\; \alpha^{\rm out}_{+}, \alpha^{\rm in}_{-} \to 0.
\end{equation}
In the real world, even the mass of the electron is extremely large compared to the cosmological constant, with $m_{\rm electron} \Lambda^{-1/2} \sim 10^{38}$, so this is a physically well motivated limit.

Henceforth, we take $\alpha$ small for the rest of the subsection.

The $\alpha$-vacuum wavefunctional $\Psi^E_\alpha[\varphi]$ on the $t = 0$ equator $E$ can be prepared via a euclidean path integral on the half-sphere with the charge inserted at the equator. The state can then be evolved to $\mIp$ with an added lorentzian path integral.

\begin{equation}\label{alphahalfdesitter}
    \Psi^E_\alpha[\varphi] = \vcenter{
    \hbox{
    \tikzset{every picture/.style={line width=0.75pt}} 

\begin{tikzpicture}[x=0.75pt,y=0.75pt,yscale=-1,xscale=1]

\draw  [draw opacity=0][line width=1.5]  (133.75,99.68) .. controls (133.75,99.68) and (133.75,99.68) .. (133.75,99.68) .. controls (133.75,101.78) and (126.74,103.48) .. (118.1,103.48) .. controls (109.45,103.48) and (102.45,101.78) .. (102.45,99.68) -- (118.1,99.68) -- cycle ; \draw  [line width=1.5]  (133.75,99.68) .. controls (133.75,99.68) and (133.75,99.68) .. (133.75,99.68) .. controls (133.75,101.78) and (126.74,103.48) .. (118.1,103.48) .. controls (109.45,103.48) and (102.45,101.78) .. (102.45,99.68) ;  
\draw  [draw opacity=0][line width=1.5]  (102.45,99.68) .. controls (102.45,97.9) and (109.45,96.46) .. (118.1,96.46) .. controls (126.74,96.46) and (133.75,97.9) .. (133.75,99.68) -- (118.1,99.68) -- cycle ; \draw  [line width=1.5]  (102.45,99.68) .. controls (102.45,97.9) and (109.45,96.46) .. (118.1,96.46) .. controls (126.74,96.46) and (133.75,97.9) .. (133.75,99.68) ;  
\draw  [draw opacity=0] (133.95,99.68) .. controls (133.95,108.37) and (126.89,115.42) .. (118.2,115.42) .. controls (109.5,115.42) and (102.45,108.37) .. (102.45,99.68) -- (118.2,99.68) -- cycle ; \draw   (133.95,99.68) .. controls (133.95,108.37) and (126.89,115.42) .. (118.2,115.42) .. controls (109.5,115.42) and (102.45,108.37) .. (102.45,99.68) ;  
\draw    (148.67,85.03) .. controls (147.81,90.44) and (143.79,93.91) .. (139.31,97.51) ;
\draw [shift={(137,99.36)}, rotate = 320.91] [fill={rgb, 255:red, 0; green, 0; blue, 0 }  ][line width=0.08]  [draw opacity=0] (8.04,-3.86) -- (0,0) -- (8.04,3.86) -- (5.34,0) -- cycle    ;

\draw (119.19,95) node [anchor=south] [inner sep=0.75pt]    {$\varphi $};
\draw (153.32,82.47) node [anchor=south] [inner sep=0.75pt]  [font=\scriptsize]  {$1+i\alpha Q_{E}^{A}$};

\end{tikzpicture}
    }}
    , \hspace{1 cm} \Psi^{\mathcal{I}^+}_\alpha[\varphi] = \vcenter{\hbox{ 
    \tikzset{every picture/.style={line width=0.75pt}} 

\begin{tikzpicture}[x=0.75pt,y=0.75pt,yscale=-1,xscale=1]

\draw    (158.17,55.83) .. controls (143.13,68.48) and (144.45,88.16) .. (143.85,87.68) ;
\draw  [draw opacity=0][line width=0.75]  (158.17,54.54) .. controls (158.17,54.54) and (158.17,54.54) .. (158.17,54.54) .. controls (158.17,58.59) and (144.67,61.87) .. (128.03,61.87) .. controls (111.38,61.87) and (97.88,58.59) .. (97.88,54.54) -- (128.03,54.54) -- cycle ; \draw  [line width=0.75]  (158.17,54.54) .. controls (158.17,54.54) and (158.17,54.54) .. (158.17,54.54) .. controls (158.17,58.59) and (144.67,61.87) .. (128.03,61.87) .. controls (111.38,61.87) and (97.88,58.59) .. (97.88,54.54) ;  
\draw  [draw opacity=0][line width=0.75]  (97.88,54.54) .. controls (97.88,51.12) and (111.38,48.35) .. (128.03,48.35) .. controls (144.67,48.35) and (158.17,51.12) .. (158.17,54.54) -- (128.03,54.54) -- cycle ; \draw  [line width=0.75]  (97.88,54.54) .. controls (97.88,51.12) and (111.38,48.35) .. (128.03,48.35) .. controls (144.67,48.35) and (158.17,51.12) .. (158.17,54.54) ;  
\draw  [draw opacity=0][line width=1.5]  (143.85,87.68) .. controls (143.85,87.68) and (143.85,87.68) .. (143.85,87.68) .. controls (143.85,89.78) and (136.84,91.48) .. (128.2,91.48) .. controls (119.55,91.48) and (112.55,89.78) .. (112.55,87.68) -- (128.2,87.68) -- cycle ; \draw  [line width=1.5]  (143.85,87.68) .. controls (143.85,87.68) and (143.85,87.68) .. (143.85,87.68) .. controls (143.85,89.78) and (136.84,91.48) .. (128.2,91.48) .. controls (119.55,91.48) and (112.55,89.78) .. (112.55,87.68) ;  
\draw  [draw opacity=0][dash pattern={on 3pt off 2.25pt}][line width=1.5]  (112.55,87.68) .. controls (112.55,85.9) and (119.55,84.46) .. (128.2,84.46) .. controls (136.84,84.46) and (143.85,85.9) .. (143.85,87.68) -- (128.2,87.68) -- cycle ; \draw  [dash pattern={on 3pt off 2.25pt}][line width=1.5]  (112.55,87.68) .. controls (112.55,85.9) and (119.55,84.46) .. (128.2,84.46) .. controls (136.84,84.46) and (143.85,85.9) .. (143.85,87.68) ;  
\draw  [draw opacity=0] (144.05,87.68) .. controls (144.05,96.37) and (136.99,103.42) .. (128.3,103.42) .. controls (119.6,103.42) and (112.55,96.37) .. (112.55,87.68) -- (128.3,87.68) -- cycle ; \draw   (144.05,87.68) .. controls (144.05,96.37) and (136.99,103.42) .. (128.3,103.42) .. controls (119.6,103.42) and (112.55,96.37) .. (112.55,87.68) ;  
\draw    (98.22,55.83) .. controls (113.27,68.48) and (111.95,88.16) .. (112.55,87.68) ;
\draw    (175.47,78.63) .. controls (174.59,84.17) and (159.58,86.09) .. (151.82,87.55) ;
\draw [shift={(149,88.16)}, rotate = 345.07] [fill={rgb, 255:red, 0; green, 0; blue, 0 }  ][line width=0.08]  [draw opacity=0] (8.04,-3.86) -- (0,0) -- (8.04,3.86) -- (5.34,0) -- cycle    ;

\draw (128.03,46.14) node [anchor=south] [inner sep=0.75pt]    {$\varphi $};
\draw (180.12,76.07) node [anchor=south] [inner sep=0.75pt]  [font=\scriptsize]  {$1+i\alpha Q_{E}^{A}$};

\end{tikzpicture}
    }}.
\end{equation}

The second equation in \eqref{alphahalfdesitter} is indeed an equation for $\Psi_\alpha^{\mathcal{I}^+}[\varphi]$, but as it is somewhat difficult to compute the lorentzian path integral from $E$ to $\mIp$, we will manipulate the expression further into an easier-to-compute form.

First, we note that the Bunch-Davies wavefunctional on $E$, which we have so far expressed using a half-sphere path integral, can also be computed by taking the Bunch-Davies wavefunctional on $\mathcal{I}^-$, denoted $\Psi^{\mathcal{I}^-}_0[\varphi]$, and evolving it forward in time from $\mathcal{I}^-$ to $E$ with a lorentzian path integral:
\begin{equation}
    \Psi^{E}_0[\varphi] = \vcenter{\hbox{
    \tikzset{every picture/.style={line width=0.75pt}} 

\begin{tikzpicture}[x=0.75pt,y=0.75pt,yscale=-1,xscale=1]

\draw  [draw opacity=0][line width=0.75]  (133.75,99.68) .. controls (133.75,99.68) and (133.75,99.68) .. (133.75,99.68) .. controls (133.75,101.78) and (126.74,103.48) .. (118.1,103.48) .. controls (109.45,103.48) and (102.45,101.78) .. (102.45,99.68) -- (118.1,99.68) -- cycle ; \draw  [line width=0.75]  (133.75,99.68) .. controls (133.75,99.68) and (133.75,99.68) .. (133.75,99.68) .. controls (133.75,101.78) and (126.74,103.48) .. (118.1,103.48) .. controls (109.45,103.48) and (102.45,101.78) .. (102.45,99.68) ;  
\draw  [draw opacity=0][line width=0.75]  (102.45,99.68) .. controls (102.45,97.9) and (109.45,96.46) .. (118.1,96.46) .. controls (126.74,96.46) and (133.75,97.9) .. (133.75,99.68) -- (118.1,99.68) -- cycle ; \draw  [line width=0.75]  (102.45,99.68) .. controls (102.45,97.9) and (109.45,96.46) .. (118.1,96.46) .. controls (126.74,96.46) and (133.75,97.9) .. (133.75,99.68) ;  
\draw  [draw opacity=0] (133.95,99.68) .. controls (133.95,108.37) and (126.89,115.42) .. (118.2,115.42) .. controls (109.5,115.42) and (102.45,108.37) .. (102.45,99.68) -- (118.2,99.68) -- cycle ; \draw   (133.95,99.68) .. controls (133.95,108.37) and (126.89,115.42) .. (118.2,115.42) .. controls (109.5,115.42) and (102.45,108.37) .. (102.45,99.68) ;  

\draw (119.19,95) node [anchor=south] [inner sep=0.75pt]    {$\varphi $};

\end{tikzpicture}
    }} = \int \mD \varphi' \Big( \vcenter{\hbox{
    \tikzset{every picture/.style={line width=0.75pt}} 

\begin{tikzpicture}[x=0.75pt,y=0.75pt,yscale=-1,xscale=1]

\draw    (158.17,82.85) .. controls (143.13,70.2) and (144.45,50.53) .. (143.85,51.01) ;
\draw  [draw opacity=0][dash pattern={on 3pt off 2.25pt}][line width=0.75]  (158.17,84.15) .. controls (158.17,84.15) and (158.17,84.15) .. (158.17,84.15) .. controls (158.17,80.1) and (144.67,76.81) .. (128.03,76.81) .. controls (111.38,76.81) and (97.88,80.1) .. (97.88,84.15) -- (128.03,84.15) -- cycle ; \draw  [dash pattern={on 3pt off 2.25pt}][line width=0.75]  (158.17,84.15) .. controls (158.17,84.15) and (158.17,84.15) .. (158.17,84.15) .. controls (158.17,80.1) and (144.67,76.81) .. (128.03,76.81) .. controls (111.38,76.81) and (97.88,80.1) .. (97.88,84.15) ;  
\draw  [draw opacity=0][line width=0.75]  (97.88,84.15) .. controls (97.88,87.57) and (111.38,90.34) .. (128.03,90.34) .. controls (144.67,90.34) and (158.17,87.57) .. (158.17,84.15) -- (128.03,84.15) -- cycle ; \draw  [line width=0.75]  (97.88,84.15) .. controls (97.88,87.57) and (111.38,90.34) .. (128.03,90.34) .. controls (144.67,90.34) and (158.17,87.57) .. (158.17,84.15) ;  
\draw  [draw opacity=0][line width=0.75]  (143.85,51.01) .. controls (143.85,51.01) and (143.85,51.01) .. (143.85,51.01) .. controls (143.85,48.91) and (136.84,47.2) .. (128.2,47.2) .. controls (119.55,47.2) and (112.55,48.91) .. (112.55,51.01) -- (128.2,51.01) -- cycle ; \draw  [line width=0.75]  (143.85,51.01) .. controls (143.85,51.01) and (143.85,51.01) .. (143.85,51.01) .. controls (143.85,48.91) and (136.84,47.2) .. (128.2,47.2) .. controls (119.55,47.2) and (112.55,48.91) .. (112.55,51.01) ;  
\draw  [draw opacity=0][line width=0.75]  (112.55,51.01) .. controls (112.55,52.78) and (119.55,54.22) .. (128.2,54.22) .. controls (136.84,54.22) and (143.85,52.78) .. (143.85,51.01) -- (128.2,51.01) -- cycle ; \draw  [line width=0.75]  (112.55,51.01) .. controls (112.55,52.78) and (119.55,54.22) .. (128.2,54.22) .. controls (136.84,54.22) and (143.85,52.78) .. (143.85,51.01) ;  
\draw    (98.22,82.85) .. controls (113.27,70.2) and (111.95,50.53) .. (112.55,51.01) ;

\draw (130.11,44.43) node [anchor=south] [inner sep=0.75pt]    {$\varphi $};
\draw (129.44,92.83) node [anchor=north] [inner sep=0.75pt]    {$\varphi '$};

\end{tikzpicture}
    }} \Big) \Psi^{\mathcal{I}^-}_0[\varphi'].
\end{equation}
Let us now take the above expression for the half-sphere path integral and plug it into \eqref{alphahalfdesitter}. The result is
\begin{equation}
    \Psi^{\mathcal{I}^+}_\alpha[\varphi] = \int \mD \varphi' \Big( \vcenter{\hbox{
    \tikzset{every picture/.style={line width=0.75pt}} 

\begin{tikzpicture}[x=0.75pt,y=0.75pt,yscale=-1,xscale=1]

\draw    (205.11,50.12) .. controls (184,72.01) and (185,89.01) .. (205.11,112.28) ;
\draw  [draw opacity=0][line width=0.75]  (205.11,48.83) .. controls (205.11,48.83) and (205.11,48.83) .. (205.11,48.83) .. controls (205.11,52.88) and (191.62,56.16) .. (174.97,56.16) .. controls (158.32,56.16) and (144.82,52.88) .. (144.82,48.83) -- (174.97,48.83) -- cycle ; \draw  [line width=0.75]  (205.11,48.83) .. controls (205.11,48.83) and (205.11,48.83) .. (205.11,48.83) .. controls (205.11,52.88) and (191.62,56.16) .. (174.97,56.16) .. controls (158.32,56.16) and (144.82,52.88) .. (144.82,48.83) ;  
\draw  [draw opacity=0][line width=0.75]  (144.82,48.83) .. controls (144.82,45.41) and (158.32,42.64) .. (174.97,42.64) .. controls (191.62,42.64) and (205.11,45.41) .. (205.11,48.83) -- (174.97,48.83) -- cycle ; \draw  [line width=0.75]  (144.82,48.83) .. controls (144.82,45.41) and (158.32,42.64) .. (174.97,42.64) .. controls (191.62,42.64) and (205.11,45.41) .. (205.11,48.83) ;  
\draw  [draw opacity=0][line width=0.75]  (205.19,112.28) .. controls (205.19,112.28) and (205.19,112.28) .. (205.19,112.28) .. controls (205.19,116.33) and (191.7,119.62) .. (175.05,119.62) .. controls (158.4,119.62) and (144.9,116.33) .. (144.9,112.28) -- (175.05,112.28) -- cycle ; \draw  [line width=0.75]  (205.19,112.28) .. controls (205.19,112.28) and (205.19,112.28) .. (205.19,112.28) .. controls (205.19,116.33) and (191.7,119.62) .. (175.05,119.62) .. controls (158.4,119.62) and (144.9,116.33) .. (144.9,112.28) ;  
\draw  [draw opacity=0][dash pattern={on 3pt off 2.25pt}][line width=0.75]  (144.82,112.28) .. controls (144.82,108.86) and (158.32,106.09) .. (174.97,106.09) .. controls (191.62,106.09) and (205.11,108.86) .. (205.11,112.28) -- (174.97,112.28) -- cycle ; \draw  [dash pattern={on 3pt off 2.25pt}][line width=0.75]  (144.82,112.28) .. controls (144.82,108.86) and (158.32,106.09) .. (174.97,106.09) .. controls (191.62,106.09) and (205.11,108.86) .. (205.11,112.28) ;  
\draw    (144.82,50.12) .. controls (165.94,72.01) and (164.94,89.01) .. (144.82,112.28) ;
\draw    (221.67,71.03) .. controls (221.67,82.31) and (213.13,81.92) .. (197.98,81.62) ;
\draw [shift={(195,81.57)}, rotate = 0.84] [fill={rgb, 255:red, 0; green, 0; blue, 0 }  ][line width=0.08]  [draw opacity=0] (8.04,-3.86) -- (0,0) -- (8.04,3.86) -- (5.34,0) -- cycle    ;
\draw  [draw opacity=0][line width=1.5]  (189.33,81.45) .. controls (189.33,81.45) and (189.33,81.45) .. (189.33,81.45) .. controls (189.33,83.39) and (182.88,84.96) .. (174.93,84.96) .. controls (166.98,84.96) and (160.53,83.39) .. (160.53,81.45) -- (174.93,81.45) -- cycle ; \draw  [line width=1.5]  (189.33,81.45) .. controls (189.33,81.45) and (189.33,81.45) .. (189.33,81.45) .. controls (189.33,83.39) and (182.88,84.96) .. (174.93,84.96) .. controls (166.98,84.96) and (160.53,83.39) .. (160.53,81.45) ;  
\draw  [draw opacity=0][dash pattern={on 3pt off 2.25pt}][line width=1.5]  (160.49,81.45) .. controls (160.49,81.45) and (160.49,81.45) .. (160.49,81.45) .. controls (160.49,79.82) and (166.94,78.49) .. (174.89,78.49) .. controls (182.85,78.49) and (189.3,79.82) .. (189.3,81.45) -- (174.89,81.45) -- cycle ; \draw  [dash pattern={on 3pt off 2.25pt}][line width=1.5]  (160.49,81.45) .. controls (160.49,81.45) and (160.49,81.45) .. (160.49,81.45) .. controls (160.49,79.82) and (166.94,78.49) .. (174.89,78.49) .. controls (182.85,78.49) and (189.3,79.82) .. (189.3,81.45) ;  
\draw  [color={rgb, 255:red, 255; green, 255; blue, 255 }  ,draw opacity=1 ][fill={rgb, 255:red, 255; green, 255; blue, 255 }  ,fill opacity=1 ] (143.31,20.34) .. controls (143.31,17.94) and (145.26,16) .. (147.66,16) .. controls (150.06,16) and (152,17.94) .. (152,20.34) .. controls (152,22.74) and (150.06,24.69) .. (147.66,24.69) .. controls (145.26,24.69) and (143.31,22.74) .. (143.31,20.34) -- cycle ;

\draw (177.79,40) node [anchor=south] [inner sep=0.75pt]    {$\varphi $};
\draw (176.79,120.81) node [anchor=north] [inner sep=0.75pt]    {$\varphi '$};
\draw (228.01,60.68) node  [font=\scriptsize]  {$1+i\alpha Q_{E}^{A}$};

\end{tikzpicture}
    }} \!\!\!\!\! \Big) \Psi^{\mathcal{I}^-}_0[\varphi'].
\end{equation}
Here comes the key step. We can now take the surface $E$ on which the insertion $1 + i \alpha Q_E^A$ is defined and deform it. We will deform $E$ to $\mathcal{I}^+$, which deforms the antipodal mirror of $E$ to $\mIm$. The non-local charge insertion will now have the effect of changing the boundary conditions of the lorentzian path integral from $\varphi$ at $\mathcal{I}^+$ and $\varphi'$ at $\mathcal{I}^-$ to $\varphi + \frac{\alpha}{2} {\varphi'}^A$ at $\mathcal{I}^+$  and $\varphi' - \frac{\alpha}{2} \varphi^A$ at $\mathcal{I}^-$.
\begin{equation} 
\vcenter{\hbox{
\tikzset{every picture/.style={line width=0.75pt}} 

\begin{tikzpicture}[x=0.75pt,y=0.75pt,yscale=-1,xscale=1]

\draw    (205.11,50.12) .. controls (184,72.01) and (185,89.01) .. (205.11,112.28) ;
\draw  [draw opacity=0][line width=0.75]  (205.11,48.83) .. controls (205.11,48.83) and (205.11,48.83) .. (205.11,48.83) .. controls (205.11,52.88) and (191.62,56.16) .. (174.97,56.16) .. controls (158.32,56.16) and (144.82,52.88) .. (144.82,48.83) -- (174.97,48.83) -- cycle ; \draw  [line width=0.75]  (205.11,48.83) .. controls (205.11,48.83) and (205.11,48.83) .. (205.11,48.83) .. controls (205.11,52.88) and (191.62,56.16) .. (174.97,56.16) .. controls (158.32,56.16) and (144.82,52.88) .. (144.82,48.83) ;  
\draw  [draw opacity=0][line width=0.75]  (144.82,48.83) .. controls (144.82,45.41) and (158.32,42.64) .. (174.97,42.64) .. controls (191.62,42.64) and (205.11,45.41) .. (205.11,48.83) -- (174.97,48.83) -- cycle ; \draw  [line width=0.75]  (144.82,48.83) .. controls (144.82,45.41) and (158.32,42.64) .. (174.97,42.64) .. controls (191.62,42.64) and (205.11,45.41) .. (205.11,48.83) ;  
\draw  [draw opacity=0][line width=0.75]  (205.19,112.28) .. controls (205.19,112.28) and (205.19,112.28) .. (205.19,112.28) .. controls (205.19,116.33) and (191.7,119.62) .. (175.05,119.62) .. controls (158.4,119.62) and (144.9,116.33) .. (144.9,112.28) -- (175.05,112.28) -- cycle ; \draw  [line width=0.75]  (205.19,112.28) .. controls (205.19,112.28) and (205.19,112.28) .. (205.19,112.28) .. controls (205.19,116.33) and (191.7,119.62) .. (175.05,119.62) .. controls (158.4,119.62) and (144.9,116.33) .. (144.9,112.28) ;  
\draw  [draw opacity=0][dash pattern={on 3pt off 2.25pt}][line width=0.75]  (144.82,112.28) .. controls (144.82,108.86) and (158.32,106.09) .. (174.97,106.09) .. controls (191.62,106.09) and (205.11,108.86) .. (205.11,112.28) -- (174.97,112.28) -- cycle ; \draw  [dash pattern={on 3pt off 2.25pt}][line width=0.75]  (144.82,112.28) .. controls (144.82,108.86) and (158.32,106.09) .. (174.97,106.09) .. controls (191.62,106.09) and (205.11,108.86) .. (205.11,112.28) ;  
\draw    (144.82,50.12) .. controls (165.94,72.01) and (164.94,89.01) .. (144.82,112.28) ;
\draw    (221.67,71.03) .. controls (221.67,82.31) and (213.13,81.92) .. (197.98,81.62) ;
\draw [shift={(195,81.57)}, rotate = 0.84] [fill={rgb, 255:red, 0; green, 0; blue, 0 }  ][line width=0.08]  [draw opacity=0] (8.04,-3.86) -- (0,0) -- (8.04,3.86) -- (5.34,0) -- cycle    ;
\draw  [draw opacity=0][line width=1.5]  (189.33,81.45) .. controls (189.33,81.45) and (189.33,81.45) .. (189.33,81.45) .. controls (189.33,83.39) and (182.88,84.96) .. (174.93,84.96) .. controls (166.98,84.96) and (160.53,83.39) .. (160.53,81.45) -- (174.93,81.45) -- cycle ; \draw  [line width=1.5]  (189.33,81.45) .. controls (189.33,81.45) and (189.33,81.45) .. (189.33,81.45) .. controls (189.33,83.39) and (182.88,84.96) .. (174.93,84.96) .. controls (166.98,84.96) and (160.53,83.39) .. (160.53,81.45) ;  
\draw  [draw opacity=0][dash pattern={on 3pt off 2.25pt}][line width=1.5]  (160.49,81.45) .. controls (160.49,81.45) and (160.49,81.45) .. (160.49,81.45) .. controls (160.49,79.82) and (166.94,78.49) .. (174.89,78.49) .. controls (182.85,78.49) and (189.3,79.82) .. (189.3,81.45) -- (174.89,81.45) -- cycle ; \draw  [dash pattern={on 3pt off 2.25pt}][line width=1.5]  (160.49,81.45) .. controls (160.49,81.45) and (160.49,81.45) .. (160.49,81.45) .. controls (160.49,79.82) and (166.94,78.49) .. (174.89,78.49) .. controls (182.85,78.49) and (189.3,79.82) .. (189.3,81.45) ;  
\draw  [color={rgb, 255:red, 255; green, 255; blue, 255 }  ,draw opacity=1 ][fill={rgb, 255:red, 255; green, 255; blue, 255 }  ,fill opacity=1 ] (143.31,20.34) .. controls (143.31,17.94) and (145.26,16) .. (147.66,16) .. controls (150.06,16) and (152,17.94) .. (152,20.34) .. controls (152,22.74) and (150.06,24.69) .. (147.66,24.69) .. controls (145.26,24.69) and (143.31,22.74) .. (143.31,20.34) -- cycle ;

\draw (177.79,40) node [anchor=south] [inner sep=0.75pt]    {$\varphi $};
\draw (176.79,120.81) node [anchor=north] [inner sep=0.75pt]    {$\varphi '$};
\draw (228.01,60.68) node  [font=\scriptsize]  {$1+i\alpha Q_{E}^{A}$};

\end{tikzpicture}
}} \!\!\!\!\! = \vcenter{\hbox{
\tikzset{every picture/.style={line width=0.75pt}} 

\begin{tikzpicture}[x=0.75pt,y=0.75pt,yscale=-1,xscale=1]

\draw    (205.11,50.12) .. controls (184,72.01) and (185,89.01) .. (205.11,112.28) ;
\draw  [draw opacity=0][line width=1.5]  (205.11,48.83) .. controls (205.11,48.83) and (205.11,48.83) .. (205.11,48.83) .. controls (205.11,52.88) and (191.62,56.16) .. (174.97,56.16) .. controls (158.32,56.16) and (144.82,52.88) .. (144.82,48.83) -- (174.97,48.83) -- cycle ; \draw  [line width=1.5]  (205.11,48.83) .. controls (205.11,48.83) and (205.11,48.83) .. (205.11,48.83) .. controls (205.11,52.88) and (191.62,56.16) .. (174.97,56.16) .. controls (158.32,56.16) and (144.82,52.88) .. (144.82,48.83) ;  
\draw  [draw opacity=0][line width=1.5]  (144.82,48.83) .. controls (144.82,45.41) and (158.32,42.64) .. (174.97,42.64) .. controls (191.62,42.64) and (205.11,45.41) .. (205.11,48.83) -- (174.97,48.83) -- cycle ; \draw  [line width=1.5]  (144.82,48.83) .. controls (144.82,45.41) and (158.32,42.64) .. (174.97,42.64) .. controls (191.62,42.64) and (205.11,45.41) .. (205.11,48.83) ;  
\draw  [draw opacity=0][line width=1.5]  (205.19,112.28) .. controls (205.19,112.28) and (205.19,112.28) .. (205.19,112.28) .. controls (205.19,116.33) and (191.7,119.62) .. (175.05,119.62) .. controls (158.4,119.62) and (144.9,116.33) .. (144.9,112.28) -- (175.05,112.28) -- cycle ; \draw  [line width=1.5]  (205.19,112.28) .. controls (205.19,112.28) and (205.19,112.28) .. (205.19,112.28) .. controls (205.19,116.33) and (191.7,119.62) .. (175.05,119.62) .. controls (158.4,119.62) and (144.9,116.33) .. (144.9,112.28) ;  
\draw  [draw opacity=0][dash pattern={on 3pt off 2.25pt}][line width=1.5]  (144.82,112.28) .. controls (144.82,108.86) and (158.32,106.09) .. (174.97,106.09) .. controls (191.62,106.09) and (205.11,108.86) .. (205.11,112.28) -- (174.97,112.28) -- cycle ; \draw  [dash pattern={on 3pt off 2.25pt}][line width=1.5]  (144.82,112.28) .. controls (144.82,108.86) and (158.32,106.09) .. (174.97,106.09) .. controls (191.62,106.09) and (205.11,108.86) .. (205.11,112.28) ;  
\draw    (144.82,50.12) .. controls (165.94,72.01) and (164.94,89.01) .. (144.82,112.28) ;
\draw    (230,37.36) .. controls (230,48.76) and (227.59,95.62) .. (213.06,107.54) ;
\draw [shift={(210.67,109.12)}, rotate = 333.07] [fill={rgb, 255:red, 0; green, 0; blue, 0 }  ][line width=0.08]  [draw opacity=0] (8.04,-3.86) -- (0,0) -- (8.04,3.86) -- (5.34,0) -- cycle    ;
\draw    (230,37.36) .. controls (230,48.64) and (226.47,50.31) .. (211.92,50.26) ;
\draw [shift={(209,50.23)}, rotate = 0.84] [fill={rgb, 255:red, 0; green, 0; blue, 0 }  ][line width=0.08]  [draw opacity=0] (8.04,-3.86) -- (0,0) -- (8.04,3.86) -- (5.34,0) -- cycle    ;

\draw (177.79,38) node [anchor=south] [inner sep=0.75pt]    {$\varphi $};
\draw (176.79,121.81) node [anchor=north] [inner sep=0.75pt]    {$\varphi '$};
\draw (227.34,23.35) node  [font=\scriptsize]  {$1+i\alpha Q_{\mIp}^{A}$};

\end{tikzpicture}
}} \!\!\!\!\!\! = \vcenter{\hbox{
\tikzset{every picture/.style={line width=0.75pt}} 

\begin{tikzpicture}[x=0.75pt,y=0.75pt,yscale=-1,xscale=1]

\draw    (205.11,50.12) .. controls (184,72.01) and (185,89.01) .. (205.11,112.28) ;
\draw  [draw opacity=0][line width=0.75]  (205.11,48.83) .. controls (205.11,48.83) and (205.11,48.83) .. (205.11,48.83) .. controls (205.11,52.88) and (191.62,56.16) .. (174.97,56.16) .. controls (158.32,56.16) and (144.82,52.88) .. (144.82,48.83) -- (174.97,48.83) -- cycle ; \draw  [line width=0.75]  (205.11,48.83) .. controls (205.11,48.83) and (205.11,48.83) .. (205.11,48.83) .. controls (205.11,52.88) and (191.62,56.16) .. (174.97,56.16) .. controls (158.32,56.16) and (144.82,52.88) .. (144.82,48.83) ;  
\draw  [draw opacity=0][line width=0.75]  (144.82,48.83) .. controls (144.82,45.41) and (158.32,42.64) .. (174.97,42.64) .. controls (191.62,42.64) and (205.11,45.41) .. (205.11,48.83) -- (174.97,48.83) -- cycle ; \draw  [line width=0.75]  (144.82,48.83) .. controls (144.82,45.41) and (158.32,42.64) .. (174.97,42.64) .. controls (191.62,42.64) and (205.11,45.41) .. (205.11,48.83) ;  
\draw  [draw opacity=0][line width=0.75]  (205.19,112.28) .. controls (205.19,112.28) and (205.19,112.28) .. (205.19,112.28) .. controls (205.19,116.33) and (191.7,119.62) .. (175.05,119.62) .. controls (158.4,119.62) and (144.9,116.33) .. (144.9,112.28) -- (175.05,112.28) -- cycle ; \draw  [line width=0.75]  (205.19,112.28) .. controls (205.19,112.28) and (205.19,112.28) .. (205.19,112.28) .. controls (205.19,116.33) and (191.7,119.62) .. (175.05,119.62) .. controls (158.4,119.62) and (144.9,116.33) .. (144.9,112.28) ;  
\draw  [draw opacity=0][dash pattern={on 3pt off 2.25pt}][line width=0.75]  (144.82,112.28) .. controls (144.82,108.86) and (158.32,106.09) .. (174.97,106.09) .. controls (191.62,106.09) and (205.11,108.86) .. (205.11,112.28) -- (174.97,112.28) -- cycle ; \draw  [dash pattern={on 3pt off 2.25pt}][line width=0.75]  (144.82,112.28) .. controls (144.82,108.86) and (158.32,106.09) .. (174.97,106.09) .. controls (191.62,106.09) and (205.11,108.86) .. (205.11,112.28) ;  
\draw    (144.82,50.12) .. controls (165.94,72.01) and (164.94,89.01) .. (144.82,112.28) ;

\draw (181.79,38) node [anchor=south] [inner sep=0.75pt]    {$\varphi +\tfrac{\alpha }{2} \varphi ^{\prime A}$};
\draw (178.79,120.81) node [anchor=north] [inner sep=0.75pt]    {$\varphi '-\tfrac{\alpha }{2} \varphi ^{A}$};

\end{tikzpicture}
}}
\end{equation}
In equations, the above manipulation can be rewritten as
\begin{equation}\label{threeds}
    \begin{aligned}
        \smatrixmeasure (1 + i \alpha   Q_{E}^A ) &= \smatrixmeasure (1 + i \alpha   Q_{\mathcal{I^+}}^A ) \\
        &= \mathcal{S}[ \varphi + \tfrac{\alpha}{2} {\varphi'}^A, \varphi' - \tfrac{\alpha}{2} \varphi^A]
    \end{aligned}
\end{equation}
where as a shorthand we have defined the S-matrix as
\begin{equation}\label{Smatrixdef}
\mathcal{S}[\varphi,\varphi'] = \int_{\eval{\phi}_{\mathcal{I}^-} = \varphi'}^{\eval{\phi}_{\mathcal{I}^+} = \varphi} \mD \phi \, e^{i S_{dS_d}[\phi]} = 
\vcenter{\hbox{
\tikzset{every picture/.style={line width=0.75pt}} 

\begin{tikzpicture}[x=0.75pt,y=0.75pt,yscale=-1,xscale=1]

\draw  [draw opacity=0][line width=0.75]  (345.92,54.54) .. controls (345.92,54.54) and (345.92,54.54) .. (345.92,54.54) .. controls (345.92,58.59) and (332.42,61.87) .. (315.78,61.87) .. controls (299.13,61.87) and (285.63,58.59) .. (285.63,54.54) -- (315.78,54.54) -- cycle ; \draw  [line width=0.75]  (345.92,54.54) .. controls (345.92,54.54) and (345.92,54.54) .. (345.92,54.54) .. controls (345.92,58.59) and (332.42,61.87) .. (315.78,61.87) .. controls (299.13,61.87) and (285.63,58.59) .. (285.63,54.54) ;  
\draw  [draw opacity=0][line width=0.75]  (285.63,54.54) .. controls (285.63,51.12) and (299.13,48.35) .. (315.78,48.35) .. controls (332.42,48.35) and (345.92,51.12) .. (345.92,54.54) -- (315.78,54.54) -- cycle ; \draw  [line width=0.75]  (285.63,54.54) .. controls (285.63,51.12) and (299.13,48.35) .. (315.78,48.35) .. controls (332.42,48.35) and (345.92,51.12) .. (345.92,54.54) ;  
\draw  [draw opacity=0][line width=0.75]  (346,117.99) .. controls (346,117.99) and (346,117.99) .. (346,117.99) .. controls (346,117.99) and (346,117.99) .. (346,117.99) .. controls (346,122.04) and (332.5,125.33) .. (315.85,125.33) .. controls (299.21,125.33) and (285.71,122.04) .. (285.71,117.99) -- (315.85,117.99) -- cycle ; \draw  [line width=0.75]  (346,117.99) .. controls (346,117.99) and (346,117.99) .. (346,117.99) .. controls (346,117.99) and (346,117.99) .. (346,117.99) .. controls (346,122.04) and (332.5,125.33) .. (315.85,125.33) .. controls (299.21,125.33) and (285.71,122.04) .. (285.71,117.99) ;  
\draw  [draw opacity=0][dash pattern={on 4.5pt off 4.5pt}][line width=0.75]  (285.63,117.99) .. controls (285.63,114.57) and (299.13,111.8) .. (315.78,111.8) .. controls (332.42,111.8) and (345.92,114.57) .. (345.92,117.99) -- (315.78,117.99) -- cycle ; \draw  [dash pattern={on 4.5pt off 4.5pt}][line width=0.75]  (285.63,117.99) .. controls (285.63,114.57) and (299.13,111.8) .. (315.78,111.8) .. controls (332.42,111.8) and (345.92,114.57) .. (345.92,117.99) ;  
\draw    (346,55.83) .. controls (324.89,77.72) and (325.89,94.72) .. (346,117.99) ;
\draw    (285.63,54.54) .. controls (306.74,76.42) and (305.74,93.42) .. (285.63,116.7) ;

\draw (315.78,46.14) node [anchor=south] [inner sep=0.75pt]    {$\varphi $};
\draw (316.85,125.39) node [anchor=north] [inner sep=0.75pt]    {$\varphi '$};

\end{tikzpicture}
}}.
\end{equation}

Using the fact that
\begin{equation}\label{HHconjugate}
    \Psi^{\mathcal{I}^-}_0[\varphi] = (\Psi^{\mathcal{I}^+}_0[\varphi])^*
\end{equation}
(which follows from the fact that $\Psi^E_0[\varphi]$ is a purely real wavefunctional) we end up with the expression, for small $\alpha$,
\begin{equation}\label{psialphasmatrix}
    \Psi^{\mathcal{I}^+}_\alpha[\varphi] = \int \mD \varphi' \mathcal{S}[ \varphi + \tfrac{\alpha}{2} {\varphi'}^A, \varphi' - \tfrac{\alpha}{2} \varphi^A] (\Psi^{\mathcal{I}^+}_0[\varphi'])^*.
\end{equation}

This expression, which we derived using the path integral, tells us how we can use $\Psi^\mIp_0[\varphi]$ to solve for $\Psi^\mIp_\alpha[\varphi]$. 

Let us now compute the S-matrix. We decompose the inputs of $\mathcal{S}[\varphi,\varphi']$ as
\begin{equation}
    \varphi(\Omega) = \sum_{\ell,m} c_{\ell m} Y_{\ell m}(\Omega), \hspace{1 cm} \varphi'(\Omega) = \sum_{\ell,m} c'_{\ell m} Y_{\ell m}(\Omega).
\end{equation}
Because we are dealing with a free field theory, the S-matrix must be a gaussian of the form
\begin{equation}
    \mathcal{S}[\varphi,\varphi'] = \mathcal{N}_{\mathcal{S}}\exp( i \sum_{\ell,m} (a_{\ell} (c_{\ell m})^2 +b_{\ell}( c_{\ell m} c'_{\ell m} ) + a_{\ell} (c'_{\ell m})^2 ))
\end{equation}
for some real coefficients $a_\ell$ and $b_\ell$ and normalization constant $\mathcal{N}_{\mathcal{S}}$.

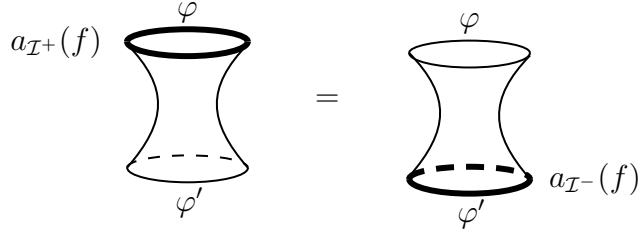
\begin{figure}
    \centering
    \tikzset{every picture/.style={line width=0.75pt}} 

\begin{tikzpicture}[x=0.75pt,y=0.75pt,yscale=-1,xscale=1]

\draw    (205.11,50.12) .. controls (184,72.01) and (185,89.01) .. (205.11,112.28) ;
\draw  [draw opacity=0][line width=2.25]  (205.11,48.83) .. controls (205.11,48.83) and (205.11,48.83) .. (205.11,48.83) .. controls (205.11,52.88) and (191.62,56.16) .. (174.97,56.16) .. controls (158.32,56.16) and (144.82,52.88) .. (144.82,48.83) -- (174.97,48.83) -- cycle ; \draw  [line width=2.25]  (205.11,48.83) .. controls (205.11,48.83) and (205.11,48.83) .. (205.11,48.83) .. controls (205.11,52.88) and (191.62,56.16) .. (174.97,56.16) .. controls (158.32,56.16) and (144.82,52.88) .. (144.82,48.83) ;  
\draw  [draw opacity=0][line width=2.25]  (144.82,48.83) .. controls (144.82,45.41) and (158.32,42.64) .. (174.97,42.64) .. controls (191.62,42.64) and (205.11,45.41) .. (205.11,48.83) -- (174.97,48.83) -- cycle ; \draw  [line width=2.25]  (144.82,48.83) .. controls (144.82,45.41) and (158.32,42.64) .. (174.97,42.64) .. controls (191.62,42.64) and (205.11,45.41) .. (205.11,48.83) ;  
\draw  [draw opacity=0][line width=0.75]  (205.19,112.28) .. controls (205.19,112.28) and (205.19,112.28) .. (205.19,112.28) .. controls (205.19,116.33) and (191.7,119.62) .. (175.05,119.62) .. controls (158.4,119.62) and (144.9,116.33) .. (144.9,112.28) -- (175.05,112.28) -- cycle ; \draw  [line width=0.75]  (205.19,112.28) .. controls (205.19,112.28) and (205.19,112.28) .. (205.19,112.28) .. controls (205.19,116.33) and (191.7,119.62) .. (175.05,119.62) .. controls (158.4,119.62) and (144.9,116.33) .. (144.9,112.28) ;  
\draw  [draw opacity=0][dash pattern={on 4.5pt off 4.5pt}][line width=0.75]  (144.82,112.28) .. controls (144.82,108.86) and (158.32,106.09) .. (174.97,106.09) .. controls (191.62,106.09) and (205.11,108.86) .. (205.11,112.28) -- (174.97,112.28) -- cycle ; \draw  [dash pattern={on 4.5pt off 4.5pt}][line width=0.75]  (144.82,112.28) .. controls (144.82,108.86) and (158.32,106.09) .. (174.97,106.09) .. controls (191.62,106.09) and (205.11,108.86) .. (205.11,112.28) ;  
\draw  [draw opacity=0][line width=0.75]  (345.92,54.54) .. controls (345.92,54.54) and (345.92,54.54) .. (345.92,54.54) .. controls (345.92,58.59) and (332.42,61.87) .. (315.78,61.87) .. controls (299.13,61.87) and (285.63,58.59) .. (285.63,54.54) -- (315.78,54.54) -- cycle ; \draw  [line width=0.75]  (345.92,54.54) .. controls (345.92,54.54) and (345.92,54.54) .. (345.92,54.54) .. controls (345.92,58.59) and (332.42,61.87) .. (315.78,61.87) .. controls (299.13,61.87) and (285.63,58.59) .. (285.63,54.54) ;  
\draw  [draw opacity=0][line width=0.75]  (285.63,54.54) .. controls (285.63,51.12) and (299.13,48.35) .. (315.78,48.35) .. controls (332.42,48.35) and (345.92,51.12) .. (345.92,54.54) -- (315.78,54.54) -- cycle ; \draw  [line width=0.75]  (285.63,54.54) .. controls (285.63,51.12) and (299.13,48.35) .. (315.78,48.35) .. controls (332.42,48.35) and (345.92,51.12) .. (345.92,54.54) ;  
\draw  [draw opacity=0][line width=2.25]  (346,117.99) .. controls (346,117.99) and (346,117.99) .. (346,117.99) .. controls (346,117.99) and (346,117.99) .. (346,117.99) .. controls (346,122.04) and (332.5,125.33) .. (315.85,125.33) .. controls (299.21,125.33) and (285.71,122.04) .. (285.71,117.99) -- (315.85,117.99) -- cycle ; \draw  [line width=2.25]  (346,117.99) .. controls (346,117.99) and (346,117.99) .. (346,117.99) .. controls (346,117.99) and (346,117.99) .. (346,117.99) .. controls (346,122.04) and (332.5,125.33) .. (315.85,125.33) .. controls (299.21,125.33) and (285.71,122.04) .. (285.71,117.99) ;  
\draw  [draw opacity=0][dash pattern={on 6.75pt off 4.5pt}][line width=2.25]  (285.63,117.99) .. controls (285.63,114.57) and (299.13,111.8) .. (315.78,111.8) .. controls (332.42,111.8) and (345.92,114.57) .. (345.92,117.99) -- (315.78,117.99) -- cycle ; \draw  [dash pattern={on 6.75pt off 4.5pt}][line width=2.25]  (285.63,117.99) .. controls (285.63,114.57) and (299.13,111.8) .. (315.78,111.8) .. controls (332.42,111.8) and (345.92,114.57) .. (345.92,117.99) ;  
\draw    (144.82,50.12) .. controls (165.94,72.01) and (164.94,89.01) .. (144.82,112.28) ;
\draw    (346,55.83) .. controls (324.89,77.72) and (325.89,94.72) .. (346,117.99) ;
\draw    (285.63,54.54) .. controls (306.74,76.42) and (305.74,93.42) .. (285.63,116.7) ;

\draw (174.97,40.43) node [anchor=south] [inner sep=0.75pt]    {$\varphi $};
\draw (315.78,46.14) node [anchor=south] [inner sep=0.75pt]    {$\varphi $};
\draw (316.85,126.39) node [anchor=north] [inner sep=0.75pt]    {$\varphi '$};
\draw (176.05,120.68) node [anchor=north] [inner sep=0.75pt]    {$\varphi '$};
\draw (133.82,48.83) node [anchor=east] [inner sep=0.75pt]    {$a_{\mathcal{I}^{+}}( f)$};
\draw (354,117.99) node [anchor=west] [inner sep=0.75pt]    {$a_{\mathcal{I}^{-}}( f)$};
\draw (244.77,78.2) node    {$=$};

\end{tikzpicture}
    \caption{\label{deforma} An insertion of $a_{\mIp}(f)$ into the lorentzian path integral can be continuously deformed to $a_{\mIm}(f)$. This creates a differential equation which can then be used to solve for $\mathcal{S}[\varphi,\varphi']$.}
\end{figure}

In order to solve for $a_\ell$ and $b_\ell$, we will use the following trick. Take any $f$ which solves $(\nabla^2 - m^2) f = 0$ and consider inserting $a_\Sigma(f)$, defined on some surface $\Sigma$, into the lorentzian S-matrix path integral. $\Sigma$ can be freely deformed freely without affecting the path integral, so we can start by setting $\Sigma = \mIp$ and then deform it to $\mIm$, as in figure \ref{deforma}. This equality creates a differential equation which can be used to solve for $\mathcal{S}[\varphi,\varphi']$. 

For our choice of $f$, we will use the ``out $-$'' function $f_{o,\ell m}$ from \eqref{fo}. Our surface-deformation trick then gives

\begin{equation}
    \begin{aligned}
        (  \frac{y_{o,\ell}(T)}{i \cosh^{d-1}(T)} & \pdv{c_{\ell m}} - \dot y_{o,\ell}(T) c_{\ell m} ) \mathcal{S}[\varphi, \varphi'] \\
        & =         (-  \frac{y_{o,\ell}(-T)}{i \cosh ^{d-1}(-T)} \pdv{c'_{\ell m}} - \dot y_{o,\ell}(-T) c'_{\ell m} ) \mathcal{S}[\varphi,\varphi'].
    \end{aligned}
\end{equation}

Because $\mathcal{S}[\varphi, \varphi']$ is the product of an independent gaussian for each $\ell$, $m$, the above differential equation can be reduced to the 1-dimensional equation
\begin{equation}\label{sinter}
\begin{aligned}
    &( \eta_\ell \pdv{c_{\ell m}} + \xi_\ell c_{\ell m}) \exp(i (a_\ell c_{\ell m}^2 + b_\ell c_{\ell m} c_{\ell m}' + a_\ell {c'_{\ell m}}^2)) \\
    &= (\eta_\ell^* \pdv{c'_{\ell m}} - \xi_\ell^* c'_{\ell m} )\exp(i (a_\ell c_{\ell m}^2 + b_\ell c_{\ell m} c_{\ell m}' + a_\ell {c'_{\ell m}}^2))
\end{aligned}
\end{equation}
for
\begin{equation}
    \eta_\ell = \frac{y_{o,\ell}(T)}{i \cosh^{d-1}(T)} = \frac{e^{i \theta_\ell - h_- T}}{i \cosh^{d-1}(T) }, \hspace{1 cm} \xi_\ell = - \dot y_{o,\ell}(T) = h_- e^{i \theta_\ell - h_- T}.
\end{equation}
Equation \eqref{sinter} reduces to
\begin{equation}
    \begin{aligned}
        2 i \eta_\ell  a_\ell + \xi_\ell = i \eta^*_\ell  b_\ell, \hspace{1 cm} i \eta_\ell  b_\ell  = 2 i \eta^*_\ell  a_\ell - \xi^*_\ell,
    \end{aligned}
\end{equation}
which has solutions
\begin{equation}
\begin{aligned}
    a_\ell = \frac{i}{2} \frac{ \eta_\ell \xi_\ell + \eta_\ell^* \xi_\ell^*}{\eta_\ell^2 - (\eta_\ell^*)^2}, \hspace{0.5 cm} b_\ell = i \frac{\eta_\ell^* \xi_\ell + \eta_\ell \xi_\ell^*}{\eta_\ell^2 - (\eta_\ell^*)^2}.
\end{aligned}
\end{equation}
The $a_\ell$, $b_\ell$ coefficients then turn out to be, in the large $T$ limit, 
\begin{equation}
\begin{aligned}
    a_\ell &= -\frac{e^{(d-1)T}}{2^d}  \left(\frac{d - 1}{2} - \mu  \cot (2 (\theta_\ell +\mu  T))\right), \\
    b_\ell &= -\mu \frac{e^{(d-1)T}}{2^{d-1}} \csc (2 (\theta_\ell +\mu  T)).
\end{aligned}
\end{equation}

We have now computed the S-matrix, aside from the normalization constant which we will ignore.

With the free S-matrix, gaussian initial states will evolve into gaussian final states in the following way. For initial and final states
\begin{equation}
    \Psi_{\rm initial}^{\mIm}[\varphi] \propto \exp( - \sum_{\ell,m} B_{{\rm initial}, \ell}^{\mIm} c_{\ell m}^2 ), \hspace{0.5 cm}
    \Psi_{\rm final}^{\mIp}[\varphi] \propto \exp( - \sum_{\ell,m} B_{{\rm final}, \ell}^{\mIp} c_{\ell m}^2 ),
\end{equation}
related  by
\begin{equation}
    \Psi_{\rm final}^{\mIp}[\varphi] = \int \mD \varphi' \mathcal{S}[\varphi, \varphi'] \, \Psi_{\rm initial}^{\mIm}[\varphi'],
\end{equation}
the 1D gaussian integral
\begin{equation}\label{1dsmatrix}
\begin{aligned}
    \int d c_{\ell m}' & \exp( i (a_\ell c_{\ell m}^2 + b_\ell c_{\ell m} c_{\ell m}' + a_{\ell} {c_{\ell m}'}^2 ) ) \exp( - B_{{\rm initial}, \ell}^{\mIm} c_{\ell m}'^2 ) \\
    &= \sqrt{ \frac{\pi}{B_{{\rm initial}, \ell}^{\mIm} - i a_\ell} }\exp(i (a_\ell - \frac{b_\ell^2}{4 a_\ell + 4 i B_{{\rm initial}, \ell}^{\mIm}}) c_{\ell m}^2 )
\end{aligned}
\end{equation}
implies that the ``$B$-coefficients'' evolve from $\mIm$ to $\mIp$ as
\begin{equation}\label{Bevolve}
    B_{{\rm initial},\ell}^{\mIm} \longrightarrow B_{{\rm final},\ell}^{\mIp} = -i \left(a_\ell - \frac{b_\ell^2}{4 a_\ell + 4 i B_{{\rm initial}, \ell}^{\mIm}} \right).
\end{equation}

Let us now perform two checks on our S-matrix calculation. For our first check, let us confirm that the Bunch-Davies wavefunctional at $\mIm$ really does evolve into the Bunch-Davies wavefunctional at $\mIp$.

If we write the Bunch-Davies wavefunctional on $\mIm$ as
\begin{equation}
    \Psi^\mIm_0[\varphi]  = \mathcal{N}^\mIm_0\exp( - \sum_{\ell,m} B^{\mIm}_{0,\ell} c_{\ell m}^2 ).
\end{equation}
then from \eqref{HHconjugate} we know
\begin{equation}
    B^{\mIm}_{0,\ell} = (B^{\mIp}_{0,\ell})^*.
\end{equation}
Plugging in our known expressions for $B^{\mIm}_{0,\ell}$ and $B^{\mIp}_{0,\ell}$ from \eqref{Bip0} into \eqref{Bevolve}, we do find that
\begin{equation}
    B^{\mIm}_{0,\ell} \longrightarrow B^{\mIp}_{0,\ell} \;\;\;\; \checkmark
\end{equation}
as expected.

For our second check, let us confirm that the ``in'' vacuum really does evolve into the ``out'' vacuum, as we know it must in odd dimensions.

Because $\alpha^{\rm out/in}_{\pm} \in \mathbb{R}$ in odd dimensions, we know that
\begin{equation}
    B^{\mIm}_{\alpha^{\rm in}_-, \ell} = ( B^{\mIp}_{\alpha^{\rm out}_+,\ell})^*
\end{equation}
and thus, using our already known expressions for $B^{\mIm}_{\alpha^{\rm in}_-,\ell}$ and $B^{\mIp}_{\alpha^{\rm out}_+,\ell}$ from \eqref{Boutell}, we can plug them into \eqref{Bevolve} and confirm that indeed
\begin{equation}
    B^{\mIm}_{\alpha^{\rm in}_-, \ell} \longrightarrow B^{\mIp}_{\alpha^{\rm out}_+,\ell} \;\;\;\; \checkmark
\end{equation}
as expected.

With our checks complete, let us now confirm that our path integral expression for the $\alpha$-vacuum wavefunctional \eqref{psialphasmatrix} really does hold when $\alpha$ is small.

Defining
\begin{equation}
\begin{aligned}
    c^\alpha_{\ell m} & \equiv c_{\ell m} + (-1)^\ell \frac{\alpha}{2} \, c'_{\ell m} \\
    {c^\alpha_{\ell m}}' & \equiv c'_{\ell m} - (-1)^\ell \frac{\alpha}{2} \, c_{\ell m}
\end{aligned}
\end{equation}
one can compute the 1D integral
\begin{equation}
\begin{aligned}
    \int d c_{\ell m}' & \exp( i (a_\ell (c^\alpha_{\ell m})^2 + b_\ell (c^\alpha_{\ell m} {c^\alpha_{\ell m}}') + a_{\ell} ({c_{\ell m}^\alpha}')^2 ) ) \exp( - B^{\mIm}_{0,\ell} c_{\ell m}'^2 ) \\
    &\propto \exp( - \left(  - i a_\ell + i (-1)^\ell \frac{\alpha}{2} b_\ell + \frac{b^2}{4 B^{\mIm}_{0,\ell} - 4 i a_\ell - 2 i (-1)^\ell \alpha b_\ell} \right) c_{\ell m}^2 )
\end{aligned}
\end{equation}
to the first order in $\alpha$ (where we have dropped the $c_{\ell m}$-independent normalization constant out front). This 1D integral computes the $\alpha$-vacuum wavefunctional at $\mIp$. Comparing the $B$-coefficients, we find that indeed
\begin{equation}
      - i a_\ell + i (-1)^\ell \frac{\alpha}{2} b_\ell + \frac{b^2}{4 B_{0,\ell}^{\mIm} - 4 i a_\ell - 2 i (-1)^\ell \alpha b_\ell} = B^{\mIp}_{\alpha,\ell} + \mathcal{O}(\alpha^2) \;\;\;\; \checkmark 
\end{equation}
which confirms that \eqref{psialphasmatrix} computes the $\alpha$-vacuum out state for small $\alpha$.

\subsection{Operator ordering considerations for finite $\alpha$}\label{sec94}

In the previous section we limited our discussion of the $\alpha$-vacuum wavefunctional at $\mIp$ to the case of small $\alpha$. This is because, in the small $\alpha$ limit,
\begin{equation}
    e^{i \alpha \hat{Q}^A} = 1 + i \alpha \hat{Q}^A
\end{equation}
and one does not have to worry about operator ordering for higher order terms like $(\hat{Q}^A)^n$ for $n \geq 2$.

Let us explain how the operator ordering of powers of conserved charges are usually dealt with in quantum path integrals. Consider $\hat Q^n$, where $\hat{Q}$ is some Noether charge of some local symmetry (unlike $\hat{Q}^A$, which is the Noether charge of a non-local symmetry). If we were to prepare $\hat{Q}^n \ket{0}$ with a path integral, we would place the $n$ $Q$-insertions into the path integral in the following way. Defining the surface $E_\epsilon$ as
\begin{equation}
    E_\epsilon = \{ (\tilde{t} = \epsilon, \Omega ) \text{ surface} \}
\end{equation}
which is the equator $E$ shifted slightly in euclidean time by $\epsilon$, we insert the charges sequentially on the $n$ slices $E_0$, $E_{-\epsilon}$, ... , $E_{-(n-1)\epsilon}$, with $\epsilon > 0$.
\begin{equation}
\begin{aligned}
    \hat{Q}^n \Psi_0^E [\varphi] &= \lim_{\epsilon \to 0^+} \int_{\eval{\phi}_E = \varphi} \mD \phi \, e^{- S_{S_{d,-}}[\phi]} Q_{E_0} Q_{E_{-\epsilon}} \ldots Q_{E_{- (n-1) \epsilon}} \\
    &= \vcenter{\hbox{
\tikzset{every picture/.style={line width=0.75pt}} 

\begin{tikzpicture}[x=0.75pt,y=0.75pt,yscale=-1,xscale=1]

\draw  [draw opacity=0][line width=1.5]  (352.66,106.18) .. controls (352.66,106.18) and (352.66,106.18) .. (352.66,106.18) .. controls (352.66,106.18) and (352.66,106.18) .. (352.66,106.18) .. controls (352.66,114.37) and (329.46,121.01) .. (300.84,121.01) .. controls (272.22,121.01) and (249.02,114.37) .. (249.02,106.18) -- (300.84,106.18) -- cycle ; \draw  [line width=1.5]  (352.66,106.18) .. controls (352.66,106.18) and (352.66,106.18) .. (352.66,106.18) .. controls (352.66,106.18) and (352.66,106.18) .. (352.66,106.18) .. controls (352.66,114.37) and (329.46,121.01) .. (300.84,121.01) .. controls (272.22,121.01) and (249.02,114.37) .. (249.02,106.18) ;  
\draw  [draw opacity=0][line width=1.5]  (249.02,106.18) .. controls (249.02,97.99) and (272.22,91.34) .. (300.84,91.34) .. controls (329.46,91.34) and (352.66,97.99) .. (352.66,106.18) -- (300.84,106.18) -- cycle ; \draw  [line width=1.5]  (249.02,106.18) .. controls (249.02,97.99) and (272.22,91.34) .. (300.84,91.34) .. controls (329.46,91.34) and (352.66,97.99) .. (352.66,106.18) ;  
\draw  [draw opacity=0] (352.66,106.18) .. controls (352.18,134.36) and (329.19,157.05) .. (300.9,157.05) .. controls (272.3,157.05) and (249.12,133.87) .. (249.12,105.28) -- (300.9,105.28) -- cycle ; \draw   (352.66,106.18) .. controls (352.18,134.36) and (329.19,157.05) .. (300.9,157.05) .. controls (272.3,157.05) and (249.12,133.87) .. (249.12,105.28) ;  
\draw  [draw opacity=0][line width=1.5]  (352.5,111.05) .. controls (352.5,111.05) and (352.5,111.05) .. (352.5,111.05) .. controls (352.5,119.17) and (329.5,125.75) .. (301.13,125.75) .. controls (272.77,125.75) and (249.77,119.17) .. (249.77,111.05) -- (301.13,111.05) -- cycle ; \draw  [line width=1.5]  (352.5,111.05) .. controls (352.5,111.05) and (352.5,111.05) .. (352.5,111.05) .. controls (352.5,119.17) and (329.5,125.75) .. (301.13,125.75) .. controls (272.77,125.75) and (249.77,119.17) .. (249.77,111.05) ;  
\draw  [draw opacity=0][dash pattern={on 5.63pt off 4.5pt}][line width=1.5]  (249.77,111.05) .. controls (249.77,111.05) and (249.77,111.05) .. (249.77,111.05) .. controls (249.77,102.93) and (272.77,96.34) .. (301.13,96.34) .. controls (329.5,96.34) and (352.5,102.93) .. (352.5,111.05) -- (301.13,111.05) -- cycle ; \draw  [dash pattern={on 5.63pt off 4.5pt}][line width=1.5]  (249.77,111.05) .. controls (249.77,111.05) and (249.77,111.05) .. (249.77,111.05) .. controls (249.77,102.93) and (272.77,96.34) .. (301.13,96.34) .. controls (329.5,96.34) and (352.5,102.93) .. (352.5,111.05) ;  
\draw  [draw opacity=0][line width=1.5]  (350.88,116.35) .. controls (350.88,116.35) and (350.88,116.35) .. (350.88,116.35) .. controls (350.88,116.35) and (350.88,116.35) .. (350.88,116.35) .. controls (350.88,124.23) and (328.58,130.61) .. (301.08,130.61) .. controls (273.57,130.61) and (251.27,124.23) .. (251.27,116.35) -- (301.08,116.35) -- cycle ; \draw  [line width=1.5]  (350.88,116.35) .. controls (350.88,116.35) and (350.88,116.35) .. (350.88,116.35) .. controls (350.88,116.35) and (350.88,116.35) .. (350.88,116.35) .. controls (350.88,124.23) and (328.58,130.61) .. (301.08,130.61) .. controls (273.57,130.61) and (251.27,124.23) .. (251.27,116.35) ;  
\draw  [draw opacity=0][dash pattern={on 5.63pt off 4.5pt}][line width=1.5]  (251.27,116.35) .. controls (251.27,108.48) and (273.57,102.09) .. (301.08,102.09) .. controls (328.58,102.09) and (350.88,108.48) .. (350.88,116.35) -- (301.08,116.35) -- cycle ; \draw  [dash pattern={on 5.63pt off 4.5pt}][line width=1.5]  (251.27,116.35) .. controls (251.27,108.48) and (273.57,102.09) .. (301.08,102.09) .. controls (328.58,102.09) and (350.88,108.48) .. (350.88,116.35) ;  
\draw  [draw opacity=0][line width=1.5]  (349.25,121.44) .. controls (349.25,121.44) and (349.25,121.44) .. (349.25,121.44) .. controls (349.25,129.09) and (327.6,135.28) .. (300.88,135.28) .. controls (274.17,135.28) and (252.52,129.09) .. (252.52,121.44) -- (300.88,121.44) -- cycle ; \draw  [line width=1.5]  (349.25,121.44) .. controls (349.25,121.44) and (349.25,121.44) .. (349.25,121.44) .. controls (349.25,129.09) and (327.6,135.28) .. (300.88,135.28) .. controls (274.17,135.28) and (252.52,129.09) .. (252.52,121.44) ;  
\draw  [draw opacity=0][dash pattern={on 5.63pt off 4.5pt}][line width=1.5]  (252.52,121.44) .. controls (252.52,121.44) and (252.52,121.44) .. (252.52,121.44) .. controls (252.52,113.79) and (274.17,107.59) .. (300.88,107.59) .. controls (327.6,107.59) and (349.25,113.79) .. (349.25,121.44) -- (300.88,121.44) -- cycle ; \draw  [dash pattern={on 5.63pt off 4.5pt}][line width=1.5]  (252.52,121.44) .. controls (252.52,121.44) and (252.52,121.44) .. (252.52,121.44) .. controls (252.52,113.79) and (274.17,107.59) .. (300.88,107.59) .. controls (327.6,107.59) and (349.25,113.79) .. (349.25,121.44) ;  

\draw (365.5,116) node [anchor=west] [inner sep=0.75pt]    {$\Bigr\} n$};

\end{tikzpicture}
}}
\end{aligned}
\end{equation}
Shifting the equators in this way naturally replicates the repeated action of $\hat{Q}$ on the quantum state.

However, this approach faces a problem when it comes to our non-local antipodal charge $\hat{Q}^A$. This surface-ordering prescription cannot be implemented with a charge that is defined on a pair of antipodal surfaces. If we attempt to deform the surface on which $Q^A_E$ is defined slightly away from the equator in a path-integral correlator, we get 
\begin{equation}
    \langle {Q}^A_{E_\epsilon} \ldots \rangle = \frac{1}{2}\int d^{d-1} \Omega \langle (\phi(-\epsilon, \Omega^A) \dot \phi(\epsilon, \Omega) + \phi(\epsilon, \Omega) \dot \phi(-\epsilon, \Omega^A) ) \ldots \rangle
\end{equation}
which can't be inserted in the half-sphere path integral because it includes insertions at both $\tilde{t} = \epsilon$ and $\tilde{t} = - \epsilon$. Furthermore, the sequential ordering of the surfaces $E_0$, $E_{-\epsilon}$, $\ldots$, $E_{-(n-1) \epsilon}$ necessary to recreate $\hat{Q}^n = \hat{Q} \ldots \hat{Q}$ within the path integral cannot be reproduced if the antipodal surfaces ``surround'' each other, with $E_{j \epsilon}$ and $E_{-j\epsilon}$ surrounding $E_{(j-1)\epsilon}$ and $E_{-(j-1)\epsilon}$, for example.

If we want to prepare $(\hat{Q}^A)^n \ket{0}$ using the path integral, we have to define a new, slightly modified insertion that we'll denote as ${Q'}^A$, given by
\begin{equation}
    \langle {Q'}^A_{E_\epsilon} \ldots \rangle \equiv \int d^{d-1} \Omega \langle \phi(\epsilon, \Omega) \dot \phi(\epsilon, \Omega^A) \ldots \rangle.
\end{equation}
By fiat, ${Q'}^A_{E_\epsilon}$ enforces all insertions are placed on $E_\epsilon$. The cost of this is that ${Q'}^A_{E_\epsilon}$ is not actually a conserved charge whose surface is able to be deformed freely. Nonetheless, it does enable us to write $(\hat{Q}^A)^n \ket{0}$ using a path integral as
\begin{equation}
    (\hat{Q}^A)^n \Psi_0^E [\varphi] = 
    \lim_{\epsilon \to 0^+} \int_{\eval{\phi}_E = \varphi} \mD \phi \, e^{- S_{S_{d,-}}[\phi]} {Q'}^A_{E_0} {Q'}^A_{E_{-\epsilon}} \ldots {Q'}^A_{E_{- (n-1) \epsilon}}.
\end{equation}

However, if one wants to evaluate the wavefunctional of the $\alpha$-vacua at $\mIp$, we are no longer allowed to deform the charges from $E$ to $\mIp/\mIm$ like we did before in the small $\alpha$ case. In other words, the steps we took to prove $\Psi^{\mIp}_{\alpha}[\varphi]$ can be evaluated with equation \eqref{threeds} cannot be replicated when $\alpha$ is finite, because operator ordering issues prevent us from deforming the charges away from $E$.

\section{Can the $\alpha$-vacua be constructed in interacting theories?}\label{sec10}

In constructing the $\alpha$-vacua, we used the existence of the symmetry variation \eqref{varA}, $\delta \phi = \epsilon \, \phi^A$, to derive an associated conserved current and Ward identity. This variation was a symmetry because the equation of motion in a free theory is linear, implying the sum of two solutions is also a solution. However, in an interacting theory, this variation is no longer a symmetry. In an interacting theory with action
\begin{equation}
    S^{\rm int}_U[\phi] = \int_U d^d \tx \sqrt{g} \left(\frac{1}{2}  ( \nabla \phi)^2 + \frac{1}{2} m^2 \phi^2 + V(\phi) \right)
\end{equation}
if we attempt to perform the Noether trick with $\delta \phi = \varepsilon \, \phi^A$, regarding $\varepsilon = \varepsilon(\tx)$ as a tiny function that vanishes on the boundary $\partial U$, then we get
\begin{equation}
    \delta S_U^{\rm int}[\phi] = \int_U d^d \tx \sqrt{g} \left(  ( \partial^\mu \varepsilon ) J^A_\mu + (\varepsilon \, \phi) ( \nabla^2 \phi^A - m^2 \phi^A)  + \varepsilon \, \phi^A V'(\phi) \right).
\end{equation}

On shell, we therefore have
\begin{equation}
\begin{aligned}
    \nabla^\mu J^{A}_\mu &= - \phi 
 (\nabla^2 \phi^A - m^2 \phi^A) + \phi^A V'(\phi) \\
 &= -\phi V'(\phi^A) + \phi^A V'(\phi)
\end{aligned}
\end{equation}
and so the current $J^A_\mu$ is not conserved in an interacting theory. At the quantum level, the conservation of the current $J^A_\mu$ in a free theory implies that $\hat{Q}^A$ commutes with the free isometry boost generator $\hat{L}^{01}_{\rm free}$. (The reason for this was explained in section \ref{sec6}. See for instance figure \ref{figcommute}.)
\begin{equation}
    [\hat{L}_{\rm free}^{01}, \hat{Q}^A] = 0.
\end{equation}
Consequently, the non-conservation of this current in an interacting theory implies that $\hat{Q}^A$ does not commute with the interacting isometry boost generator $\hat{L}^{01}_{\rm int}$.
\begin{equation}
    [\hat{L}^{01}_{\rm int}, \hat{Q}^A] \neq 0.
\end{equation}
This fact can also be seen in the operator formalism. The interacting boost generator is
\begin{equation}
    \hat{L}^{01}_{\rmint} = \int d^{d-1} \Omega \, X^1(0,\Omega) \left( \frac{1}{2} \hat{\pi}^2(0,\Omega) + \frac{1}{2} m^2 \hat{\phi}^2(0,\Omega)+ V(\hat{\phi}(0,\Omega)) \right)
\end{equation}
where $X^1(0,\Omega)$ satisfies $X^1(0,\Omega^A) = - X^1(0,\Omega)$, and its commutator with $\hat{Q}^A$ is non-zero in the presence of interactions:
\begin{equation}\label{LQcomm}
    [ \hat{L}^{01}_{\rmint}, \hat{Q}^A] = -i \int d^{d-1} \Omega \, X^1(0,\Omega) \, \hat{\phi}(0,\Omega)\,  V'(\hat{\phi}(0,\Omega^A)) .
\end{equation}
This shows that there is a tension between interactions and the dS-invariance of $\hat{Q}^A$, suggesting that it is impossible to define $SO(1,d)$ invariant $\alpha$-vacua in an interacting theory.

So, is it possible to define $\alpha$-vacua in an interacting theory? In order to address this question directly, we first need to specify what properties an $\alpha$-vacuum should even have in an interacting theory, and then determine if there are any such states that have these properties.

The first property an interacting $\alpha$-vacuum should have is that all of its $n$-point functions should be linear combinations of euclidean $n$-point functions in the same way that the free $\alpha$-vacuum $n$-point functions are linear combinations of free euclidean $n$-point functions. The second property an interacting $\alpha$-vacuum should have is that it should be $SO(1,d)$ invariant. We are now going to argue that these two properties are in contradiction with each other.

Let's start with the first property. Consider, for instance, the free euclidean 2-point function which is defined by
\begin{equation}
    G_0^{\rm free}(\Omega_1, \Omega_2) \equiv \bra{0} \hat{\phi}(0,\Omega_1) \hat{\phi}(0,\Omega_2) \ket{0}.
\end{equation}
The $\alpha$-vacuum 2-point function is given by
\begin{equation}
    G_\alpha^{\rm free}(\Omega_1, \Omega_2) \equiv \bra{\alpha} \hat{\phi}(0,\Omega_1) \hat{\phi}(0,\Omega_2) \ket{\alpha}.
\end{equation}
However, from the algebra
\begin{equation}
\begin{aligned}
    \bra{\alpha} \hat{\phi}(0,\Omega_1) \hat{\phi}(0,\Omega_2) \ket{\alpha} &= \bra{0} e^{-i \alpha \hat{Q}^A} \hat{\phi}(0,\Omega_1) e^{i \alpha \hat{Q}^A} e^{-i \alpha \hat{Q}^A}\hat{\phi}(0,\Omega_2) e^{i \alpha \hat{Q}^A}\ket{0} \\
    &= \bra{0} ( \cosh \alpha \, \hat{\phi}(0,\Omega_1) - \sinh \alpha \,\hat{\phi}(0,\Omega_1^A) \\
    & \;\;\;\;\;\;\;\;\;\;\; ( \cosh \alpha \, \hat{\phi}(0,\Omega_2) - \sinh \alpha \, \hat{\phi}(0,\Omega_2^A)\ket{0}
\end{aligned}
\end{equation}
we know that the $\alpha$-vacuum 2-point function is a linear combination of euclidean 2-point functions via
\begin{equation}
\begin{aligned}
    G_\alpha^{\rm free}(\Omega_1, \Omega_2)
    &= (\cosh^2 \alpha) G_0^{\rm free}(\Omega_1, \Omega_2) + (\sinh^2 \alpha) G_0^{\rm free}(\Omega_1^A, \Omega_2^A) \\
    & - (\cosh \alpha \sinh \alpha ) G_0^{\rm free}(\Omega_1^A , \Omega_2) - (\cosh \alpha \sinh \alpha ) G_0^{\rm free}(\Omega_1 , \Omega_2^A).
\end{aligned}
\end{equation}
A similar statement also holds for $n$-point functions. The free euclidean and $\alpha$-vacuum $n$-point functions are defined by 
\begin{equation}
\begin{aligned}
    G_0^{\rm free}(\Omega_1, \ldots, \Omega_n) &\equiv \bra{0} \hat{\phi}(0,\Omega_1) \ldots \hat{\phi}(0,\Omega_n) \ket{0} \\
    G_\alpha^{\rm free}(\Omega_1, \ldots, \Omega_n) &\equiv \bra{\alpha} \hat{\phi}(0,\Omega_1) \ldots \hat{\phi}(0,\Omega_n) \ket{\alpha}
\end{aligned}
\end{equation}
and similarly, we have 
\begin{equation}
\begin{aligned}
    G_\alpha^{\rm free}(\Omega_1, \ldots, \Omega_n) &= \bra{\alpha} \hat{\phi}(0,\Omega_1) \ldots \hat{\phi}(0,\Omega_n) \ket{\alpha} \\
    &= \sum_{i_1 \in \{ 0,1\} } \ldots \sum_{i_n \in \{ 0,1\} } (\cosh \alpha)^{i_1 + \ldots + i_n}(- \sinh \alpha)^{n - i_1 - \ldots - i_n} \\
    & \;\;\;\;\;\;  \times G^{\rm free}_0\left( \begin{cases} \Omega_1 & \text{ if } i_1 = 1 \\ \Omega_1^A & \text{ if } i_1 = 0 \end{cases}, \ldots, \begin{cases} \Omega_n & \text{ if } i_n = 1 \\ \Omega_n^A & \text{ if } i_n = 0 \end{cases} \right).
\end{aligned}
\end{equation}

In the interacting theory, we will denote the interacting euclidean vacuum as $\ket{0, \rmint}$ with $n$-point function
\begin{equation}
    G_0^{\rmint}(\Omega_1, \ldots, \Omega_n) = \bra{0,\rmint} \hat{\phi}(0,\Omega_1) \ldots \hat{\phi}(0,\Omega_n) \ket{0,\rmint}.
\end{equation}

Next we are going to \textit{define} the $\alpha$-vacuum $n$-point function as a linear combination of interacting euclidean $n$-point functions via
\begin{equation}\label{Galpha}
\begin{aligned}
    G_\alpha^{\rmint}(\Omega_1, \ldots, \Omega_n) &\equiv \sum_{i_1 \in \{ 0,1\} } \ldots \sum_{i_n \in \{ 0,1\} } (\cosh \alpha)^{i_1 + \ldots + i_n}(- \sinh \alpha)^{n - i_1 - \ldots - i_n} \\
    & \;\;\;\;\;\;  \times G^{\rm int}_0\left( \begin{cases} \Omega_1 & \text{ if } i_1 = 1 \\ \Omega_1^A & \text{ if } i_1 = 0 \end{cases}, \ldots, \begin{cases} \Omega_n & \text{ if } i_n = 1 \\ \Omega_n^A & \text{ if } i_n = 0 \end{cases} \right).
\end{aligned}
\end{equation}

Now, a quantum state is uniquely determined by the collection of all of its $n$-point functions. Therefore, if we can find a state $\ket{\alpha, \rmint}$ such that 
\begin{equation}
    G_\alpha^{\rmint}(\Omega_1, \ldots, \Omega_n) = \bra{\alpha, \rmint} \hat{\phi}(0,\Omega_1) \ldots \hat{\phi}(0,\Omega_n) \ket{\alpha, \rmint}
\end{equation}
then we know that it is the unique state satisfying equation \eqref{Galpha}. However, it is easy to find this state. It is nothing but
\begin{equation} \label{alphaint}
    \ket{\alpha, \rmint} = e^{i \alpha \hat{Q}^A} \ket{0, \rmint}
\end{equation}
where $\hat{Q}^A$ is just the free antipodal charge on the $t = 0$ slice, given in equation \eqref{QAhat}.

So, to recap, if we want the interacting $\alpha$-vacuum to satisfy \eqref{Galpha}, then $\ket{\alpha, \rmint}$ must be given by \eqref{alphaint}.

We shall now ask: is this state $\ket{\alpha, \rmint}$ $SO(1,d)$ invariant? Certainly it is invariant under spatial rotations, but is it invariant under boosts?

Well, we know that the euclidean state is invariant under boosts, even in an interacting theory.
\begin{equation}
    \hat{L}^{01}_{\rmint} \ket{0,\rmint} = 0
\end{equation}
But what about $\ket{\alpha, \rmint}$? Is it annihilated by $\hat{L}^{01}_{\rmint}$? Well,
\begin{equation}
    \hat{L}^{01}_{\rmint} \ket{\alpha, \rmint} = [\hat{L}^{01}_{\rmint}, e^{i \alpha \hat{Q}^A}] \ket{0,\rmint} 
\end{equation}
so asking if $\ket{\alpha, \rmint}$ is $SO(1,d)$ invariant is equivalent to asking if $[\hat{L}^{01}_{\rmint}, e^{i \alpha \hat{Q}^A}]$ annihilates $\ket{0, \rmint}$. Because $[\hat{L}^{01}_{\rmint}, \hat{Q}^A] \neq 0$ this will not be the case for generic $\alpha$. For instance, if $\alpha$ is small, then from \eqref{LQcomm} we know that $[ \hat{L}^{01}_{\rmint}, 1 + i \alpha \hat{Q}^A ] \ket{0, \rmint} \neq 0$. For this reason, the two properties we'd like an interacting $\alpha$-vacuum to have are in tension. It could potentially be the case that  interacting $\alpha$-vacua could be constructed for special values of $\alpha$. Certainly, though, if anybody would like to claim that interacting $\alpha$-vacua exist, the onus would be on them to provide a construction.

\section{Discussion}\label{secdiscussion}

The purpose of this work has been to study the de Sitter $\alpha$-vacua using the quantum path integral. Our main object of interest has been a certain antipodal charge operator $\hat{Q}^A$ which, when inserted into the path integral, is defined on a pair of antipodal codimension-1 surfaces instead of a single codimension-1 surface like most charges.

Due to this feature, the antipodal charge has unique properties. Because it is defined on a pair of antipodal surfaces, it cannot be contracted on the euclidean half-sphere, meaning that it does not annihilate the Bunch-Davies vacuum and $\hat{Q}^A \ket{0} \neq 0$. However, because the charge is still nonetheless conserved, when it is inserted on the equator in the full euclidean sphere, said equator can be rotated at will, implying that the charge commutes with all the isometry generators and $\hat{Q}^A \ket{0}$ is invariant under the full de Sitter isometry group just as $\ket{0}$ is. Therefore, the fact that the $\alpha$-vacua are both $SO(1,d)$ invariant and non-vanishing is due to this quirk of the antipodal charge.

We also studied the wavefunctional of the $\alpha$-vacua at $\mathcal{I}^\pm$, denoted $\Psi^{\mIp}_\alpha[\varphi]$. For small $\alpha$, this wavefunctional can be prepared with a path integral over the euclidean half-sphere, followed by an insertion of the antipodal charge $1 + i \alpha Q_E^A$ on the equator, followed by the lorentzian path integral from $t=0$ to $t = \infty$. Using the fact that the antipodal charge is conserved, we could then push the equator $E$ to $\mIp$ (with $\mIm$ being its antipodal surface) and compute the wavefunctional at $\mIp$ as a path integral over the full lorentzian de Sitter spacetime, with $\alpha$-rotated boundary conditions, convolved with the Bunch-Davies wavefunctional at $\mIm$. We computed the wavefunctional using this path integral strategy to confirm that it did indeed equal $\Psi^{\mIp}_\alpha[\varphi]$ as we argued it should. Unfortunately, if $\alpha$ is not small, but is instead a finite number, operator ordering issues preclude us from using this argument to compute $\Psi^{\mIp}_\alpha[\varphi]$.

Finally, because our antipodal symmetry charge is a Noether charge for the symmetry variation $\delta \phi = \epsilon \, \phi^A$ which only exists in the free theory, we argued that $\alpha$-vacua cannot be constructed in interacting field theories. More precisely, we argued one cannot construct an $SO(1,d)$ invariant state with initial time slice $n$-point functions that can be expressed as linear combinations of the euclidean $n$-point functions in the way that the free theory $\alpha$-vacuum $n$-point functions are linear combinations of the Bunch-Davies $n$-point functions. We have not ruled out the possibility that a sufficiently inspired analog of the $\alpha$-vacua, defined differently, defined in some special theory, or defined for only special values of $\alpha$, could still be constructed in the presence of interactions.

\section*{Acknowledgments}

This work benefited from conversations with Alek Bedroya, Alex Cohen, Alexander Maloney, and Zimo Sun. The author gratefully acknowledges support from the Sivian Fund at the Institute for Advanced Study and DOE grant DE-SC0009988, along with the Princeton Gravity Initiative at Princeton University.

\appendix

\section{On shifting the spacetime boundary of a path integral}\label{appA}

In this appendix we review how shifting the spacetime boundary of a path integral along a vector field $\xi^\mu$ amounts to inserting the integral of $-\xi^\nu T_{\mu \nu}$ in the path integral on the boundary, where $T_{\mu \nu}$ is the stress energy tensor.

This fact was implicitly used when discussing the boost-invariance of the Bunch-Davies vacuum in section \ref{sec3}, where we took the vector field $\xi^\mu$ to be the analytic continuation of the boost vector field $L^{01}$. We use euclidean signature in this appendix, and denote euclidean spacetime points as $\tx$.

\begin{figure}[h]
    \centering
    \input{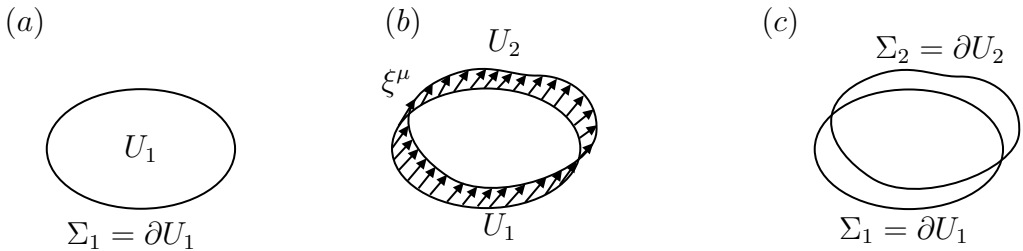}
    \caption{\label{figu} (a): A region of spacetime $U_1$ with boundary $\Sigma_1 = \partial U_1$. (b) and (c): When we shift $U_1$ along a vector field $\xi^\mu$ by an amount $\epsilon$, we end up with $U_2$ with boundary $\Sigma_2 = \partial U_2$.}
\end{figure}

Say $U_1$ is a spacetime volume with a boundary $\Sigma_1 = \partial U_1$ and $\xi^\mu$ is some vector field. Say we shift every point in $U_1$ by $\tx^\mu \mapsto \tx^\mu +  \epsilon \, \xi^\mu$ for tiny constant $\epsilon$, and this map sends $U_1 \mapsto U_2$, with boundary $\Sigma_2 = \partial U_2$. See figure \ref{figu}.

\begin{figure}[t]
    \centering
    \includegraphics[width=0.49\textwidth]{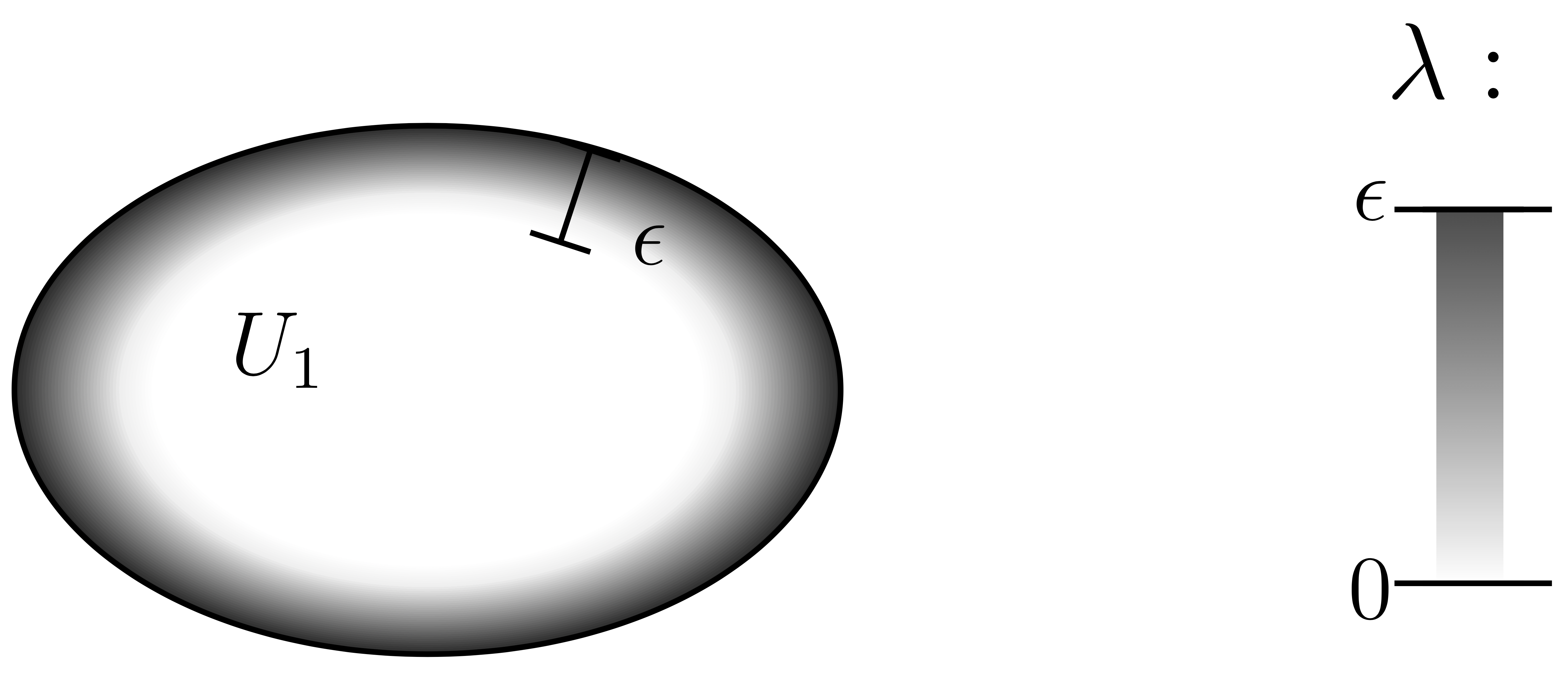}
    \caption{\label{figshade} A sketch of the function $\lambda(\tx)$ on the patch $U_1$. It is equal to $\epsilon$ on $\Sigma_1$ and is equal to zero in most of the interior, smoothly interpolating between the two in a small volume of width $\epsilon$. The value of the function $\lambda$ is displayed with a gradient, with black being $\epsilon$ and white being $0$.}
\end{figure}

Say $\phi$ is some classical scalar field on the spacetime, not necessarily satisfying the equations of motion. Consider the variation
\begin{equation}
    \delta \phi = -\lambda \,  \xi^\mu \partial_\mu \phi
\end{equation}
where $\lambda = \lambda(\tx)$ is a smooth spacetime bump function defined on $U_1$ that behaves in the following way:
\begin{enumerate}
    \item For all $\tx \in \Sigma_1$, $\lambda(\tx) = \epsilon$.
    \item For any $\tx$ in $U_1$ that is further than $\epsilon$ away from $\Sigma_1$ (in geodesic distance), $\lambda(\tx) = 0$.
    \item $\lambda(\tx)$ interpolates smoothly from being $0$ in the interior of $U_1$ to being $\epsilon$ on the boundary, within a small region of size $\epsilon$.
\end{enumerate}
This function is depicted in figure \ref{figshade}.

Notice that $\phi$, when restricted to $\Sigma_1$, has the same ``boundary conditions'' as $\phi + \delta \phi$ when restricted to $\Sigma_2$, where points on $\Sigma_1$ and $\Sigma_2$ are identified in the natural way by ``pushing'' points from $\Sigma_1$ to $\Sigma_2$ along $\xi^\mu$.

We are interested in computing
\begin{equation}
    \frac{1}{\epsilon}\delta S = \frac{1}{\epsilon} \left( S_{U_2}[\phi + \delta \phi] - S_{U_1}[\phi] \right)
\end{equation}
with an action of the form $S_U[\phi] = \int_U d^d x \sqrt{g} L(\phi,\partial \phi)$. Let us denote the (signed) difference in spacetime volumes between $U_2$ and $U_1$ as $\Delta U = U_2 - U_1$. Integrating the spacetime volume measure over $\Delta U$ is equivalent to integrating the surface area of $\Sigma_1$ multiplied by $\epsilon \, n^\mu \xi_\mu$, where $n^\mu$ is an outward pointing unit normal vector to $\Sigma_1$. Denoting the area element normal of $\Sigma_1$ as $\sqrt{h} n^\mu$, we have
\begin{equation}
    \int_{\Delta U} d^d x \sqrt{g} = \epsilon \int_{\Sigma_1} d^{d-1} x \sqrt{h}  n^\mu \xi_\mu = \epsilon \int_{\Sigma_1} d \Sigma_1^\mu \xi_\mu.
\end{equation}
Therefore we have
\begin{equation}
    \begin{aligned}
        \frac{1}{\epsilon}\delta S &= \frac{1}{\epsilon} \left( S_{\Delta U}[\phi] + S_{U_1}[\phi + \delta \phi] - S_{U_1}[\phi] \right) \\
        &= \frac{1}{\epsilon} \Bigg( \epsilon \int_{\Sigma_1} d \Sigma_1^\mu \xi_\mu L + \underbrace{\int_{U_1} d^d x \sqrt{g} \left( \frac{\partial L}{\partial \phi} - \partial_\mu \frac{\partial L}{\partial \partial_\mu \phi} \right)\delta \phi}_{\sim \epsilon^2} + \int_{\Sigma_1} d \Sigma_1^\mu  \frac{\partial L}{\partial \partial^\mu \phi} \delta \phi \Bigg).
    \end{aligned}
\end{equation}
The middle term disappears in the $\epsilon \to 0$ limit, because the function $\lambda$ has a maximum value of $\epsilon$ and only attains it for a total volume proportional to $\epsilon$.

Defining $T_{\mu \nu} = -\frac{\partial L}{\partial \partial^\mu \phi} \partial_\nu \phi + g_{\mu \nu} L$, we therefore have
\begin{equation}
    \delta S = \epsilon \int_{\Sigma_1} d \Sigma_1^\mu \, \xi^\nu \, T_{\mu \nu}.
\end{equation}

Now let us move to the quantum story. If $\varphi$ is a function on the surface $\Sigma_1$, let $\varphi'$ be the function on $\Sigma_2$ naturally defined as the pullback of $\varphi$ from $\Sigma_2$ to $\Sigma_1$ induced by the infinitesimal vector flow along $\xi$. (That is, $\varphi'$ is just $\varphi$ after being pushed along $\xi^\mu$.)

The difference between the path integral over $U_2$ with boundary conditions $\eval{\phi}_{\Sigma_2} = \varphi'$ and the path integral over $U_1$ with boundary conditions $\eval{\phi}_{\Sigma_1} = \varphi$ can be computed by identifying the two measures of integration over fields on $U_2$ and $U_1$ by infinitesimally shifting one of them in the tubular neighborhood of the boundary. We then find that the difference is simply given by $-\xi^\nu T_{\mu \nu}$ integrated over the boundary, which is exactly what we wanted to show:

\begin{equation}
\begin{aligned}
    &\int_{\eval{\phi}_{\Sigma_2} = \varphi'} \mD \phi \, e^{-S_{U_2}[\phi]} - \int_{\eval{\phi}_{\Sigma_1} = \varphi} \mD \phi \, e^{-S_{U_1}[\phi]} \\
    =& \int_{\eval{\phi}_{\Sigma_1} = \varphi} \mD (\phi + \delta \phi) \, e^{-S_{U_2}[\phi + \delta \phi]} - \int_{\eval{\phi}_{\Sigma_1} = \varphi} \mD \phi \, e^{-S_{U_1}[\phi]} \\
    =& - \epsilon \int_{\eval{\phi}_{\Sigma_1}  = \varphi} \mD \phi \int_{\Sigma_1} d \Sigma_1^\mu \, \xi^\nu \, T_{\mu \nu} e^{-S_{U_1}[\phi]}.
\end{aligned}
\end{equation}
Here we are using the fact that no diffeomorphism anomaly \cite{Alvarez-Gaume:1983ihn,Fujikawa:1984kq} appears  in the path integral measure, so $\mD (\phi + \delta \phi) = \mD \phi$.

\bibliography{alpha_bib.bib}

@article{Alvarez-Gaume:1983ihn,
    author = "Alvarez-Gaume, Luis and Witten, Edward",
    editor = "Salam, A. and Sezgin, E.",
    title = "{Gravitational Anomalies}",
    reportNumber = "HUTP-83/A039",
    doi = "10.1016/0550-3213(84)90066-X",
    journal = "Nucl. Phys. B",
    volume = "234",
    pages = "269",
    year = "1984"
}

@article{Fujikawa:1984kq,
    author = "Fujikawa, Kazuo and Tomiya, Mitsuyoshi and Yasuda, Osamu",
    title = "{COMMENT ON GRAVITATIONAL ANOMALIES}",
    reportNumber = "RRK-84-13a",
    doi = "10.1007/BF01575737",
    journal = "Z. Phys. C",
    volume = "28",
    pages = "289",
    year = "1985"
}

@article{Banks:2002nv,
    author = "Banks, T. and Mannelli, L.",
    title = "{De Sitter vacua, renormalization and locality}",
    eprint = "hep-th/0209113",
    archivePrefix = "arXiv",
    doi = "10.1103/PhysRevD.67.065009",
    journal = "Phys. Rev. D",
    volume = "67",
    pages = "065009",
    year = "2003"
}

@article{Mottola:1984ar,
    author = "Mottola, E.",
    title = "{Particle Creation in de Sitter Space}",
    reportNumber = "NSF-ITP-84-123",
    doi = "10.1103/PhysRevD.31.754",
    journal = "Phys. Rev. D",
    volume = "31",
    pages = "754",
    year = "1985"
}

@article{Allen:1985ux,
    author = "Allen, Bruce",
    title = "{Vacuum States in de Sitter Space}",
    reportNumber = "UCSB-TH-3-1985",
    doi = "10.1103/PhysRevD.32.3136",
    journal = "Phys. Rev. D",
    volume = "32",
    pages = "3136",
    year = "1985"
}

@article{Burges:1984qm,
    author = "Burges, C. J. C.",
    title = "{The De Sitter Vacuum}",
    reportNumber = "BRX-TH-158",
    doi = "10.1016/0550-3213(84)90562-5",
    journal = "Nucl. Phys. B",
    volume = "247",
    pages = "533--543",
    year = "1984"
}

@article{Akhmedov:2013vka,
    author = "Akhmedov, E. T.",
    title = "{Lecture notes on interacting quantum fields in de Sitter space}",
    eprint = "1309.2557",
    archivePrefix = "arXiv",
    primaryClass = "hep-th",
    reportNumber = "ITEP-TH-32-13",
    doi = "10.1142/S0218271814300018",
    journal = "Int. J. Mod. Phys. D",
    volume = "23",
    pages = "1430001",
    year = "2014"
}

@article{Kaloper:2002cs,
    author = "Kaloper, Nemanja and Kleban, Matthew and Lawrence, Albion and Shenker, Stephen and Susskind, Leonard",
    title = "{Initial conditions for inflation}",
    eprint = "hep-th/0209231",
    archivePrefix = "arXiv",
    reportNumber = "SLAC-PUB-9533, BRX-TH-505, SU-ITP-02-02",
    doi = "10.1088/1126-6708/2002/11/037",
    journal = "JHEP",
    volume = "11",
    pages = "037",
    year = "2002"
}

@article{Bousso:2001mw,
    author = "Bousso, Raphael and Maloney, Alexander and Strominger, Andrew",
    title = "{Conformal vacua and entropy in de Sitter space}",
    eprint = "hep-th/0112218",
    archivePrefix = "arXiv",
    doi = "10.1103/PhysRevD.65.104039",
    journal = "Phys. Rev. D",
    volume = "65",
    pages = "104039",
    year = "2002"
}

@article{Ng:2012xp,
    author = "Ng, Gim Seng and Strominger, Andrew",
    title = "{State/Operator Correspondence in Higher-Spin dS/CFT}",
    eprint = "1204.1057",
    archivePrefix = "arXiv",
    primaryClass = "hep-th",
    doi = "10.1088/0264-9381/30/10/104002",
    journal = "Class. Quant. Grav.",
    volume = "30",
    pages = "104002",
    year = "2013"
}

@article{Chernikov:1968zm,
    author = "Chernikov, N. A. and Tagirov, E. A.",
    title = "{Quantum theory of scalar fields in de Sitter space-time}",
    journal = "Ann. Inst. H. Poincare A Phys. Theor.",
    volume = "9",
    pages = "109",
    year = "1968"
}

@inproceedings{schomblond1976conditions,
  title={Conditions d’unicit{\'e} pour le propagateur $\Delta^1(x,y)$ du champ scalaire dans l’univers de de Sitter},
  author={Schomblond, Christiane and Spindel, Philippe},
  booktitle={Annales de l'institut Henri Poincar{\'e}. Section A, Physique Th{\'e}orique},
  volume={25},
  number={1},
  pages={67--78},
  year={1976}
}

@article{deBoer:2004nd,
    author = "de Boer, Jan and Jejjala, Vishnu and Minic, Djordje",
    title = "{Alpha-states in de Sitter space}",
    eprint = "hep-th/0406217",
    archivePrefix = "arXiv",
    reportNumber = "ITFA-2004-24, VPI-IPPAP-04-04",
    doi = "10.1103/PhysRevD.71.044013",
    journal = "Phys. Rev. D",
    volume = "71",
    pages = "044013",
    year = "2005"
}

@article{Lagogiannis:2011st,
    author = "Lagogiannis, Philip and Maloney, Alexander and Wang, Yi",
    title = "{Odd-dimensional de Sitter Space is Transparent}",
    eprint = "1106.2846",
    archivePrefix = "arXiv",
    primaryClass = "hep-th",
    month = "6",
    year = "2011"
}

@article{Kanno:2014lma,
    author = "Kanno, Sugumi and Murugan, Jeff and Shock, Jonathan P. and Soda, Jiro",
    title = "{Entanglement entropy of $\alpha$-vacua in de Sitter space}",
    eprint = "1404.6815",
    archivePrefix = "arXiv",
    primaryClass = "hep-th",
    reportNumber = "KOBE-TH-14-04, QGASLAB-14-02",
    doi = "10.1007/JHEP07(2014)072",
    journal = "JHEP",
    volume = "07",
    pages = "072",
    year = "2014"
}

@mastersthesis{Goldstein:2005re,
    author = "Goldstein, Kevin",
    title = "{de Sitter space, interacting quantum field theory and alpha vacua}",
    reportNumber = "UMI-31-74611",
    type = "Other thesis",
    year = "2005"
}

@article{Goldstein:2003ut,
    author = "Goldstein, Kevin and Lowe, David A.",
    title = "{A Note on alpha vacua and interacting field theory in de Sitter space}",
    eprint = "hep-th/0302050",
    archivePrefix = "arXiv",
    reportNumber = "BROWN-HET-1341",
    doi = "10.1016/j.nuclphysb.2003.07.014",
    journal = "Nucl. Phys. B",
    volume = "669",
    pages = "325--340",
    year = "2003"
}

@article{Polyakov:2007mm,
    author = "Polyakov, A. M.",
    title = "{De Sitter space and eternity}",
    eprint = "0709.2899",
    archivePrefix = "arXiv",
    primaryClass = "hep-th",
    reportNumber = "PUPT-2244",
    doi = "10.1016/j.nuclphysb.2008.01.002",
    journal = "Nucl. Phys. B",
    volume = "797",
    pages = "199--217",
    year = "2008"
}

@article{Melton:2023dee,
    author = "Melton, Walker and Niewinski, Filip and Strominger, Andrew and Wang, Tianli",
    title = "{Hyperbolic vacua in Minkowski space}",
    eprint = "2310.13663",
    archivePrefix = "arXiv",
    primaryClass = "hep-th",
    doi = "10.1007/JHEP08(2024)046",
    journal = "JHEP",
    volume = "08",
    pages = "046",
    year = "2024"
}

@article{Bunch:1978yq,
    author = "Bunch, T. S. and Davies, P. C. W.",
    title = "{Quantum Field Theory in de Sitter Space: Renormalization by Point Splitting}",
    doi = "10.1098/rspa.1978.0060",
    journal = "Proc. Roy. Soc. Lond. A",
    volume = "360",
    pages = "117--134",
    year = "1978"
}

@article{Chen:2024ckx,
    author = "Chen, Pisin and Lin, Kuan-Nan and Lin, Wei-Chen and Yeom, Dong-han",
    title = "{A possible origin of the $\alpha$-vacuum as the initial state of the Universe}",
    eprint = "2404.15450",
    archivePrefix = "arXiv",
    primaryClass = "gr-qc",
    month = "4",
    year = "2024"
}

@article{Hartle:1983ai,
    author = "Hartle, J. B. and Hawking, S. W.",
    editor = "Fang, Li-Zhi and Ruffini, R.",
    title = "{Wave Function of the Universe}",
    reportNumber = "PRINT-83-0937 (CAMBRIDGE)",
    doi = "10.1103/PhysRevD.28.2960",
    journal = "Phys. Rev. D",
    volume = "28",
    pages = "2960--2975",
    year = "1983"
}

@article{Einhorn:2002nu,
    author = "Einhorn, Martin B. and Larsen, Finn",
    title = "{Interacting quantum field theory in de Sitter vacua}",
    eprint = "hep-th/0209159",
    archivePrefix = "arXiv",
    reportNumber = "MCTP-02-47",
    doi = "10.1103/PhysRevD.67.024001",
    journal = "Phys. Rev. D",
    volume = "67",
    pages = "024001",
    year = "2003"
}

@article{Spradlin:2001nb,
    author = "Spradlin, Marcus and Volovich, Anastasia",
    title = "{Vacuum states and the S matrix in dS / CFT}",
    eprint = "hep-th/0112223",
    archivePrefix = "arXiv",
    reportNumber = "PUTP-2017",
    doi = "10.1103/PhysRevD.65.104037",
    journal = "Phys. Rev. D",
    volume = "65",
    pages = "104037",
    year = "2002"
}

@article{Einhorn:2003xb,
    author = "Einhorn, Martin B. and Larsen, Finn",
    title = "{Squeezed states in the de Sitter vacuum}",
    eprint = "hep-th/0305056",
    archivePrefix = "arXiv",
    reportNumber = "MCTP-03-24",
    doi = "10.1103/PhysRevD.68.064002",
    journal = "Phys. Rev. D",
    volume = "68",
    pages = "064002",
    year = "2003"
}

@article{Danielsson:2002kx,
    author = "Danielsson, Ulf H.",
    title = "{A Note on inflation and transPlanckian physics}",
    eprint = "hep-th/0203198",
    archivePrefix = "arXiv",
    reportNumber = "UUITP-01-02",
    doi = "10.1103/PhysRevD.66.023511",
    journal = "Phys. Rev. D",
    volume = "66",
    pages = "023511",
    year = "2002"
}

@article{Danielsson:2002mb,
    author = "Danielsson, Ulf H.",
    title = "{On the consistency of de Sitter vacua}",
    eprint = "hep-th/0210058",
    archivePrefix = "arXiv",
    reportNumber = "UUITP-12-02",
    doi = "10.1088/1126-6708/2002/12/025",
    journal = "JHEP",
    volume = "12",
    pages = "025",
    year = "2002"
}

@article{Collins:2003zv,
    author = "Collins, Hael and Holman, R. and Martin, Matthew R.",
    title = "{The Fate of the alpha vacuum}",
    eprint = "hep-th/0306028",
    archivePrefix = "arXiv",
    reportNumber = "CMU-HEP-03-04",
    doi = "10.1103/PhysRevD.68.124012",
    journal = "Phys. Rev. D",
    volume = "68",
    pages = "124012",
    year = "2003"
}

@article{Goldstein:2003qf,
    author = "Goldstein, Kevin and Lowe, David A.",
    title = "{Real time perturbation theory in de Sitter space}",
    eprint = "hep-th/0308135",
    archivePrefix = "arXiv",
    reportNumber = "BROWN-HET-1375",
    doi = "10.1103/PhysRevD.69.023507",
    journal = "Phys. Rev. D",
    volume = "69",
    pages = "023507",
    year = "2004"
}

@article{Danielsson:2002qh,
    author = "Danielsson, Ulf H.",
    title = "{Inflation, holography, and the choice of vacuum in de Sitter space}",
    eprint = "hep-th/0205227",
    archivePrefix = "arXiv",
    reportNumber = "UUITP-05-02",
    doi = "10.1088/1126-6708/2002/07/040",
    journal = "JHEP",
    volume = "07",
    pages = "040",
    year = "2002"
}

@article{Easther:2001fi,
    author = "Easther, Richard and Greene, Brian R. and Kinney, William H. and Shiu, Gary",
    title = "{Inflation as a probe of short distance physics}",
    eprint = "hep-th/0104102",
    archivePrefix = "arXiv",
    doi = "10.1103/PhysRevD.64.103502",
    journal = "Phys. Rev. D",
    volume = "64",
    pages = "103502",
    year = "2001"
}

@article{Easther:2001fz,
    author = "Easther, Richard and Greene, Brian R. and Kinney, William H. and Shiu, Gary",
    title = "{Imprints of short distance physics on inflationary cosmology}",
    eprint = "hep-th/0110226",
    archivePrefix = "arXiv",
    reportNumber = "CU-TP-1039, UPR-952-T",
    doi = "10.1103/PhysRevD.67.063508",
    journal = "Phys. Rev. D",
    volume = "67",
    pages = "063508",
    year = "2003"
}

@article{Easther:2002xe,
    author = "Easther, Richard and Greene, Brian R. and Kinney, William H. and Shiu, Gary",
    title = "{A Generic estimate of transPlanckian modifications to the primordial power spectrum in inflation}",
    eprint = "hep-th/0204129",
    archivePrefix = "arXiv",
    reportNumber = "CU-TP-1055",
    doi = "10.1103/PhysRevD.66.023518",
    journal = "Phys. Rev. D",
    volume = "66",
    pages = "023518",
    year = "2002"
}

@article{Collins:2005cm,
    author = "Collins, Hael and Holman, R.",
    title = "{An Effective theory of initial conditions in inflation}",
    eprint = "hep-th/0507081",
    archivePrefix = "arXiv",
    reportNumber = "UMHEP-461, CMU-HEP-05-07",
    month = "7",
    year = "2005"
}

@article{Nguyen:2017ggc,
    author = "Nguyen, K\'evin",
    title = "{De Sitter-invariant States from Holography}",
    eprint = "1710.04675",
    archivePrefix = "arXiv",
    primaryClass = "hep-th",
    doi = "10.1088/1361-6382/aae86b",
    journal = "Class. Quant. Grav.",
    volume = "35",
    number = "22",
    pages = "225006",
    month = "10",
    year = "2017"
}

@article{Joung:2006gj,
    author = "Joung, E. and Mourad, J. and Parentani, R.",
    title = "{Group theoretical approach to quantum fields in de Sitter space. I. The Principle series}",
    eprint = "hep-th/0606119",
    archivePrefix = "arXiv",
    doi = "10.1088/1126-6708/2006/08/082",
    journal = "JHEP",
    volume = "08",
    pages = "082",
    year = "2006"
}

@article{Joung:2007je,
    author = "Joung, Euihun and Mourad, Jihad and Parentani, Renaud",
    title = "{Group theoretical approach to quantum fields in de Sitter space. II. The complementary and discrete series}",
    eprint = "0707.2907",
    archivePrefix = "arXiv",
    primaryClass = "hep-th",
    doi = "10.1088/1126-6708/2007/09/030",
    journal = "JHEP",
    volume = "09",
    pages = "030",
    year = "2007"
}

@article{Chopping:2024oiu,
    author = "Chopping, Alistair J. and Sleight, Charlotte and Taronna, Massimo",
    title = "{Cosmological correlators for Bogoliubov initial states}",
    eprint = "2407.16652",
    archivePrefix = "arXiv",
    primaryClass = "hep-th",
    doi = "10.1007/JHEP09(2024)152",
    journal = "JHEP",
    volume = "09",
    pages = "152",
    year = "2024"
}

@article{Ansari:2024pgq,
    author = "Ansari, Arhum and Banerjee, Pinak and Dhivakar, Prateksh and Jain, Sachin and Kundu, Nilay",
    title = "{Inflationary non-Gaussianities in alpha vacua and consistency with conformal symmetries}",
    eprint = "2403.10513",
    archivePrefix = "arXiv",
    primaryClass = "hep-th",
    doi = "10.1007/JHEP10(2024)147",
    journal = "JHEP",
    volume = "10",
    pages = "147",
    year = "2024"
}

@article{Cotler:2023xku,
    author = "Cotler, Jordan and Strominger, Andrew",
    title = "{Cosmic ER=EPR in dS/CFT}",
    eprint = "2302.00632",
    archivePrefix = "arXiv",
    primaryClass = "hep-th",
    month = "2",
    year = "2023"
}

@article{Akhmedov:2022uug,
    author = "Akhmedov, E. T. and Kochergin, I. V. and Milovanova, M. N.",
    title = "{Isometry invariance of exact correlation functions in various charts of Minkowski and de Sitter spaces}",
    eprint = "2210.10119",
    archivePrefix = "arXiv",
    primaryClass = "hep-th",
    doi = "10.1103/PhysRevD.107.105015",
    journal = "Phys. Rev. D",
    volume = "107",
    number = "10",
    pages = "105015",
    year = "2023"
}
\bibliographystyle{jhep}

\end{document}